\documentclass[aps,pre,twocolumn,groupedaddress,10pt]{revtex4-2}

\usepackage[english]{babel}

\usepackage{amssymb} 
\usepackage{amsmath}
\usepackage{bm}
\usepackage{bbm}
\usepackage{dsfont}
\usepackage{graphicx}
\usepackage{subfigure}

\usepackage{tikz}
\usepackage{setspace}
\usepackage{rotating}
\usepackage{xr}

\begin{document}

\title{Indecision and accuracy under social information across groups sizes}
\author{Andrew M. Bate}
\email{a.m.bate@leeds.ac.uk}
\author{Charlie Pilgrim}
\author{Richard P. Mann}
\affiliation{School of Mathematics, University of Leeds, Leeds, UK}

\begin{abstract}
Observing the decisions and actions of others provides social information that can inform decisions such as whether to follow. We consider a model where agents simultaneously gather stochastic private information, each deciding once sufficiently confident. Observed decisions and indecision provide social information that triggers discrete waves of collective response: a first decision causes others to update and potentially follow, whose decisions in turn provide further social information, generating successive waves. We explore this model across a range of group sizes and report three main findings. First, social information leads to faster and more accurate decisions than individual decision-making, but agent-level accuracy is maximised at a finite optimal group size. This contrasts with the accuracy of the majority choice, which increases monotonically with the number of agents. Second, waves frequently fail to resolve collective indecision, particularly for smaller groups and when the first decision is incorrect, leaving a subgroup of agents unconvinced. Third, these remaining undecided agents are systematically biased and make less accurate subsequent decisions, with this inaccuracy growing with group size.
\end{abstract}

\maketitle

\section{Introduction}

Access to information inferred by observing the decisions and actions of others (i.e. social information) is one of many benefits from being in a group \citep{KrauseRuxtonBook2002,SumpterBook2010}. These benefits include both faster and more accurate decisions that increase with group size \citep{KingCowlishaw2007,SumpterPratt2009}, a phenomenon also found experimentally \citep{Powell1974,MagurranEtAl1985,WardEtAl2011,PacherEtAl2025}. This means that social information from being part of a larger group shifts the normal tradeoff between time and accuracy in decision-making \citep{ChittkaEtAl2009}. 

Both the accuracy and timing of decisions matter: correct decisions made too late (such as avoiding a predator) can be much more costly than an early poor decision \citep{KaoEtAl2024}. The time taken to make a decision is also potentially informative for those observing it \citep{vanDeCalseydeEtAl2014}. Humans can use response times to infer hidden preferences of others \citep{BavardEtAl2024}, whereas humans as consumers can defer choices if both choices appear unattractive, or if the choices are too complex \citep{Dhar1997} (the latter is consistent with slow/difficult information gathering, a fixed decision period and a high threshold of decision). Additionally, difficult decisions can result in indecisiveness \citep{CheekGoebel2020} and such inability to come to a quick decision (i.e. indecision) can be overcome by dynamically reducing the decision threshold \citep{MalhotraEtAl2017}.  
This means that exploring delay, deferral and indecision is of much interest.

Much of the existing modelling literature with respect to social information considers agents deciding given what others have done, either based statically using an aggregation of decisions by others \citep{KingCowlishaw2007,ArgandaEtAl2012}, or a sequence of static decisions from a set list of agents which respond in turn after seeing some or all of the prior decisions \citep[such as ][]{Perez-EscuderoDePolavieja2011,Mann2018, Mann2021, TumpEtAl2020}, sometimes with further rounds or mechanisms of information exchange to explore collective consensus decisions \citep{KaoEtAl2014,VicentePageEtAl2018}. Common in all these models is that deferring, delaying or ongoing indecision is not included, nor is timing as they are static, forced decisions.

Recent models in social decision-making have incorporated temporal dynamics using drift-diffusion models to represent each agent's private information \citep{KaramchedEtAl2020,TumpEtAl2020,TumpEtAl2024,KurodaEtAl2025}. This expands on the growing use of such models for binary choice models in exploring the timing of decisions \citep[e.g.][]{BogaczEtAl2006,FudenbergEtAl2020,SticklerEtAl2023,LinnEtAl2024}. Within this emerging literature there are two distinct conceptual approaches. The more common considers social information as an additional drift term, creating a ``social drift-diffusion" model (henceforth ``SDD model") that provides a steady ongoing stream of social information \citep{TumpEtAl2020,TumpEtAl2024,KurodaEtAl2025}. The alternative approach explicitly models social information as agents inferring the expected private information of others based on their observed decisions (or lack thereof) and decision history \citep{KaramchedEtAl2020}. This is based on the internally-consistent assumption that other agents acquire information according to the same process as oneself. These decisions are observed as discrete and instantaneous events that lead to jumps in social information; that what is seen cannot be unseen or forgotten. These jumps result in a sequence of similarly discrete responses termed `waves' (henceforth ``waves model"). A wave is constituted by the synchronous decisions of several agents responding to a newly observed piece of social information. 

In this study we use drift-diffusion models to understand the dynamics of social decision-making in groups of rational agents. In selecting between the approaches above note that SDD models are easier to compute than the waves model, but the additional social drift term needs an appropriate function and parameterisation that should be consistent with the underlying rationale around information gathering. This is more difficult to reconcile with internal consistency and agent rationality. SDD models are also Markovian: the social drift at any point of time is only based on the current decision counts and thus ignores decision order and history, whereas the waves model can consider decision history and the influence of early decisions on later decisions. SDD models may therefore be appropriate where agents lack the working memory to remember sources of prior information, or if decisions are difficult to observe quickly. As these are not the focus of this paper, we will expand instead on the waves model, where the consistent decision history, calculable social information, fewer functions and parameters, and a clear rationale for the impact of decisions (and remaining undecided) on others, provide a good model to explore indecision and deferring in groups.

An interesting feature of the waves model is that in very large groups all agents decide within two instantaneous waves after observing the first decision \citep{KaramchedEtAl2020}. However, the waves model assumes that all agents can observe everyone else instantaneously, an assumption that becomes increasingly unrealistic for large groups. For example, being within large groups can lead to restricted vision by limiting an agent's field-of-view \citep{DavidsonEtAl2021}, or agents instead focus on the behaviour of neighbours \citep[]{Herbert-ReadEtAl2011}.
Furthermore, we are interested in  mapping the predictions of the model onto the behaviour of real social groups, and many real social groups are too small for this large scale limit to be relevant. Explaining natural social decision-making through these models thus requires analysing the behaviour in small and medium group sizes.

In this paper, we explore the behaviour of a wave model of social decision-making in small and medium sized groups. We will demonstrate that social decisions are generally both faster and more accurate than individual decision-making, even for small groups. We then explore the conditions needed for agents and thus groups to come to a decision within these waves of social information, such as group size and the impact of a poor first decision. Finally, we consider the properties of those left undecided after these waves and what decisions these undecided agents ultimately come to.

\section{Model}

The waves model involves $N$ animals, people or agents (henceforth, agents) who have to decide between two options, $H_+$ and $H_-$. These agents are not aware which option is `better' and instead all agents simultaneously wait while continuously gathering private information over time until they become sufficiently confident to make an irreversible decision. This private information is stochastic and weighted towards $H_+$, which we assume is the unknown ``better option" for all agents. These agents also observe and make inferences about the (in)decisions of others near instantaneously (by updating where they expect the private information of others is given the (in)decisions they observe), until they are sufficiently confident that an option is the best; at which point they will decide accordingly and this decision is observed as social information for other undecided agents. The model/simulation ends when all agents have made a decision, either $H_+$ or $H_-$. 

\begin{figure}
    \centering
  \begin{subfigure}[][]{\includegraphics[width=0.45\textwidth]{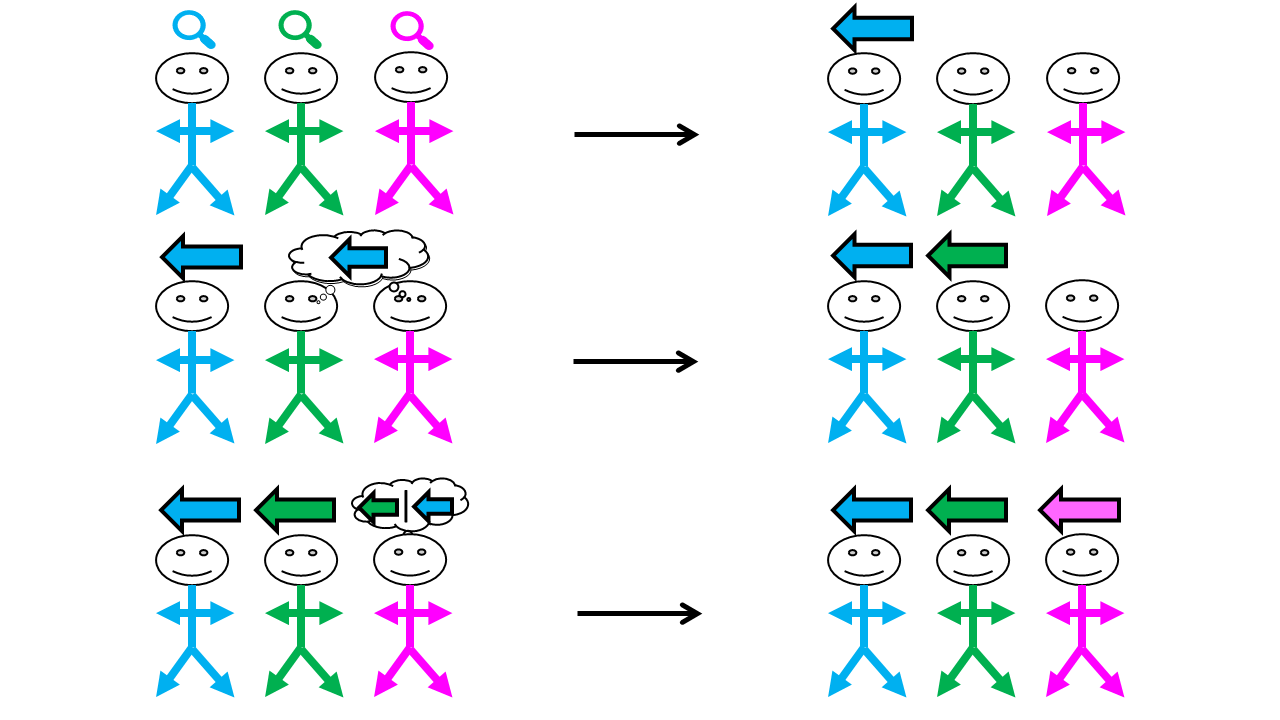}\label{fig:3AgentExampleSketch}}
  \end{subfigure}
    \begin{subfigure}[][]{\includegraphics[width=0.45\textwidth]{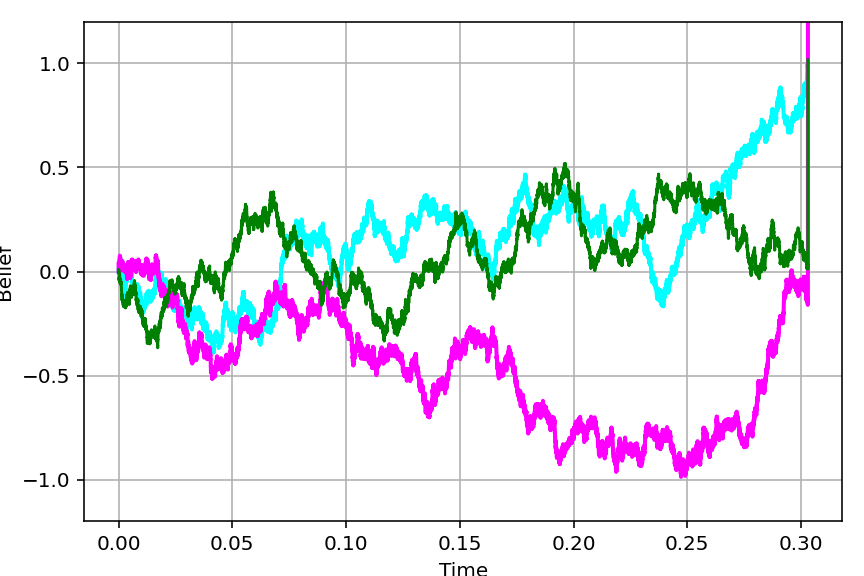}\label{fig:3AgentExampleTime}}
		\end{subfigure}     
    \caption{Example of model simulation: (a) A schematic diagram of the stages and decisions of 3 Agents, with the top row corresponding to the dynamic information gathering phase, whereas the bottom two rows correspond to two instantaneous waves of social information; (b) a corresponding time series of the beliefs of the 3 Agents. For the belief updates across the waves of social information following Blue's decision, see Table~\ref{tab:Example}. All agents have $\theta=1$, $\alpha=1$ and $D=1$.}
    \label{fig:3AgentExample}
\end{figure}

An illustrative example of the model (Figure~\ref{fig:3AgentExample}) demonstrates these processes. Three identical and initially-naive agents, Blue, Green and Pink, acquire private information over time, leading to their beliefs evolving as a drift-diffusion process (Figure~\ref{fig:3AgentExampleTime}). After some time, Blue's belief reaches a certainty threshold (here taken to be $\text{Belief}=1$), at which point he irreversibly selects option $H_+$. At the same instant, Green and Pink observe this decision, and know that Blue has $\text{Belief}=1$: this is new information which they add to their observations, resulting in the discrete jumps seen in their Belief values. This immediately takes Green beyond their certainty threshold and so Green also irreversibly chooses $H_+$. This is the first `wave' of the social response. Pink's value of Belief initially remains below 1 based on the social information from Blue's decision alone. However, Pink can now observe that Green also chose $H_+$, and can thus infer that Green had a value of $\text{Belief} > 0$ just before Blue decided (since adding 1 to their Belief pushes over their threshold). This is additional social information that further increases Pink's value of Belief above 1, creating a second `wave' in which Pink also chooses $H_+$ and the decision is complete for all agents.

Exactly how each agent's private information evolves over time, and how much social information is provided by each decision, is determined by the precise formulation of the drift-diffusion model, which we describe below. We will refer back to this illustrative example to help demonstrate the model. For ease of language, private belief/information refers to the value from the drift-diffusion process; and that decisions are based on total belief, the sum of private belief and acquired social information.

\subsection{Model: Dynamic information gathering phase}
We define the Belief $y_i$ of agent $i$ to be the log-likelihood ratio assigned by that agent to whether $H_+$ or $H_-$ are true, conditioned on their private information $\bm{\xi_i}$:
$$y_i = \log \frac{P(\bm{\xi_i} \mid H_+)}{P(\bm{\xi_i} \mid H_-)}.$$
We assume that each agent continuously gathers private information, such that their belief evolves according to a drift-diffusion model \citep[used in several decision models, e.g.][]{TumpEtAl2020,SticklerEtAl2023,TumpEtAl2024}. This is based on the assumption that private information is formed of a steady stream of small independent noisy measurements  \citep{SuppMat2026} and results in the following Stochastic Differential Equation (SDE):
\begin{equation}
    dy_i= \alpha dt + \sqrt{2 D} dW_i \label{eq:SDE}
\end{equation}
where $dW_i$ are increments of independent standard Wiener Processes, $\alpha$ is the drift and $D$ is the diffusion coefficient. These parameters are the same for all agents, meaning they gather information of equal reliability at the same rate. This also assumes that everyone shares the same `correct' choice and that, without loss of generality, this is $H^+$; thus every agent has a positive drift term (`$+\alpha dt$'). Agents know that the correct choice is the same for everyone, but we stress that the agents themselves do not know which choice is correct. They therefore need to consider both the cases where $H_+$ is true and where $H_-$ is true (i.e. where every agent has a negative drift term, `$-\alpha dt$'). Additionally, we assume all agents are unbiased at the start, thus $y_i(0)=0$. 

Agents make a decision if the resulting accumulated information is sufficiently in favour of $H_+$ or $H_-$. This means that agent $i$ will decide once their belief breaches a decision threshold at $\pm \theta$  (i.e. if $y_i(t)\geq \theta$, agent $i$ will decide $H_+$, and likewise if $y_i\leq -\theta$, agent $i$ will decide $H_-$)). The implication of this rule is that agent $i$ being undecided at time $t$, together with the assumption that decisions are irreversible, means that agent $i$ was undecided for all prior times too. 

\begin{figure}
    \centering
    \begin{subfigure}[][]{\includegraphics[width=0.45\textwidth]{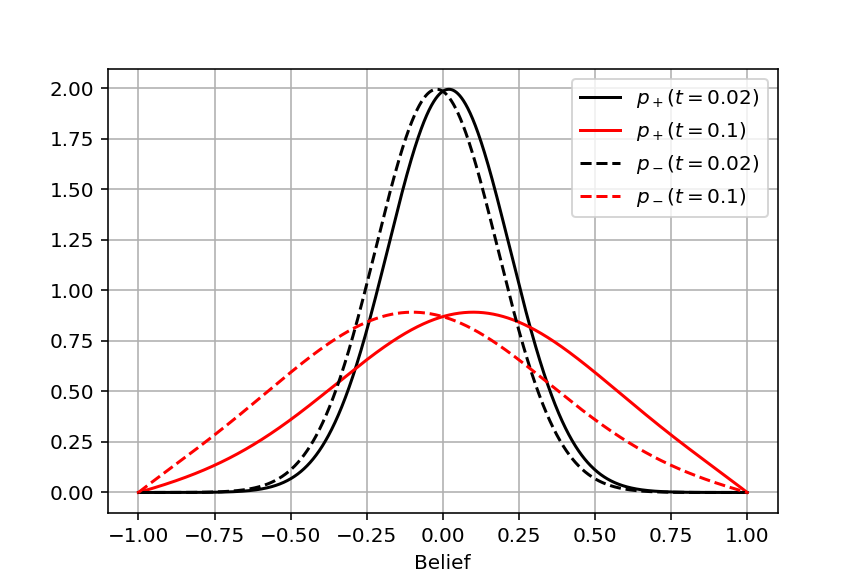}\label{fig:p(t)}}
	\end{subfigure}
    \begin{subfigure}[][]{\includegraphics[width=0.45\textwidth]{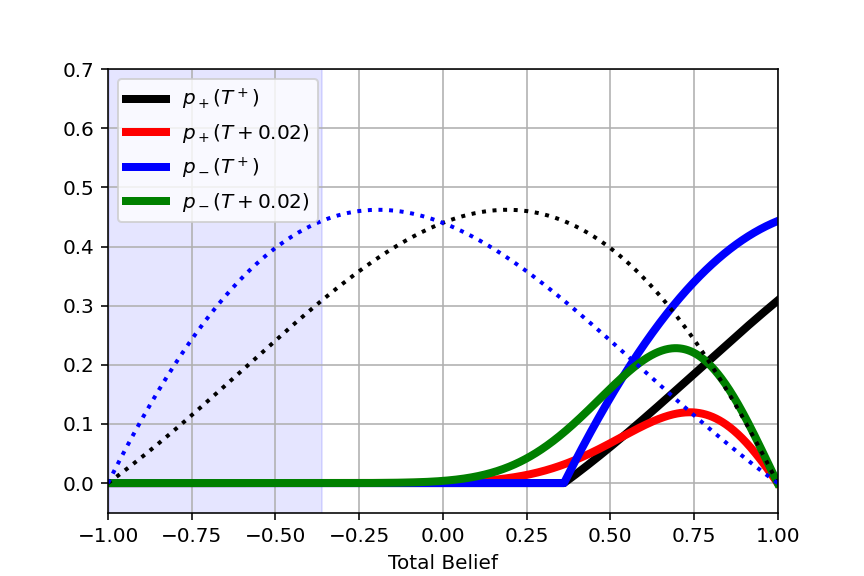}\label{fig:p_(T+t)}}
	\end{subfigure}
    \caption{Various profiles of $p_+(t)$ and $p_-(t)$ as a function of private information. (a) $p_+(t)$ (solid lines) and $p_-(t)$ (dashed lines) for various values of $t$ before the first decision. (b) $p_+$ and $p_-$ for undecided agents both at reset immediately after the waves process is finished ($T^+$) and some time afterwards ($T+0.02$), using the timings and shiftings of the  Illustrative example (Table 1)). In (b), the shaded region is the interval of indecision after the waves, the dotted lines represent $p_\pm(T)$ at time of decision; restricting $p_\pm(T)$ to the interval of indecision and shifted by the social information received  results in $p_\pm(T^+)$, the total belief distribution at the restart.  
    }
    \label{fig:p(T)}
\end{figure}

Equation~(\ref{eq:SDE}) leads to two probability density functions, $p_{i+}(x,t)=P(y_i(t)=x|H_+)$ and $p_{i-}(x,t)=P(y_i(t)=x|H_-)$; these correspond to worlds where $H_+$ or $H_-$ are true, respectively. These probability density functions evolve using the following PDE:
\begin{equation}
\frac{\partial p_{i\pm}}{\partial t}=\mp\alpha\frac{\partial p_{i\pm}}{\partial x} +D\frac{\partial^2 p_{i\pm}}{\partial x^2}, \label{eq:p}
\end{equation}
with an unbiased initial condition ($p_{i\pm }(x,0)=\delta(x)$, where $\delta(x)$ is the Dirac delta function) and destructive boundary conditions corresponding to irreversible decisions when agents hit thresholds ($p_{i\pm}(\pm\theta,t)=0$). Analytic approximations of Equation~\ref{eq:p} are available for $p_+(t)$ and $p_-(t)$ using the Method of Images \citep[][]{SuppMat2026}. Figure~\ref{fig:p(t)} demonstrates that these two distributions are mirror images of each other, which flatten and shrink over time with peaks moving in the direction of their underlying drift.

\subsection{Model: Instantaneous social information phase}

Once the first agent has made a decision, all remaining agents enter a sequence of discrete waves of social information. During this phase, agents temporarily stop gathering private information and update their beliefs only by observing the decisions (or continued indecision) of others. Each wave corresponds to a refinement, or lack thereof, of the set of private beliefs that are consistent with the observed social behaviour.

At the time of the first decision (say at time $T$), any agent who has not yet decided must have a private belief in the interval:
$$I_0 =(-\theta,\theta)$$
the initial \emph{interval of indecision}. This interval represents the set of private beliefs consistent with the agent having remained undecided up to time $T$, given irreversible decisions at thresholds $\pm \theta$.
Crucially, all agents are assumed to know the model parameters and inference rules, and therefore all agents share a common understanding that any undecided agent's belief lies within $I_0$. Each subsequent wave of social information may restrict this interval further. Let $I_k$ denote the interval of indecision for undecided agents at the start of wave $k$.

Social information is acquired by observing whether other agents decide or remain undecided in each wave. This information is quantified using the log‑likelihood ratio (LLR) between the worlds in which $H_+$ or $H_-$ is true. By assumption, the shared parameters of all agents are common knowledge, so each can calculate $p_+(t)$ and $p_-(t)$ representing the distribution of beliefs held by others (formal expressions are given in Appendix~\ref{app:SocInfo}).
\begin{itemize}
    \item {\bf Observing a decision.} If an agent decides $H_+$ (respectively $H_-$) in wave $k$, this reveals that their total belief has crossed $+\theta$ (respectively $-\theta$). In wave 0, this provides social information of $\pm \theta$ to all observers; more generally at wave $k$ it reveals that their private belief now lies outside the current interval $I_k$ (moreover, this private belief lies within $I_{k-1}\setminus I_{k}$).

    \item {\bf Observing indecision.} If an agent remains undecided in wave $k$, this reveals that their private belief lies within the current interval $I_k$, which also provides social information, but typically of smaller magnitude and opposite sign to that obtained from a decision.
\end{itemize}

Because all agents observe all (in)decisions, each undecided agent in wave $k$ aggregates the social information obtained from every other agent's behaviour in wave $k-1$

Consider the first decision made at time $T$. This decision provides social information of $\pm \theta$ to every other agent (Appendix equations 6-7). Each undecided agent then compares this social information with their private belief $y_i(T)$. If the combined information crosses a decision threshold (i.e. $y_i(T)+\theta>\theta$ for $H_+$, $y_i(T)-\theta<-\theta$ for $H_-$), the agent decides immediately in wave 1. Otherwise, the agent remains undecided. That is, agent $i$  will decide to follow an $H_+$ ($H_-$) first decision if their private information is $y_i(T)\in(0,\theta)$ ($y_i(T)\in(-\theta,0)$) and will remain undecided if their private information is $y_i(T)\in(-\theta,0)$ ($y_i(T)\in(0,\theta)$). For example, in our illustrative example (middle row of Figure~\ref{fig:3AgentExampleSketch}, wave 1), Green decided to follow Blue as Green's private information was positive  ($y_G\in (0,1)$), whereas Pink remains undecided as Pink's private information was negative  ($y_P\in (-1,0)$) (Table~\ref{tab:Example}).

As a result, wave 1 splits the original interval of indecision $I_0$ into two sub‑intervals: one corresponding to agents who decide in wave 1, and one corresponding to agents who remain undecided. The latter defines a new interval of indecision $I_1 \subseteq I_0$.

If some agents remain undecided after wave 1, each undecided agent observes all the agents that decided to follow and all the other agents that still remain undecided. This social information is gathered using inference around the private information each agent has, knowing that their private information is within a set interval. From the Appendix~\ref{app:SocInfo}, we have that the social information gathered from the observing agent $j$ following the first $H_+$ decision is:
\begin{equation}
   S^{j+}_{i1}(T)= \frac{D}{\alpha}\ln\left( \frac{\int^{\theta}_{0} p_+(x,T) dx}{\int^{\theta}_{0} p_-(x,T) dx}\right)=c(T),
\end{equation}
whereas if agent $j$ remained undecided following $H_+$ then the social information is $S^{j0}_{i1}(T)=-c(T)$. For our illustrative example, the social information for Pink observing that Green decided $H_+$ in wave 1, and thus inferring that Green actually has positive private information is: 
\begin{align}
    S^+_{G1}&=\ln \left(\frac{P(\text{Green picks $H_+ |$ Blue picks $H_+$ \& $H_+$})}{P(\text{Green picks $H_+ |$ Blue picks $H_+$ \& $H_-$})}\right)\nonumber\\
    &=\ln \left(\frac{P(\text{$y_G\in (0,1)|y_G\in (-1,1)$ and $H_+$)})}{P(\text{$y_G\in (0,1)|y_G\in (-1,1)$ and $H_-$)})}\right)\nonumber\\
    &=
    \ln \left(\frac{\int^{1}_{0} p_+(x,T) dx}{\int^{1}_{0} p_-(x,T) dx}\right)\approx0.361,
\end{align}
where $T\approx0.303$.

Since an undecided agent $i$ sees what all agents did in wave 1, the social information gathered in total from observing wave 1 is the sum of the social information from each of these actions: 
$S^{\Omega}_{i1}=\sum_{j\in D_{+1}} S^{j+}_{i1}(T)+\sum_{j \in U_{1i}} S^{j0}_{i1}(T)$ (if $H_+$, whereas if the first decision was $H_-$, then $S^{\Omega}_{i1}=\sum_{j\in D_{-1}} S^{j-}_{i1}(T)+\sum_{j \in U_{1i}} S^{j0}_{i1}(T)$), where $D_{+1}$ ($D_{-1}$) and $U_{1i}$ are the sets of agents that decided $H_+$ ($H_-$) or remained undecided after wave 1 (excluding agent $i$), respectively. This social information is added to the social information acquired in the previous wave, and decisions are then made using this aggregated social information and their private information (e.g. remain undecided if $y_i(T)+ S^{\Omega}_{i 1 }+S^{1\pm}_{i0} \in(-\theta,\theta)$, where $S^{1\pm}_{i0}=\pm \theta$ is the social information from the first decision from agent 1).

As long as some agents remain undecided, further waves may occur. At the start of wave $k$, each undecided agent knows that every other undecided agent's private belief lies within $I_{k-1}$. By observing which agents decide or remain undecided in wave $k-1$, each agent updates their belief and infers a new interval $I_k$. Using terminology inspired by epistemic defeaters \citep{PollockBook1974,Melis2014}, we categorise the aggregated social information from a wave into one of four qualitatively distinct outcomes:
\begin{enumerate}
    \item {\bf Supporting information:} $I_k$ shrinks in the direction of the original decision, potentially causing more agents to follow in the next wave.
    \item {\bf Overriding information:} $I_k$ shrinks in the opposite direction, potentially causing agents to make the opposite decision.
    \item {\bf Overwhelming information:} The interval shrinks to nothing ($I_k= \varnothing$), so all remaining undecided agents must decide in the next wave in the direction of the shrinking. An extreme version of the first two outcomes.
    \item {\bf Undermining information:} The interval does not shrink ($I_k = I_{k-1}$) as new information reverses some (but not all) of the prior social information (or is zero). No further social information can be inferred (e.g. if both $y(T)\in(-\theta,\theta)$ and $y(T)+\theta\in(-\theta,\theta)$ are known to be true by previous actions, then $y+\frac{\theta}{2}\in(-\theta,\theta)$ must also be true).
\end{enumerate}
Outcomes 1 and 2 may lead to further waves. The waves phase ends with either outcome 3 or 4. That is, either all agents have made a decision, or no new social information can be obtained because the interval of indecision no longer shrinks. In this last case undecided agents must resume gathering private information, as described in the next section. Table~\ref{tab:Wave2Size} demonstrates what social information is need for each outcome in wave 1.

In our illustrative example, Green's decision to follow Blue provided Pink after wave 1 with Supporting information, which shrinks the interval of indecision from $I_1=(-1,0)$ to $I_2=(-1,-0.361)$. Pink's private belief was outside $I_2$ (because $y_P(T)\in(-0.361,0)$) so Pink now decides to follow the earlier $H_+$ decisions. However, if instead Pink's private belief was in $I_2$ (i.e. $y(T)\in(-1,-0.361)$, Pink would have remained undecided, with no remaining sources of social information available and so would restart the dynamic information gathering.

\subsection{Model: Return to dynamic information gathering phase}. 

With the wave phase ended there are one of three scenarios possible: (i) no undecided agents, (ii) 1 undecided agent, and (iii) 2 or more undecided agents. If (i), then the simulation ends as every agent has already come to a decision. If (ii), then the sole undecided agent will restart collecting private information as per the drift-diffusion model until their total information (i.e. the sum of the private information gathered and total social information) hits their threshold to decide ($\theta$ for $H_+$, and $-\theta$ for $H_-$), at which point there will be no remaining undecided agents and the simulation ends. In case (iii), where two or more agents remain undecided after the waves phase, the dynamics are more subtle. These agents resume gathering private information, but with three important differences compared to the initial phase:
\begin{itemize}
\item {\bf Biased and truncated beliefs}. Each undecided agent's private belief is known to lie within the final interval of indecision $I_k$ produced by the waves phase, resulting in private belief distributions ($p_+(x)\mathbbm{1}_{I_k}(x)$ and $p_-(x)\mathbbm{1}_{I_k}(x)$) that are truncated and biased. These distributions become the initial condition for the resumed drift–diffusion process (Figure \ref{fig:p_(T+t)}).
\item {\bf Social information from new decisions}.
When an undecided agent eventually makes a decision after the restart, this decision carries social information for the other undecided agents. This new social information needs to filter out what is already known about the decision, both prior information about the deciding agent's private information  and all the social information that the deciding agent has received previously. Consequently, the social information from a post‑restart decision is relatively small unless the decision is against prior bias.  
\item {\bf Social information from indecision}.
With a biased initial condition, $p_+$ and $p_-$ no longer evolve symmetrically over time \citep[][]{SuppMat2026}. This raises the prospect of social information gathered in a time interval based on the lack of decision.
\end{itemize}
Despite these complications, the resumed dynamics retain the same qualitative structure as the original process. A new ``first decision'' among the remaining undecided agents eventually occurs, triggering another sequence of waves of social information. Following this new first decision, the same principles as previous apply: the first decision provides social information that leads to refinements of the interval of indecision via waves of decisions and indecisions, which terminate either when everyone has come to a decision or the waves of new social information are exhausted leading to another return to dynamic information gathering. The full mathematical details of this phase are considered within \citep{SuppMat2026}.

\begin{table}[bt]
    \centering
    \begin{tabular}{|c|c|c|c|c|}
    \hline
     Wave & Time & $y_{B}$ & $y_{G}$ & $y_{P}$  \\
     \hline
     0th & 0.303 & 1.00 ($H_+$) & 0.0160 & -0.135 \\
     1st & - & - & 1.02 ($H_+$) & 0.865 \\
     2nd & - & - & - & 1.23 ($H_+$)\\ \hline 
     
\end{tabular}
    \caption{The evolution of total beliefs (private plus social) during the instantaneous waves phase of the simulation in Figure~\ref{fig:3AgentExample}. }
    \label{tab:Example}
\end{table}

\section{Results}
\subsection{Simulation parameters}
For our focal analyses we assume all agents have the same underlying parameter values. For all figures within this paper we will use $\alpha=D=1$ and $\theta =1$ (matching values used in \cite{KaramchedEtAl2020}). Choosing $\alpha =D$ corresponds to assuming that agents' beliefs are well-founded based on their private information; e.g. an agent that assigns a probability of 90\% to $H_+$ will be correct 90\% of the time. Varying the common value of $\alpha$ and $D$ simply represents a scaling of the arbitrary time scale. The certainty level for decisions set by $\theta =1$ implies that an individual choosing based only on their private information will correctly chooses $H_+$ approximately $73\%$ of the time (since the underlying drift-diffusion model in Equation \ref{eq:SDE}  has the property $P(\text{choose $H_+$})=\frac{1}{1+\exp\left(-\frac{\alpha\theta}{D}\right)}$ \citep{BogaczEtAl2006}; a value consistent with the accuracy of private decisions of humans within a ``stay-or-escape" decision experiment \citep{TumpEtAl2020}), which means that individuals are good at choosing $H_+$ in reasonable time, but that there is still much room for improvement and thus value in gathering social information. In the Supplementary Material \citep{SuppMat2026}, equivalent figures are available where each of these parameters are changed one-by-one to $\theta=0.2$, $\alpha=0.5$ and $D=0.5$. These values correspond to accuracy of $55\%$ \citep[][]{WardEtAl2011} (the approximate decision accuracy found in an experiment involving mosquitofish and a cryptic predator), $62\%$ and $88\%$, respectively. We will consider group sizes ($N$) covering all integers from 1-20, as well as 25, 30, 35, 40, 50, 60, 70, 80, 90 and 100 (except Figure~\ref{fig:Accuracy500} where values up to 500 are included); this allows us to consider the range of small to medium sized groups. Unless otherwise stated, results will be based on 10000 simulations for each value of $N$.

\subsection{Time and accuracy}
Theory and experimental evidence suggest that decisions are both faster and more accurate in groups. Here we quantify these improvements in the waves model across the group sizes we study.

\begin{figure}
    \centering
    \includegraphics[width=0.45\textwidth]{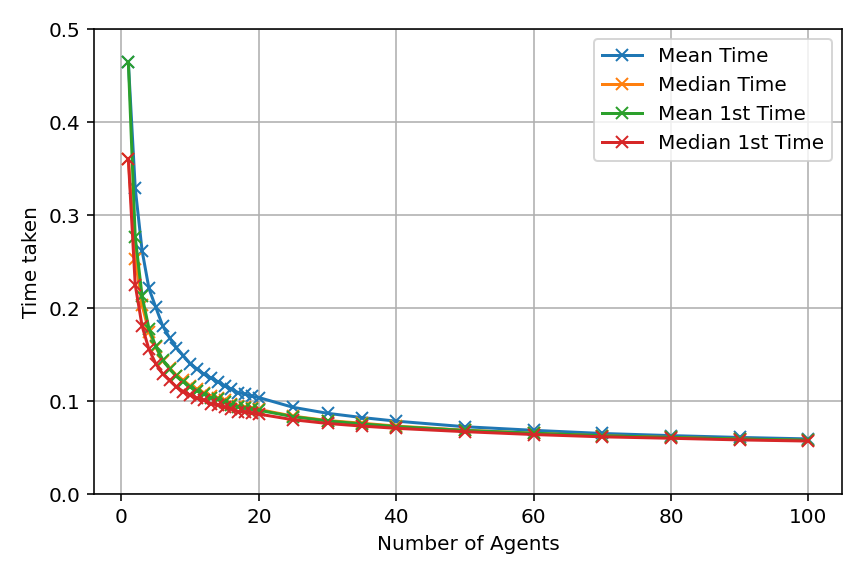}
    \caption{Average (mean and median) times for both the first decision in each simulation and of all agents, across various group sizes ($N$). 
    }
    \label{fig:Time}
\end{figure}

In the absence of social information (i.e. $N=1$), agents have a mean decision time of around 0.47, and median decision time of 0.36 (Figure~\ref{fig:Time}), with this mean time matching the analytic result ($\tanh(0.5)$) within \citet{BogaczEtAl2006} . Including social information, these decision times decline dramatically as the group size increases, with both measures between 0.11 and 0.14 by groups of 10 and down to around 0.07 (both mean and median) for groups of 100. Additionally, typical decision times converge with first decision times as groups get larger suggesting first decisions and the resulting waves become more decisive. This means that decisions are much quicker for agents within larger groups. 

\begin{figure}
    \centering
        \begin{subfigure}[][]{\includegraphics[width=0.45\textwidth]{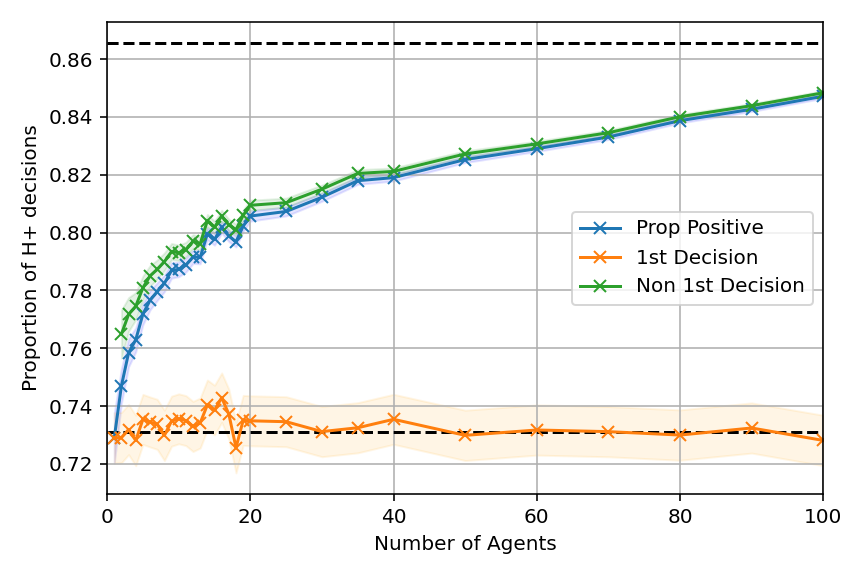}\label{fig:AccuracyTheta1}}
		\end{subfigure}   
          \begin{subfigure}[][]{\includegraphics[width=0.45\textwidth]{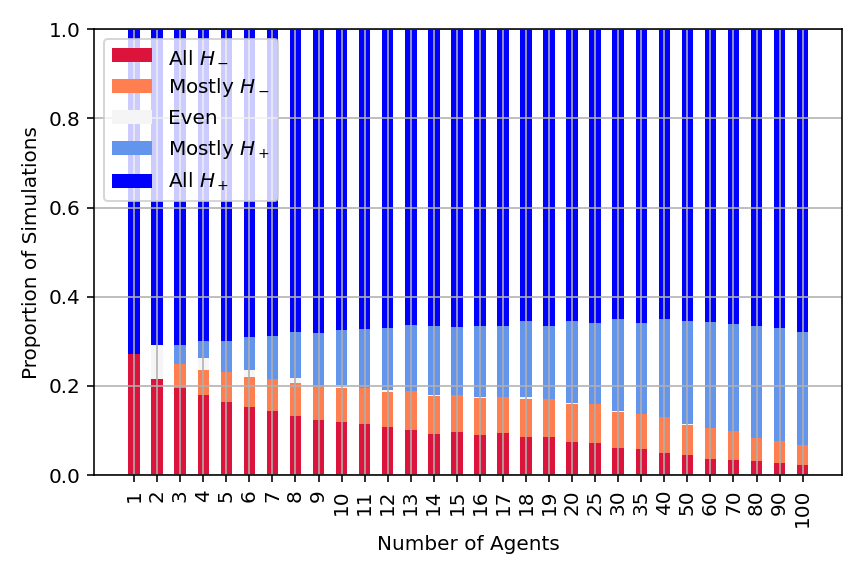}\label{fig:AccuracyGroupTheta1}}
		\end{subfigure}       
    \caption{Accuracy (proportion of $H_+$ decisions) by (a) agents and (b) the group as a whole. Here, ``mostly" means one or more agents disagree with the majority of the group after everyone comes to a decision, which is typically the majority disagreeing with the first decision. Dashed lines in (a) represent analytic results for single decisions  and the benchmark for infinite groups (approximately 73.1\% and 86.6\%, respectively).
    }
    \label{fig:Accuracy}
\end{figure}

In terms of accuracy, we see that first/fastest agents decide $H_+$ in the same proportion as the analytic results of 73\% (yellow lines, Figure~\ref{fig:AccuracyTheta1}). The accuracy for all other agents beats this 73\% and gets progressively better with group size (green lines, Figure~\ref{fig:AccuracyTheta1}). This should plateau around 87\% for very large $N$, using the argument \citep{KaramchedEtAl2020} that an $H_+$ first decision will lead to everyone following in two waves (73\% of the time) and an $H_-$ decision first decision leads to just under half following $H_-$ in wave 1 with the rest deciding $H_+$ in wave 2 $(\approx 0.5\times27\%)$. Since we find that groups as small as 15 have 80\% accuracy (and thus half way between the singleton and the near infinite group), we find that much of the accuracy benefits can be gathered from modest sized groups, and that increasing group size has diminishing accuracy benefits. However, the accuracy overshoots the accuracy of an infinite group, suggesting that near infinite group does not provide the maximum accuracy and instead it is consistently beaten by sufficiently large groups. This overshooting is actually shown in Figure~\ref{fig:Accuracy500} (in Appendix), where groups over $N\approx 200$  have higher accuracy than infinite groups under default parameters, whereas for $D=0.5$ this benchmark accuracy is beaten by even medium-sized groups \citep[over $N\approx 20$, Figure~3(e) ][]{SuppMat2026}. This means that infinite sized groups do not provide optimal accuracy for agents and thus there is an optimal finite group size for accuracy since inaccuracy within large groups is dominated by agents rushed into following an $H_-$ decision in wave 1 (Appendix~\ref{app:FiniteOptimal}). Furthermore, this suggests that groups of agents with high individual accuracy (i.e. large $\frac{\alpha\theta}{D}$) are more likely to exceed the accuracy of an infinite group.  

With respect to the group, simulations where everyone decides $H_-$ decrease steadily as $N$ increases. A similar but weaker pattern occurs for agreement for $H_+$, although this appears to start reversing for larger $N$ (Figure~\ref{fig:AccuracyGroupTheta1}). For unanimity to occur, the group must agree with the first decider (which is not a certainty for $N>1$). However, the social information for larger $N$ becomes consistently positive, reducing the likelihood that the group will disagree with an $H_+$ first decider and increasing the likelihood the group disagrees with an $H_-$ first decider.

\subsection{When agents and groups decide}
As illustrated above, larger groups have, on average, faster and more accurate decisions; whereas the agents in large groups have faster decisions with decision accuracy increasing until some optimal group size is reached.. In the wave model, as $N \rightarrow \infty$ decisions tend to occur at arbitrarily short time scales and with a well established limiting accuracy per agent \cite{KaramchedEtAl2020}. However, our core interest is in the dynamics of biologically-relevant small and medium sized groups, and in particular the time scales and dynamics of resolving collective indecision. In this section we analyse when and why waves succeed or fail in resolving indecision across these group sizes.

To understand when waves lead to further decisions rather than collapse into indecision, we first quantify the social information conveyed by a single (in)decision in wave 1. We find that this social information is surprisingly consistent despite the wide range of decision times. In particular, the typical social information from  each (in)decision in very small groups is around 37\% of a first decision, whereas increasing the group size to 100 agents only decreases this to around 28\% and even the fastest agent (of 100000) only drops this social information to 20\%. 

With this consistency in social information per (in)decision, the aggregated social information from all the (in)decisions observed from wave 1 (excluding the first decider and the observing undecided agent) is also relatively consistent. This means that the composition (especially the size and sign of the majorities) for the different scenarios of aggregated social information in Table~\ref{tab:Wave2Size} are relatively consistent. For example, an undecided agent seeing a slim majority (up to 2 or 3, depending on how quick the first decision was) of agents deciding to not follow in wave 1 gives aggregate social information that contradicts, but is weaker than the first decision. This leads to ``Undermining" social information that results in no shrinking of the interval of indecision, leading to no further social information from further waves and thus all undecided agents will need to start gathering private information again. Conversely, large (in)decision majorities (over say 6 to 8, depending on how quick the first decision was) lead to ``Overwhelming" social information, where the interval of indecision disappears and everyone must decide with this social information, irrespective to the original decision. Majorities between these extremes can be overwhelming if supporting the original decision, or override the original if opposing; the former leading to everyone deciding in wave 2, whereas the latter will either lead to further waves or end in wave 2 if everyone decides.

\begin{figure}
    \centering         
            \begin{subfigure}[][]{\includegraphics[width=0.45\textwidth]{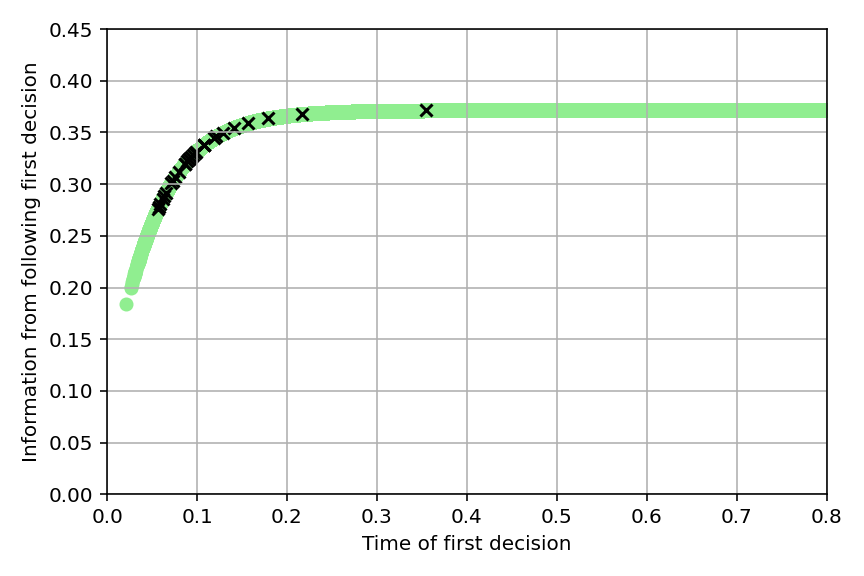}\label{fig:FirstDecisionTiming100000}}
		\end{subfigure} 
          \begin{subfigure}[][]{\includegraphics[width=0.45\textwidth]{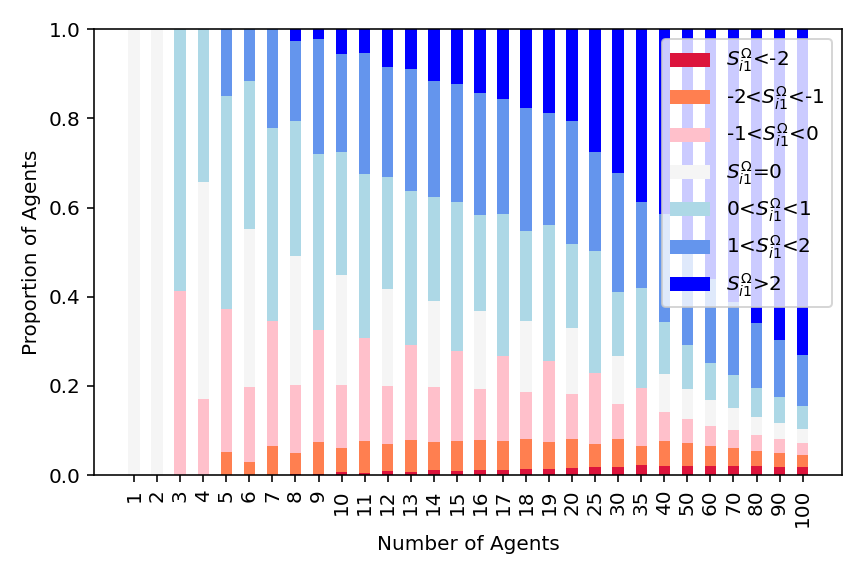}\label{fig:Wave1SocInfoCalc}}    
		\end{subfigure}       
    \caption{(a) Social Information gathered from observing an agent's (in)decision in wave 1 by time of decision as a function of time, (b) probability of total social information from (in)decisions from wave 1 in within key value in Table~\ref{tab:Wave2Size}. (a) is generated by 100000 single agent simulations for time of decision, with 1000 samples of $N$ agents then taken to gather with their median time for agents (black crosses, going from top right ($N=1$) to bottom left ($N=100$)). (b) takes these samples and together with the binomial calculations of number of deciding and undecided agents to gather the total social information within wave 1 (denoted $c$ here). 
    }
    \label{fig:Combinatorics}
\end{figure}

\begin{table*}[]
    \centering
    \begin{tabular}{|c|c|c|}
    \hline
       Social Information & $H_+$ & $H_-$ \\ \hline
       $S^{\Omega}_{i1}<-2\theta$  & Overriding \& Overwhelming  & Supporting \& Overwhelming \\ 
       $S^{\Omega}_{i1}\in (-2\theta,-\theta)$  & Overriding & Supporting \& Overwhelming \\
       $S^{\Omega}_{i1}\in (-\theta,0)$  & Undermining & Supporting \\
         $S^{\Omega}_{i1}=0$ & Neutral & Neutral \\
         $S^{\Omega}_{i1}\in (0,\theta)$  & Supporting & Undermining \\
         $S^{\Omega}_{i1}\in (\theta,2\theta)$  & Supporting \&  Overriding &  Overriding \\
         $S^{\Omega}_{i1} >  2\theta$ & Supporting \& Overriding & Overriding \& Overwhelming \\ \hline
    \end{tabular}
    \caption{Key to Figure \ref{fig:Wave1SocInfoCalc} around the social information gathered in wave 1. 
    }
    \label{tab:Wave2Size}
\end{table*}

\begin{figure*}
    \centering
    \begin{subfigure}[][]{\includegraphics[width=0.3\textwidth]{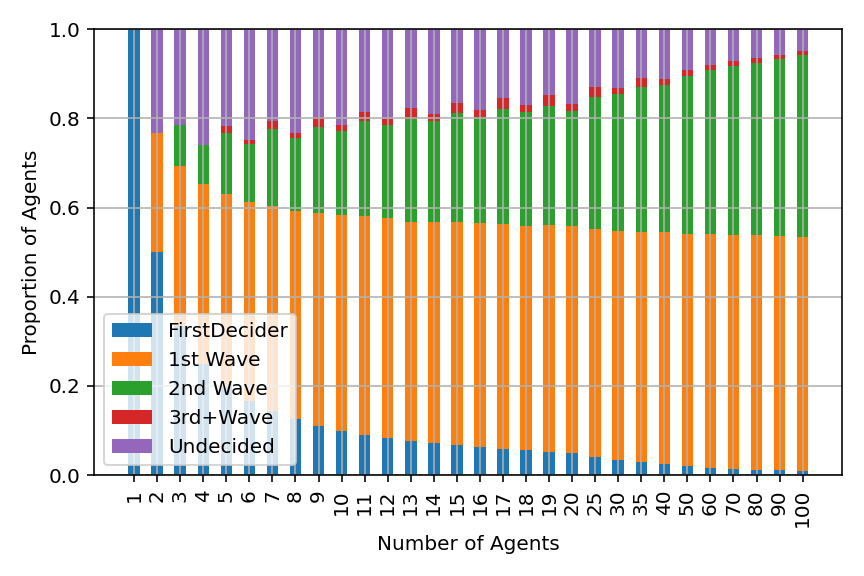}\label{fig:BarchartWaveAgent}}
		\end{subfigure}
    \begin{subfigure}[][]{\includegraphics[width=0.3\textwidth]{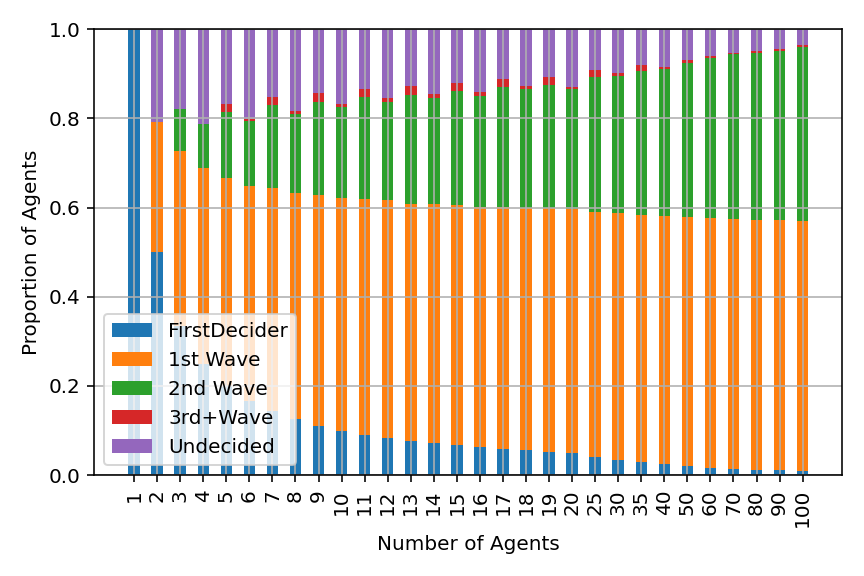}\label{fig:BarchartWaveAgentPlus}}
		\end{subfigure}
        \begin{subfigure}[][]{\includegraphics[width=0.3\textwidth]{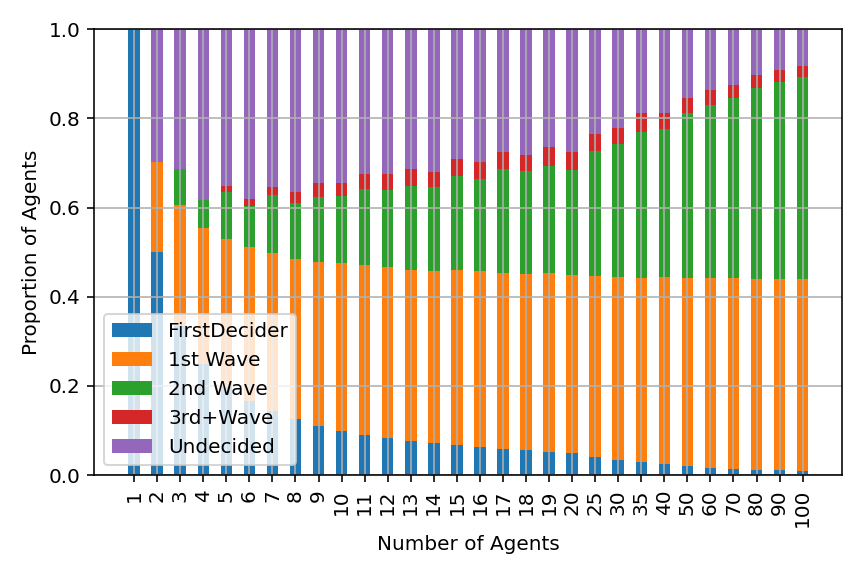}\label{fig:BarchartWaveAgentMinus}}
		\end{subfigure}
    \\
         \begin{subfigure}[][]{\includegraphics[width=0.3\textwidth]{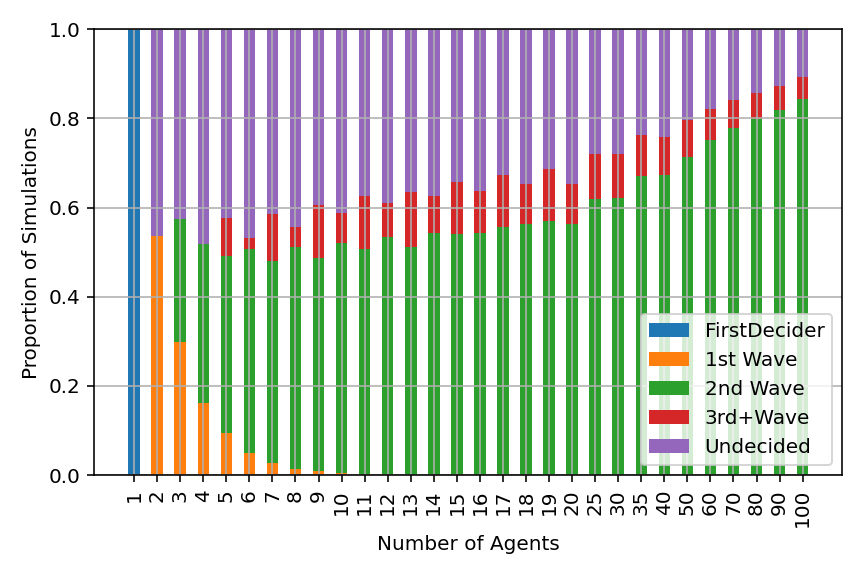}\label{fig:BarchartWaveSeed}}
		\end{subfigure}  
        \begin{subfigure}[][]{\includegraphics[width=0.3\textwidth]{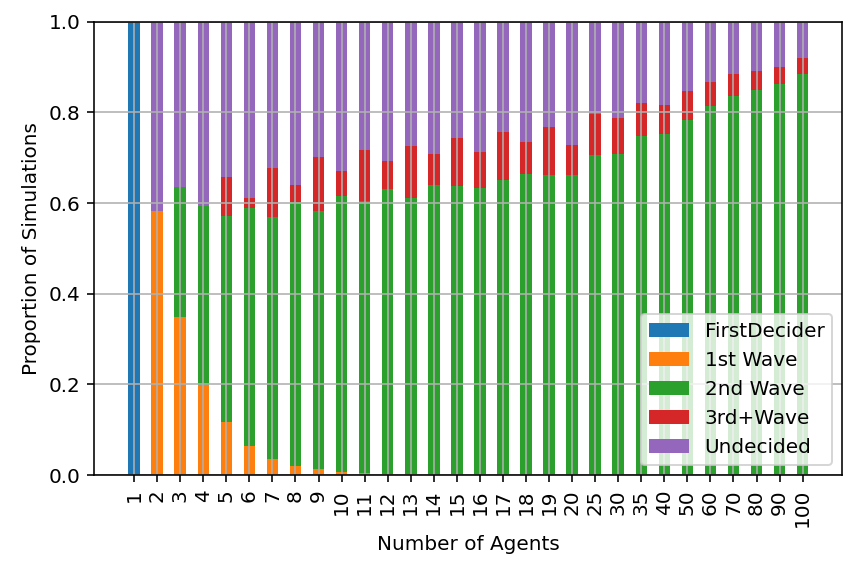}\label{fig:BarchartWaveSeedPlus}}
		\end{subfigure}  
        \begin{subfigure}[][]{\includegraphics[width=0.3\textwidth]{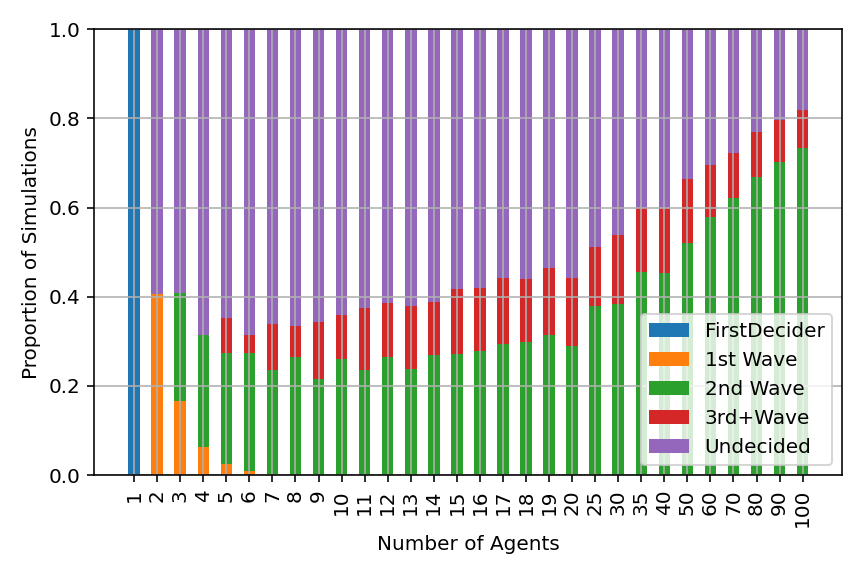}\label{fig:BarchartWaveSeedMinus}}
		\end{subfigure}          
    \caption{Proportion of (a),(b),(c) agents by when they come to a decision (d),(e),(f) simulations by when the last agent comes to a decision. (a),(d) consider all simulations, whereas (b),(e) consider only simulations with an $H_+$ first decision and (c),(f) considers only simulations with an $H_-$ first decision.}
    \label{fig:BarchartWave}
\end{figure*}

Figure~\ref{fig:BarchartWave} demonstrates that the illustrative example (Figure~\ref{fig:3AgentExample}) is not unique, with around 22\% of agents remaining undecided within groups of three, with 45\% of groups of three having one or more undecided agents; a phenomenon still fairly common even for large groups. In between deciding by wave 2 and remaining undecided, those that decide in wave 3 or beyond only starts to occur for groups of 5 or larger. These decisions in wave 3 or beyond are rather rare at any agent level, although the number of simulations that finish in wave 3 or beyond is still noticeable, with around 5-10\% of simulations ending in wave 3 or beyond (and is higher for odd $N$). 

If we consider the influence of the first decision, comparing Figures~\ref{fig:BarchartWaveAgentPlus}\&\subref{fig:BarchartWaveAgentMinus} we see that agents seeing an $H_-$ first decision are less likely to follow in wave 1, and more likely to decide in wave 3 or beyond or remain undecided after the waves. This effect is more clear at the group level (comparing Figures~\ref{fig:BarchartWaveSeedPlus}\&\subref{fig:BarchartWaveSeedMinus}), where a majority of simulations are not finished in 2 waves after an $H_-$ first decision for all group sizes smaller than 50, and a majority of simulations having undecided agents after the waves for groups up to around 30. This is considerably different from the case where the first decision is $H_+$, where around 60\% of these simulations are finished by wave 2 for all smaller groups sizes, this percentage increases steadily to nearly 90\% for $N=100$. 

\subsection{Properties and fate of the undecided}
In the previous section we demonstrated that indecision can remain present and unresolved after a wave phase; that there remain agents who are unconvinced for either option after all available social information has been `used up' (i.e. no further shrinkage of the intervals of indecision). In this section we interrogate the properties of this unresolved indecision: what is the typical number of undecided agents, what beliefs do they hold and what subsequently happens to their decision-making time and accuracy?

\begin{figure}
    \centering    
        \begin{subfigure}[][]{\includegraphics[width=0.45\textwidth]{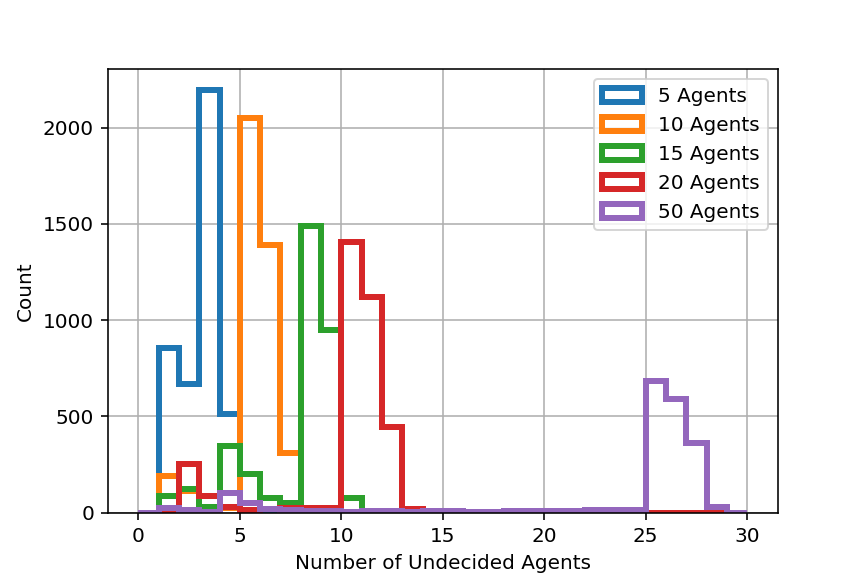}\label{fig:HistUndecidedTheta1}}
		\end{subfigure}  
    \begin{subfigure}[][]{\includegraphics[width=0.45\textwidth]{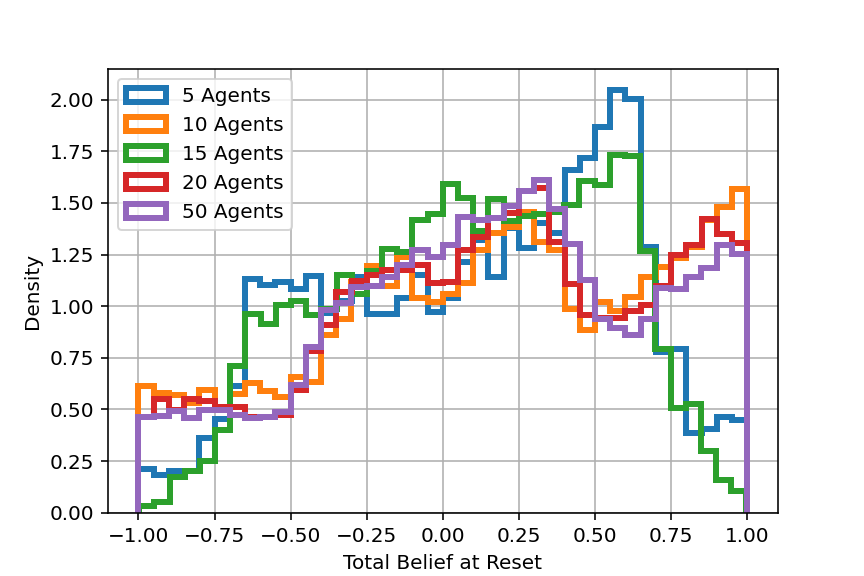}\label{fig:HistUndecidedTotalBeliefWeightedVertTheta1}}
	\end{subfigure}         
    \caption{Undecided agents across selected group sizes over 10000 simulations. (a) gives the non-zero number of undecided agents in the group at reset, whereas (b) gives the total belief (private plus social) at reset for undecided agents across the 10000 simulations. 2D-Heatmaps covering all group sizes (as well as equivalent figures for private and social information of undecided agents) are available in the Supplementary Material \citep[Figures~10-13, particularly Figures~10(a)\&13(a)]{SuppMat2026}.
    }
    \label{fig:HistUndecided}
\end{figure}

The size of the remaining undecided groups is typically just above $N/2$, corresponding to a slim undecided majority that does not follow the first decision that leads to undermining social information in wave 2 (Figure~\ref{fig:HistUndecidedTheta1}). There are no peaks of larger groups above this peak, although some smaller peaks exist for smaller undecided groups, corresponding to undecided agents that experience 3 or more waves of social information. At reset, these undecided agents have a relatively evenly distributed  total belief  (Figure~\ref{fig:HistUndecidedTotalBeliefWeightedVertTheta1}), especially compared to the private belief of all agents ($p_+(t)$, see Figure~\ref{fig:p(t)}) or the private belief of undecided agents \citep[][Figure~11(a)]{SuppMat2026}. This total belief distribution is skewed towards $H_+$, but to a lesser extent than $p_+$; in particular, negative total belief is more common than in $p_+$. Even-sized groups are more likely to be close to the decision thresholds compared to odd-sized groups, likely due to ties in wave 1 that are only possible for even $N$. In addition to the distribution being wide and relatively flat over 10000 simulations, agents within the same group/simulation will have similar total belief as they have received the same social information and have private belief within the same interval of indecision.


\begin{figure}
    \centering
            \begin{subfigure}[][]{\includegraphics[width=0.45\textwidth]{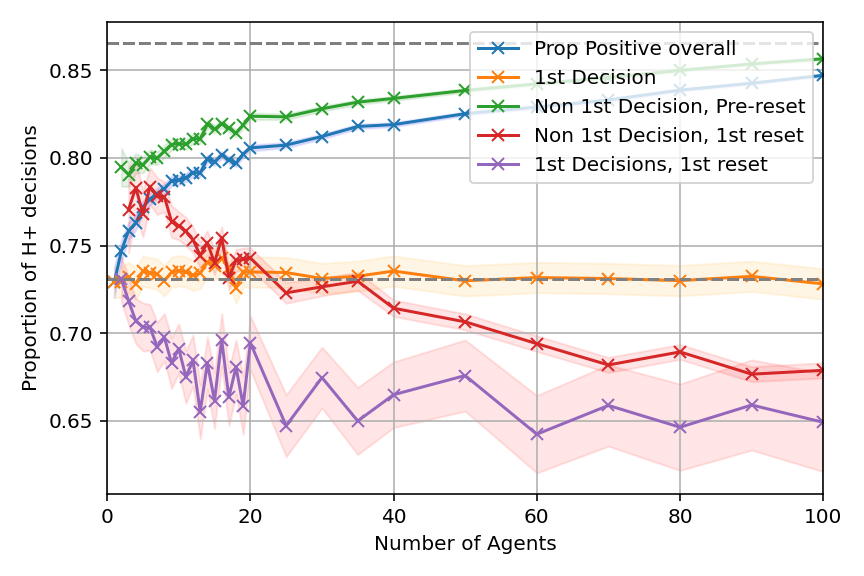}\label{fig:AccuracyPostReset1stReset}}
		\end{subfigure}  
        \begin{subfigure}[][]{\includegraphics[width=0.45\textwidth]{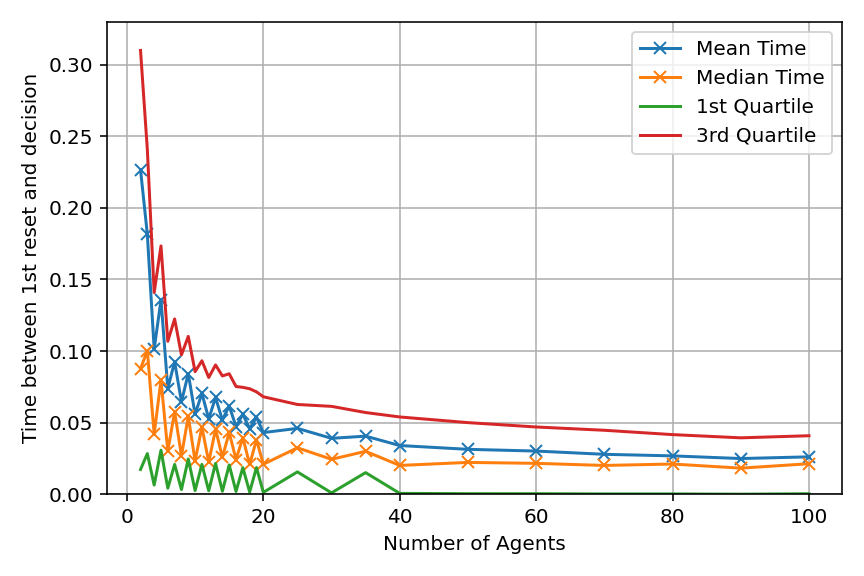}\label{fig:TimeBetweenResets}}
		\end{subfigure} 
    \caption{First decision after reset: (a) The accuracy of this first decision of reset and all other undecideds (see Figure~\ref{fig:AccuracyTheta1}). (b) Time between the reset and this first decision after reset. In (b), the 1st quartile (green line) is very small but non-zero for larger, even $N$ (between $t=10^{-3}$ and $t=10^{-4}$).
}
    \label{fig:PostReset}
\end{figure}

The first decision following the restart of dynamic information gathering by undecided agents results in less accurate decisions than previously (Figure~\ref{fig:PostReset}). This accuracy also declines with group size, from the normal 73\% to around 65\% for larger groups. Note that an agent left to their own devices will actually come to the same overall accuracy as before (an effect demonstrated by undecided agents for $N=2$), meaning that this under-accuracy is not related to prior social information and private information per se since such information is all independent. 

This reduced accuracy follows directly from properties of the undecided subgroup already established above. The undecided agents are a group with a clustered bias in total belief (skewing less towards $H_+$) that increases in number approximately linearly with increasing $N$. Applying a similar principle to \citep{LinnEtAl2024} where the fastest decisions are typically among the most biased agents, here the first decision after reset is likely to be made by the agents with the most biased/extreme beliefs. These highly-biased agents are expected to be more biased/extreme and more numerous as the number of undecided agents increases, explaining why overall accuracy of the first decision decreases as group size increases.

An odd-even pattern also appears, with faster first decision times for even $N$, especially at the lower quartile. This is consistent with the difference in total belief distributions for undecided agents between odd and even $N$ likely due to ties (Figure~\ref{fig:HistUndecidedTotalBeliefWeightedVertTheta1}). 

The other (non-first) undecided agents are still more accurate for small-to-medium sized groups, but this accuracy decreases with group size, and ultimately becomes less accurate. The social information received from first decisions is often small, especially in cases where biases are strong (unless the decision is somehow against all prior social information and hits the opposite threshold, in which case the social information is very large), and so the social information in these waves and responses are smaller and further resets occur rather frequently.

\section{Discussion}

We analysed the behaviour of the waves model of social decision-making in `small to medium' group sizes of 1 to 100 agents. In common with several models of collective decision-making we found that agents' decisions improved on average in both speed and accuracy with increasing group sizes\citep{KingCowlishaw2007,SumpterPratt2009}, and is consistent with social information altering the time-accuracy trade-off \citep{ChittkaEtAl2009}. The accuracy improvement found within small groups agrees very well with empirical results from a human ``stay-or-escape" decision experiment \citep[][Fig~2A]{TumpEtAl2020}; where decisions improved from 74\% for private decisions to 79\% for decisions within a blend of small groups (from 3 to 17 participants), whereas our results in Figure~\ref{fig:Accuracy} match this well, starting at 73\% and reaching around 78\% for groups of sizes 7-10. 

We have demonstrated that the average accuracy of agents can exceed that of the infinite group; that at the agent level accuracy has a finite optimal group size. This phenomenon is present but overlooked within \citep[][]{KaramchedEtAl2020}) (see Figure S3 of their Supplementary Material). The rationale for this finite optimal group size is that inaccuracy becomes dominated by those following an $H_-$ first decision in wave 1 as the group size increases (all other cases of inaccuracy become increasingly rare). However, increasing group size also results in faster first decisions. These faster decisions increase the likelihood of having somewhat poor private information that would result in following an $H_-$ first decision in wave 1 (i.e. $P(y(T)<0|H_+)$), whereas the social information behind these wave 1 decisions is fixed and does not change with the group size and timing. Consequently, the dominant source of inaccuracy in large groups increases with group size as agents are rushed into decisions before they have gathered quality private information. 

Looking at accuracy of the group as a whole, we find that the group (either all agents or the majority) become more accurate with increasing group size, as in \citep{KaramchedEtAl2020}. However, in our model we commonly found groups split by decisions, especially for small groups and cases where the first decision is wrong. The model does not consider any mechanism that would enforce group cohesion or consensus decision-making under social information \citep{MillerEtAl2013,NielsenEtAl2024, Mann2024}; that following the group is not preferable per se other than that collective actions of the group provide social information that enables better informed decisions.

Although we do not consider the costs and benefits of fast and accurate decisions  (such as opportunity costs), we think it is likely that if such costs and benefits were included, the optimal level of certainty needed to make a decision would vary with group sizes and impact decision times. For example, \cite{TumpEtAl2022} found that increasing group size resulted in faster decisions in cooperative groups, but slower decisions in competitive groups. In both cases, the decision threshold increases with group size but to a greater extent for competitive groups as they wait for others to make the first decision. 

Interrogating the properties of the groups where indecision persisted, we consistently found that the group of undecided agents after the waves are completed typically consists of just over half the group, as a result of ties and undermining social information within wave 2 from the (in)actions in wave 1, in a mechanism similar to the indecision from ties in Condorcet like problems \citep{Gillett1977}. We did not find any cases where a strong majority remained undecided, and cases of smaller minorities remaining undecided after more than three waves were rare. 

The cascading of first decisions is found in many static, sequential models, where the first few decisions can lead to an ever-enforcing and overwhelming consensus, be it the static sequential models \citep{Mann2018}, or to a lesser extent social drift-diffusion model \citep{TumpEtAl2024Earlier}. Making decisions under ``Overwhelming'' information with disappearing intervals of indecision are reminiscent of information cascades common in sequential and network decision models, where the first few decisions lead to everyone else following \citep{McCormickEtAlAl2024,BikhchandaniEtAl1992}. These information cascades are also present within SDD models \citep{TumpEtAl2020,TumpEtAl2022,TumpEtAl2024}. However, overwhelming information is stronger since the resulting decision must occur, and not just highly likely or even "almost surely" that applies to most limit cases of information cascades. Consequently, decisions made under overwhelming information provide no new social information (it is impossible to refuse). Similarly, indecision following ``Undermining" information provides no new social information. 

The destiny of the undecided agents is more complex. We found that those undecided after the waves of social information are typically less accurate when they finally arrive at a decision, with accuracy decreasing as groups get larger. This is due to these undecided agents now having a clear bias and that the next agent to decide will likely be among the most biased. This is consistent with the findings in \citep{LinnEtAl2024}, where they demonstrate that fast decisions represent biased initial conditions. In practice, this phenomenon of bias and poor decisions after a period of indecision could materialise as those deciding after some delay after a cluster of decisions become less accurate than others. Additionally, this relative inaccuracy after a period of indecision is more pronounced in larger groups. This bias and inaccuracy suggests that the first/fastest decisions after reset are no longer independent (breaking an implicit assumption in our model); and that it might be possible to factor in this dependency within the first decision into the calculation of the resulting social information. In addition to factoring in this dependency, alternative approaches to the fate of undecided agents could be to include a proxy for this dependency (such as time since the previous decision), or even `go at it alone' by ignoring further social information \citep{MorinEtAl2021}. These alternatives will have different combinations of time, accuracy, computation complexity and cognitive difficulty/constraint profiles.

Agents can have conflicting preferences \citep{Conradt2012}, for example hungry agents would prioritise food whereas those who have just eaten have little to no such need. Although we assumed everyone had the same preference, our results should be straightforwardly extended to cases where preferences are not aligned if all underlying preferences are common knowledge. In this case an individual can treat the decision for $H_+$ of those with conflicting preferences as being equivalent to decisions for $H_-$ by those with aligned preferences. More complex is the case where preference alignment is unknown, though the model could likely be extended in line with similar sequential models \cite{Mann2020, Mann2022}.
 
Our assumption that decisions and social information come in discrete waves requires compatible timescales for gathering private information, decision-making, (in)decisions becoming observable (in)actions, and social information gathering and processing. In particular, observing (in)action means they have seen all previous (in)decisions and their observed (in)action is based on a common and current state of social information; that this is not just a slow or delayed decision based on older social information.

\section*{Acknowledgements}
This work was supported by a UK Research and Innovation Future Leaders Fellowship (MR/X036863/1) and the Templeton World Charity Foundation Inc. (TWCF-2021-20647).

\appendix
\section*{Appendix}
\subsection{Social information notation and calculations}\label{app:SocInfo}
The broad definition for social information received by agent $j$ from observing event $E_i$ from agent $i$ given the sequence of prior events $\hat{E_i}$:
\begin{align}
    S^{E_i}_{j}&(T)=\frac{D}{\alpha}LLR\left(\frac{ P(E_i \mid \hat{E_i} \& H_{+}) }{P(E_i \mid \hat{E_i}  \& H_{-}) }\right)\nonumber \\
    =&\frac{D}{\alpha}\ln \left( \frac{P(y_i(T)\in I_{E_i}\mid y_i(T)\in I_{\hat{E_i}}\& H_+)  }{P(y_i(T)\in I_{E_i}\mid y_i(T)\in I_{\hat{E_i}}\& H_-)  }\right)\nonumber \\
     =&\frac{D}{\alpha}\ln \left( \frac{\frac{\int^{I_{E_i}} p_+(x,T) dx}{\int^{I_{\hat{E_i}}} p_+(x,T) dx}}{\frac{\int^{I_{E_i}} p_-(x,T) dx}{\int^{I_{\hat{E_i}}} p_-(x,T) dx}} \right)\nonumber\\
     =&\frac{D}{\alpha}\left(\ln \left( \int^{I_{E_i}} p_+(x,T) dx \right) - 
    \ln\left( \int^{I_{E_i}} p_-(x,T) dx\right)\right.\nonumber\\ 
     &     -\left.\ln\left(\int^{I_{\hat{E_i}}} p_+(x,T) dx\right)+\ln\left(\int^{I_{\hat{E_i}}} p_-(x,T) dx \right)\right) 
\end{align}
where $I_{E_i}$ and $I_{\hat{E_i}}$ are the inferred intervals of possible private beliefs and  $\frac{D}{\alpha}$ is a scaling factor. This definition differs from \citet{KaramchedEtAl2020} where an aggregated/cumulative notation is used. These are equivalent by a summing/telescoping argument, but our formulation gives the information from each individual (in)decision (which will be summed together) and makes clearer the underlying decision and information history.

The first decision (say at time $T$ by agent 1) is observed by all other agents. The social information agent $i$ gathers from agent 1's $H_+$ decision (for wave 0) is:  
\begin{align}
    S^{1+}_{i0}(T)&=\frac{D}{\alpha}LLR\left(\frac{ P(\text{$H_+$ at }T\mid\text{was undecided } \& H_{+}) }{P(\text{$H_+$ at }T \mid\text{was undecided } \& H_{-}) }\right)\nonumber \\
    &=\frac{D}{\alpha}\ln\left(\frac{\frac{1}{1+\exp\left(-\frac{\alpha\theta}{D}\right)}}{\frac{\exp\left(-\frac{\alpha\theta}{D}\right)}{1+\exp\left(-\frac{\alpha\theta}{D}\right)}}\right)= \theta, \label{eq:SocH+} 
\end{align}
using calculations within \citep{BogaczEtAl2006}.
Conversely, the social information agent $i$ gathers from an $H_-$ first decision is:
\begin{align}
    S^{1-}_{i0}(T)&=\frac{D}{\alpha}LLR\left(\frac{ P(\text{$H_-$ at }T\mid \text{was undecided } \& H_{+}) }{P(\text{$H_-$ at }T \mid \text{was undecided } \&  H_{-}) }\right)\nonumber\\
    &=\frac{D}{\alpha}\ln\left(\frac{\frac{\exp\left(-\frac{\alpha\theta}{D}\right)}{1+\exp\left(-\frac{\alpha\theta}{D}\right)}}{\frac{1}{1+\exp\left(-\frac{\alpha\theta}{D}\right)}}\right)= -\theta. \label{eq:SocH-}
\end{align} 

\subsubsection{The first wave}

In the first wave, agents demonstrate the sign of their private belief when they decide to follow the first decision or remaining undecided. In particular, the social information received by an undecided agent $i$ observing agent $j$  deciding to follow an original $H_+$ decision in wave 1 would gather the following social information: 
\begin{align}
   S^{j+}_{i1}&(j\text{ decides }H_+|j \text{ was undecided})\nonumber\\&=\frac{D}{\alpha}LLR(y_j(T)\in (0,\theta)| y_j(T)\in (-\theta,\theta)) \nonumber \\
    &=\frac{D}{\alpha}\ln \left( \frac{P(y_j(T)\in(0,\theta)\mid H_+\& y_j(T)\in (-\theta,\theta))  }{P(y_i(T)\in (0,\theta) \mid H_- \& y_i(T)\in (-\theta,\theta))}\right)\nonumber \\
     &=\frac{D}{\alpha}\ln \left( \frac{\frac{\int^{\theta}_{0} p_+(x,T) dx}{\int^{\theta}_{-\theta} p_+(x,T) dx}}{\frac{\int^{\theta}_{0} p_-(x,T) dx}{\int^{\theta}_{-\theta} p_-(x,T) dx}} \right) \nonumber \\
     & = \frac{D}{\alpha}\ln\left( \frac{\int^{\theta}_{0} p_+(x,T) dx}{\int^{\theta}_{0} p_-(x,T) dx}\right):=c(T),\label{eq:Soc+ji1}
\end{align}
by symmetry (i.e. $p_+(x,t)=p_-(-x,t)$). The same calculation is used for the social information from refusing to follow an $H_-$ first decision  ($S^{j0}_{i1}$). Likewise, the social information received by an undecided agent $i$ observing agent $j$ refusing to follow an original $H_+$ decision is:
\begin{align}
    S^{j0}_{i1}&(j\text{ still undecided}|j \text{ was undecided})\nonumber \\&=\frac{D}{\alpha}LLR(y_j(T)\in (-\theta,0)| y_j(T)\in (-\theta,\theta))\nonumber
    \\ &= \frac{D}{\alpha}\ln \left(\frac{\int^{0}_{-\theta} p_+(x,T) dx}{\int^{0}_{-\theta} p_-(x,T) dx}\right)=-c(T).\label{eq:Soc0ji1}
\end{align}
This value also applies to $S^{j-}_{i1}$ for observing those who follow an $H_-$ first decision.

\subsection{Finite group size for optimal accuracy}\label{app:FiniteOptimal}

For an optimal finite group size for accuracy to exist, it is sufficient to demonstrate that the accuracy of agents within infinite groups is approached from above. Key results within \citep{KaramchedEtAl2020} are that all agents almost surely decide within two waves, and this is approached quickly (Figure~\ref{fig:BarchartWave}). Additionally, for large $N$, time for the first decision is so fast that agents are clustered around 0, although this is approached considerably slower than finishing in two waves (see Figure~\ref{fig:FirstDecisionTiming100000}). So for large $N$, we can consider agents being slightly more likely to be positive at the first decision time, and thus we define $P(y(T)\in(0,\theta))=\frac{1}{2}+\epsilon(N)$, where $\epsilon(N)>0$ converges to zero as $N\rightarrow \infty$. 

For very large $N$, all agents will almost surely follow an $H_+$ first decision in two waves; whereas for $H_-$ first decisions, only agents in the overwhelming and overriding second wave decide $H_+$ (i.e. agents with $y(T)\in(0,\theta)$). Thus, the average accuracy for very large groups is: 
\begin{align}
    P(\text{$H_+$ decision|not 1st})&=p\times 1+(1-p)\left(\frac{1}{2}+\epsilon(N)\right)\nonumber\\
    &=\frac{1+p}{2}+(1-p)\epsilon(N),
\end{align}
where  $p=P(\text{choose $H_+$})=\frac{1}{1+\exp\left(-\frac{\alpha\theta}{D}\right)}$ \citep{BogaczEtAl2006} is the probability of a correct first decision.
Since $1-p>0$ and $\epsilon(N)>0$, this is greater than the infinite limit $\frac{1+p}{2}$, and thus there must exist a finite group size that maximises the decision accuracy of agents. 

\begin{figure}
    \centering
    \includegraphics[width=0.45\textwidth]{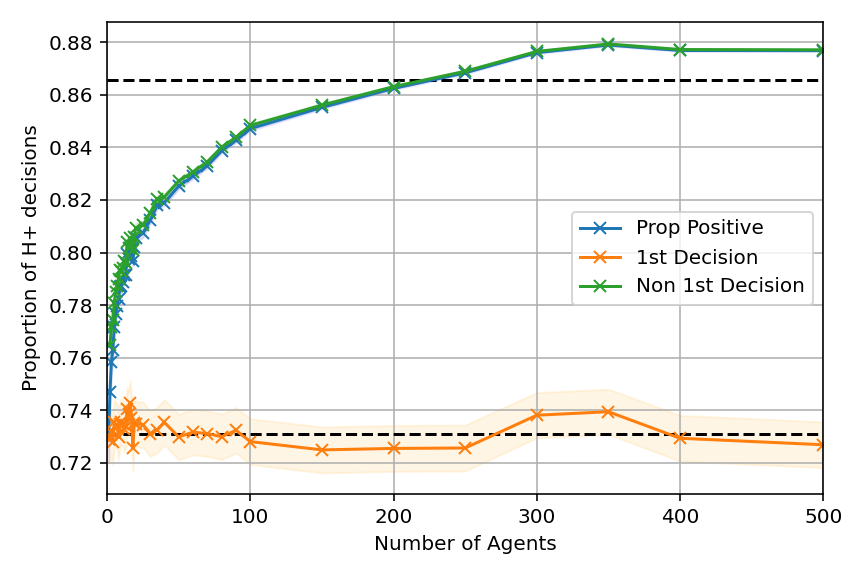}
    \caption{Accuracy (proportion of $H_+$ decisions) for a larger range of $N$ than Figure~\ref{fig:Accuracy}. Notice that overall accuracy comfortably exceeds the upper dashed line, meaning agents in finite groups can be more accurate than the infinite limit, and thus there must be one or more optimal group sizes beyond which accuracy will decline. 
    }
    \label{fig:Accuracy500}
\end{figure}

\bibliography{sample}
\end{document}


\title{Supplementary Material of ``Indecision and accuracy under social information across groups sizes"}
\author{Andrew M. Bate, Charlie Pilgrim \& Richard P. Mann \\ School of Mathematics, University of Leeds, Leeds UK}

\maketitle

\section{Deriving the Stochastic Differential Equation}

Each agent gathers private information for their decision in a stream of independent, noisy measurements $\xi_{1:n}=(\xi_1,\xi_2,..., \xi_n)$ at equally spaced times $t_{1:n}=(t_1,t_2,...,t_n)$. Let $f_+:=P(H^+\mid \xi_{1:n})$ and $f_-:=P(H^-\mid \xi_{1:n})$. The ratio of the probabilities:
\begin{equation*}
R_n:=\frac{P(H_+\mid \xi_{1:n})}{P(H_-\mid \xi_{1:n})}=\frac{f_+(\xi_1)f_+(\xi_2)\dots f_+(\xi_n)P(H_+)}{f_-(\xi_1)f_-(\xi_2)\dots f_-(\xi_n)P(H_-)}
\end{equation*} 
when $P(H_+)$ and $P(H_-)$ are prior distributions (which we will assume are equal). This can be written recursively, such that $R_n=\left(\frac{f_+(\xi_n)}{f_-(\xi_n)}\right)R_{n-1}$, with $R_0=\frac{P(H_+)}{P(H_-)}$. Writing this in terms of the Log-Likelihood-Ratio (LLR) $y_n:=\ln (R_n)$ gives $y_n=y_{n-1} + \ln (\frac{f_+(\xi_n)}{f_-(\xi_n)})=\ln \left(\frac{P(H^+|\xi_{1:t}^i)}{P(H^-|\xi_{1:t}^i)}\right)$.

If these private observations are rapid (i.e. $t_{i}-t_{i-1}=\Delta t\rightarrow 0$) and uncorrelated in time, then the dynamics of each agents private belief can be expressed with the Stochastic Differential Equation (SDE):
\begin{equation}
    dy_i=\pm \alpha dt + \sqrt{2D} dW_i \label{eq:SDE1}
\end{equation}
when $W_i(t)$ are independent standard Wiener processes.

\section{Method of Images approximation}

From the main text, we will use the Method of Images to approximate solutions to the following PDEs that governs the probability density functions $p_+$ and $p_-$:
\begin{equation}
\frac{\partial p_{\pm}}{\partial t}=\mp\alpha\frac{\partial p_{\pm}}{\partial x} +D\frac{\partial^2 p_{\pm}}{\partial x^2}, \label{eq:p}
\end{equation}

The Method of Images involves taking solutions for Equation~\ref{eq:p} that satisfy the initial condition (but not the boundary condition), then using shifted versions of these solutions that match and cancel out the boundary conditions. This leads to an infinite series as these corrections lead to smaller errors at the other boundary. 

For an infinite domain, the solution for \ref{eq:p} with initial condition $p_\pm(x,0)=\delta(x)$ must satisfy:
\begin{equation}
p_\pm(x,t)=\frac{1}{\sqrt{4\pi D t}}\exp \left( -(x\mp \alpha t)^2/4tD  \right)   
\end{equation}
However, this solution does not satisfy the boundary condition $p_\pm(\pm \theta,t)=0$. To make the solution satisfy the boundary conditions, we will use other shifted solutions to Equation~\ref{eq:p} that satisfy $p(x,0)=0$ for all $x\in(-\theta_i,\theta_i)$ (and thus when added any multiple of these has no impact on the initial condition), and use multiples to cancel the boundary condition. 
These shifted solutions are of the form  $\Psi_\pm(x,t,\xi)=\frac{1}{\sqrt{4\pi D t}}\exp \left( -(x\mp \alpha t-\xi)^2/4Dt   \right)$, where $\xi$ is the shift (and thus if $\Psi_\pm(x,t,0)=p_\pm(x,t)$).

Thus, by repeated matching of boundary conditions, the analytic solution that matches the boundary conditions is:
\begin{align}
p_{\pm}(x,t)&=\Psi_{\pm}(x,t,0) +\sum_{n\in \mathbb{N}}(-1)^{n}\left(\exp^{\pm n\theta\frac{\alpha}{D}}\Psi_{\pm}(x,t,2n\theta)+\exp^{\mp n\theta\frac{\alpha}{D}}\Psi_{\pm}(x,t,-2n\theta)\right). \label{eq:MethodOfImages}
\end{align}

\begin{figure}
    \centering
        \includegraphics[width=0.7\textwidth]{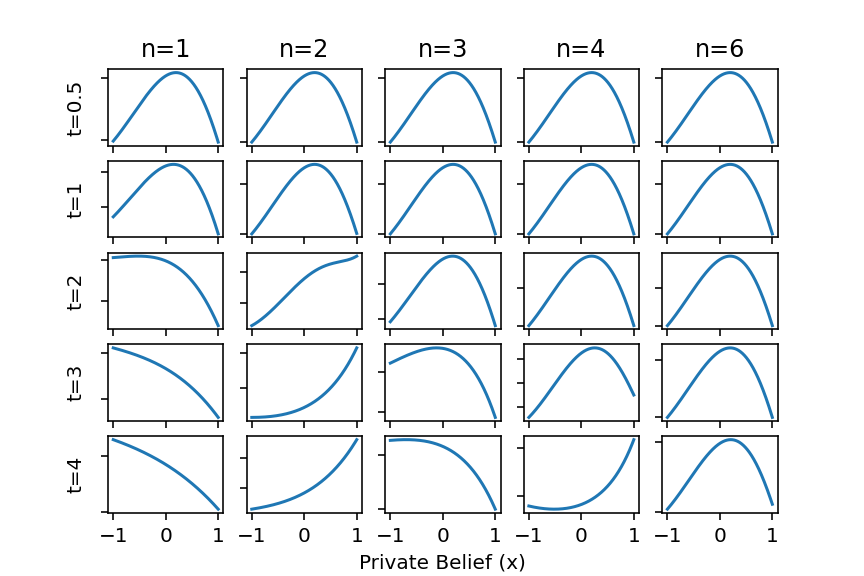}
    \caption{Profiles of the $N$-th (columns) order approximations of the Method of Images solutions of $p_+$ at selected times (rows). The accuracy of these approximations breaks when either endpoint deviates from zero; in particular both endpoints must be level with each other. 
    }
    \label{MethodOfImages}
\end{figure}

Numerical solutions cannot handle infinite sums, so we used truncated versions Equation~\ref{eq:MethodOfImages} using only terms up to and including $n=6$ (the 6th order approximation). Using the 6th order approximation has both a ``positive error" (i.e. the 7th order terms are negative, meaning the 7th order errors do not lead to negative probability densities, at least at first instance). This 6th order approximation is good until around $t=3.5$ (for $\theta,\alpha,D=1$, see Figure~\ref{MethodOfImages})  when the solution becomes visibly largely than zero around $\pm\theta$, and thus starting to break the boundary conditions (and still decent approximation at $t=4$). Only 5 out of 100000 single agent simulations used in Figure~5 in the main paper exceed $t=3.5$, the slowest being at $t=4.15$.

\section{Two or more undecided agents after reset}

With two or more undecided agents after all the waves are finished, there are three major complications: the resetting of $y_i(t)$, $p_+(x,t)$ and  $p_-(x,t)$; the social information from indecision; and the social information gathered at the next decision. 

The distribution of $p_+(x,t)$ and $p_-(x,t)$ are still governed by Equation~\ref{eq:p}, the underlying process of gathering private information is unchanged. 
However, the interval of indecision during the waves only applied at that time, and that during the dynamic phase undecided agents now  have different thresholds for their private information that incorporate the social information received during the waves; namely $H^+$ now requires $y_i(t^+)\geq \theta-\sum S_i$ (and $H^-$ requires $y_i(t^+)\leq -\theta-\sum S_i$), where $\sum S_i$ is the sum of all social information agent $i$ acquired across all prior waves. This means that the interval of indecision for all undecided agents becomes $y_i(t^+)\in (-\theta-\sum S_i,\theta-\sum S_i)$.

Conceptually, an ongoing rate of social information gathered by agent $i$ from agent $j$ remaining undecided existed in the model prior to the first decision (see Section~\ref{Sect:DynamicIndecision}). Prior to the first decision this term is zero (and thus ignored) due to the initial condition being unbiased, leading to the symmetry between $p_+$ and $p_-$ before the first decision. 
Resetting leads to a new biased initial condition that breaks the symmetry between $p_+$ and $p_-$ and thus ongoing social information from indecision is no longer zero. However, for simplicity, we will not consider this social information from indecision as a dynamic term (which would need to be recalculated at each time step, leading to a much slower and complex simulations as well as potential numerically instability), and instead consider this social information as part of a lump sum that is incorporated within the first decision and first wave.

Lastly, the first decision from the undecided agents in this new dynamic phase (at time $T_1$) will trigger analogous social information from the first decision and responding waves of decisions and social information to what happened with the original decision prior to reset (at time $T_0$). However, this value is not automatic and instead needs to understand the principle that the social information from each (in)decision is to update where you expect their private information is. An $H_+$ or $H_-$ decision by an agent (say agent 2) means that their private information is at the new threshold, which includes all the social information received in prior waves. However, all other agents have received prior social information from agent 2's previous indecision, and so already expected agent 2 to have a particular private belief based on all of agent 2's prior indecisions. The resulting social information from this first decision is:
\begin{equation}
    S^{2}_{i}(\text{Decide }H_\pm)= \pm\theta -\stackrel{\text{SocInfo 2 Received}}{\sum S_2}-\stackrel{\text{SocInfo 2 Emitted to $i$}}{\sum_{m} S^{2,0}_{im}}
\end{equation}
Note that our simplifying assumption around ignoring dynamic social information means that the emitted social information term does not include this dynamic social information (i.e. including dynamic social information would lead to it be cancelled out at a decision, ignoring interacting affects), nor does the social information received.

\subsection{After new decision}

At decision time, all undecided agents $i$ start with an interval of indecision of $y_i(T_2)\in (-\theta-\sum S_i,\theta-\sum S_i)$. Seeing the decision by agent 2, these undecided agents receive $S^{2}_{i}(\text{Decide }H_\pm)$ and potentially can decide to follow based on this information. This leads to a shrinking of this interval of indecision $I_{i0}(T_2)$, if   $S^{2}_{i}(\text{Decide }H_\pm)>0$ (as it should for $H_+$), then this shrinks from above, whereas $I_{i0}(T_2)$ shrinks from below if $S^{2}_{i}(\text{Decide }H_\pm)<0$ (as it should for $H_-$).
However, we will incorporate the accumulated social information of indecision here by considering the (in)decision is conditional on $y_i(T_1)$ (the state at the reset after the end of the prior waves):
\begin{align}
    S\left( \text{Undecided in Wave 1 at $T_2$ $\mid$ Undecided at $T_1$}\right)\nonumber&=\ln\left(\frac{P\left(\text{undecided $\mid$ $H_+$, undecided at $T_1$}\right)}{P\left(\text{undecided $\mid$ $H_-$, undecided at $T_1$}\right)}\right) \nonumber \\
    &=\ln\left( \frac{\frac{\int^{I_{i0}(T_2)} p_+(x,T_2) dx}{\int^{\theta-\sum S_i}_{-\theta-\sum S_i} p_+(x,T_1) dx}}{\frac{\int^{I_{i0}(T_2)} p_-(x,T_2) dx}{\int^{\theta-\sum S_i}_{-\theta-\sum S_i} p_-(x,T_1) dx}}\right) \label{eq:ResetFirstDecision}
\end{align}
Analogous to what was done in the main paper for the first waves, the social information gathered from $H_+$ or $H_-$ decisions in wave 1 are calculated using $({-\theta-\sum S_i},{\theta-\sum S_i} )\setminus  I_{i0}(T_2)$ instead. Then the social information gathered by undecided agents from the (in)decisions of wave 1 by summing of all the social information from all the other agents deciding or not in wave 1. The process continues like the first set of waves (pre-reset) for wave 2 and onwards, with analogous processes around the shrinking of the interval of indecision, social information, decisions, whether another waves occur, and again if necessary what happens after restart. In particular, if some agents are still undecided after the waves, then they go back to gathering private information until another agent makes a decision and then go through this process again (and noting that it is this reset time $T_2$ becomes the new benchmark for accumulating dynamic social information from indecision, as such information from before $T_2$ has already been observed and incorporated),

\subsection{Method of Images: Distribution after reset}

Let $p_\pm^*(z,T)$  be the distribution of total belief $z=x+\sum S\in (-\theta,\theta)$ (where $x$ is the private belief) immediately after reset. This distribution is a truncated and shifted version of $p_\pm(x,T)$. Then, assuming no dynamics social information from indecision then the underlying dynamics are defined by \ref{eq:p} and thus functions made by combinations of $\Psi_\pm(z,t,\xi)$ still satisfy the underlying  dynamics. For solutions at time $t=\tau+T$, we must construct functions $f_\pm(z,\tau,K)$, where $K$ signifies the position for a point source release as an initial condition (i.e. $f_\pm(z,T,K)=\delta(K)$), that also satisfy boundary conditions $f_\pm(t,z,-\theta)=f_\pm(t,z,\theta)=0$. The resulting Method of Images solution is:
\begin{align}
f_{\pm}(z,\tau,K)=\Psi_{\pm}(z,\tau,K) +\sum_{n\in \mathbb{N}}&(-1)^{n}\left(  \exp^{\pm\frac{\alpha}{D}(n\theta-KI(n) )}\Psi_{\pm}(x,\tau,2n\theta+(-1)^nK) \right.\nonumber
\\ &  \left. +\exp^{\mp\frac{\alpha}{D}( n\theta+KI(n))}\Psi_{\pm}(x,\tau,-2n\theta+(-1)^nK) \right),
\end{align}
where $I(n)=-\sum_{i=0}^{n-1}(-1)^{i}$, which alternates between 1 (odd $n$) and 0 (even $n$). Like before we will use all terms up to and including $n=6$ as an approximation. With this function $f_\pm$ approximated, we can finally calculate the distributions $p_\pm(z,t)$ for any time after a reset (most notably at the time of next decision):
\begin{equation}
p_\pm(z,t)=\int_{-\theta}^{\theta} f^\pm(z,t-T,K)p_{\pm}^*(K,T) dK.
\end{equation}

\section{Social information from indecision}\label{Sect:DynamicIndecision}

The social information gathered by agent $j$ from the ongoing indecision by agent $i$ during the small time period: 
\begin{align*}
    S^{j0}_{i}(T+\delta t)&=LLR\left(\frac{ P(\text{undecided at }T+\delta t \mid \text{undecided at }T \& H_{+}) }{P(\text{undecided at }T+\delta t \mid \text{undecided at }T \& H_{-}) }\right)\\
    &=\ln\left(\frac{\frac{\int^{\theta}_{-\theta}p_+(x) \left( (1- P(\text{decide $H_+$ at $T+\delta t \mid x$ at $T$ \& $H_+$})- P(\text{decide $H_-$ at $T+\delta t \mid x$ at $T$ \& $H_+$})  \right) dx}{\int^{\theta}_{-\theta} p_+(x) dx}}{\frac{\int^{\theta}_{-\theta}p_-(x) \left( (1- P(\text{decide $H_+$ at $T+\delta t \mid x$ at $T$ \& $H_-$})- P(\text{decide $H_-$ at $T+\delta t \mid x$ at $T$ \& $H_-$})  \right) dx }{\int^{\theta}_{-\theta} p_-(x) dx}} \right)
\end{align*}

Before the first decision, the initial condition is centered around zero and thus $p_+(x)$ and $p_-(x)$ are mirror images of each other by symmetry (see Figure~2(a) in main paper) and thus this becomes zero. However, after the decisions of the waves, the new initial condition is no longer symmetrically centered around zero and is instead biased. Consequently, these terms do not generally cancel and that is ongoing social information from indecision.  However, calculating this dynamically requires calculating and incorporating this term at every timestep and can lead to numerical instability, so for simplicity and we will ignore dynamic social information from indecision. Instead, we will calculate social information from the next decision and the first wave of social information that follows using the state at the previous reset. This incorporates all the social information from indecision in between the reset and the next first decision as follows:
\begin{align}
   S&(\text{Decide at }T_1|\text{Undecided at }T_0)\nonumber\\&=
   S(\text{Decide at }T_1|\text{Undecided at }T_1-\delta t)+\sum_{T_0+n\delta t\leq T_1-\delta t} S^{j0}_{i}(T_0+n\delta t).
\end{align}
This means that all this social information from indecision is accounted for after wave 1, with the exception of the first decider after reset who is already decided at this point and thus does not receive any further social information.

\newpage
\section{Replicating figures in paper with different parameter values}

\begin{figure}[h]
    \centering
        \begin{subfigure}[][]{\includegraphics[width=0.45\textwidth]{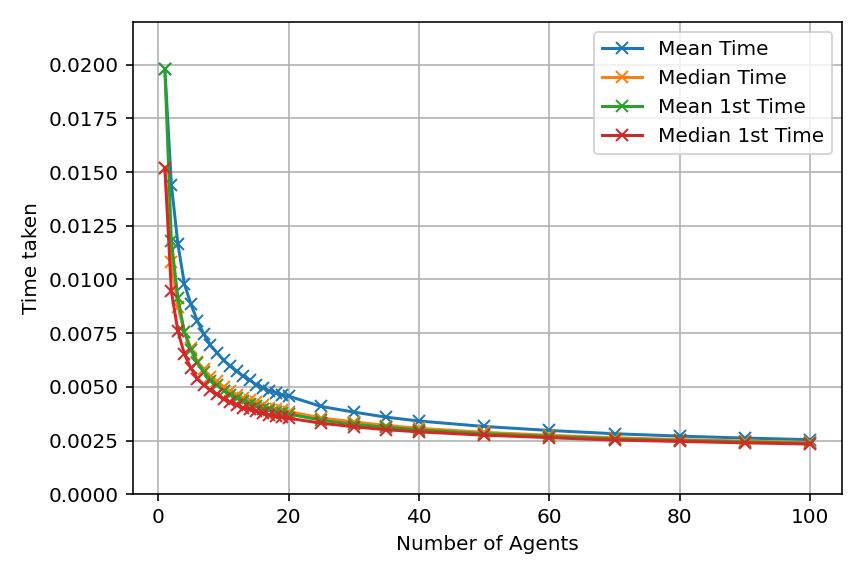}}\label{fig:TimeTheta0p2}
		\end{subfigure} 
        \begin{subfigure}[][]{\includegraphics[width=0.45\textwidth]{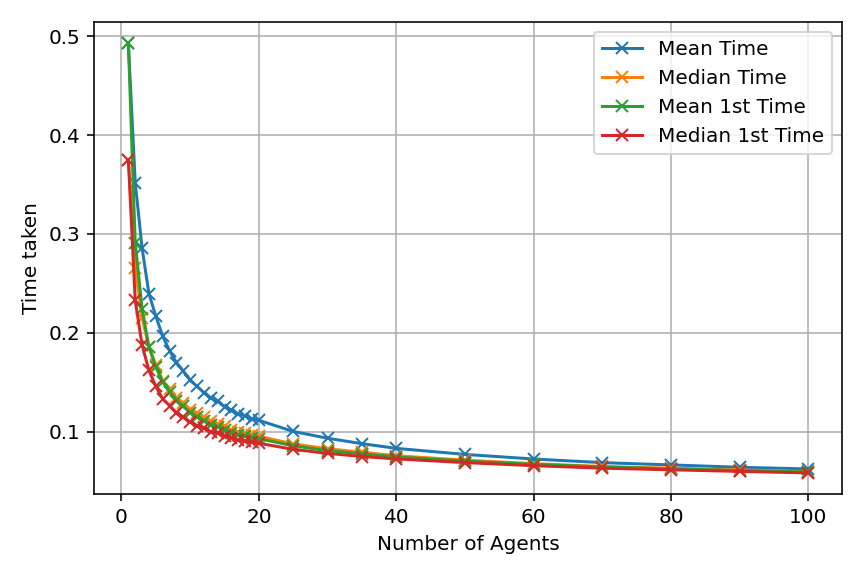}}\label{fig:TimeAlpha}
		\end{subfigure} \\
         \begin{subfigure}[][]{\includegraphics[width=0.45\textwidth]{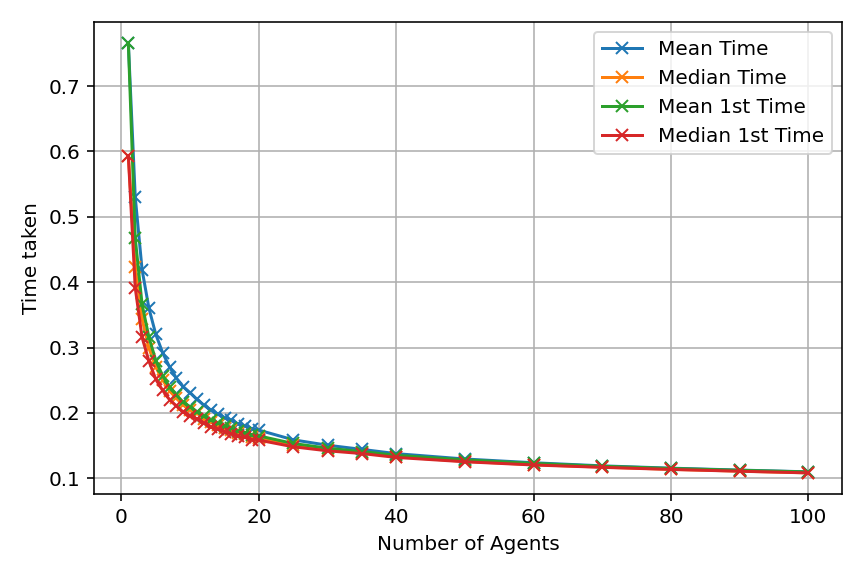}}\label{fig:TimeD}
		\end{subfigure} 
    \caption{Various times as a function (a) $\theta=0.2$, (b) $\alpha=0.5$ and (c) $D=0.5$. 10000 simulations for each of a range of $N$. Equivalent to Figure~3 in the main text.
    }
    \label{fig:TimeOther}
\end{figure}

\begin{figure}
    \centering
        \begin{subfigure}[][]{\includegraphics[width=0.45\textwidth]{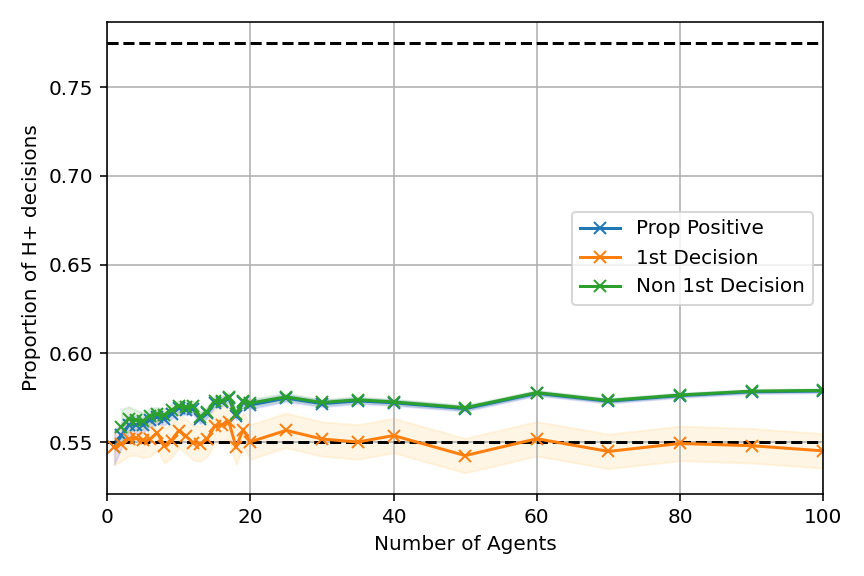}}\label{fig:AccuracyTheta0p2}
		\end{subfigure}   
        \begin{subfigure}[][]{\includegraphics[width=0.45\textwidth]{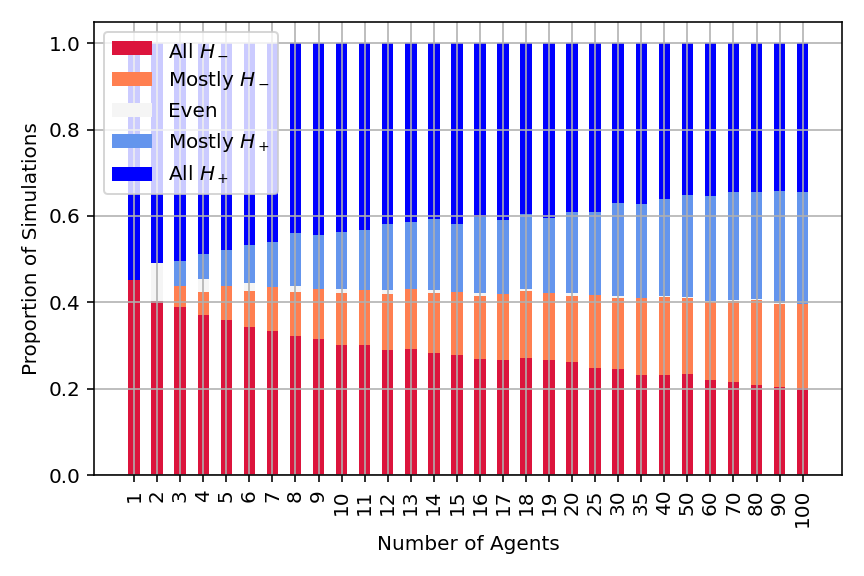}}\label{fig:AccuracyGroupTheta0p2}
		\end{subfigure}          
        \\
        \begin{subfigure}[][]{\includegraphics[width=0.45\textwidth]{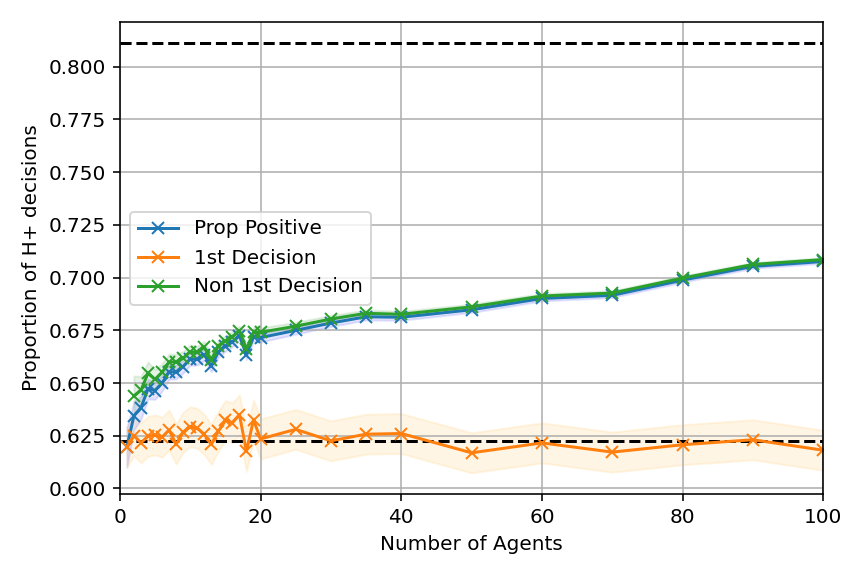}}\label{fig:AccuracyAlpha}
		\end{subfigure} 
        \begin{subfigure}[][]{\includegraphics[width=0.45\textwidth]{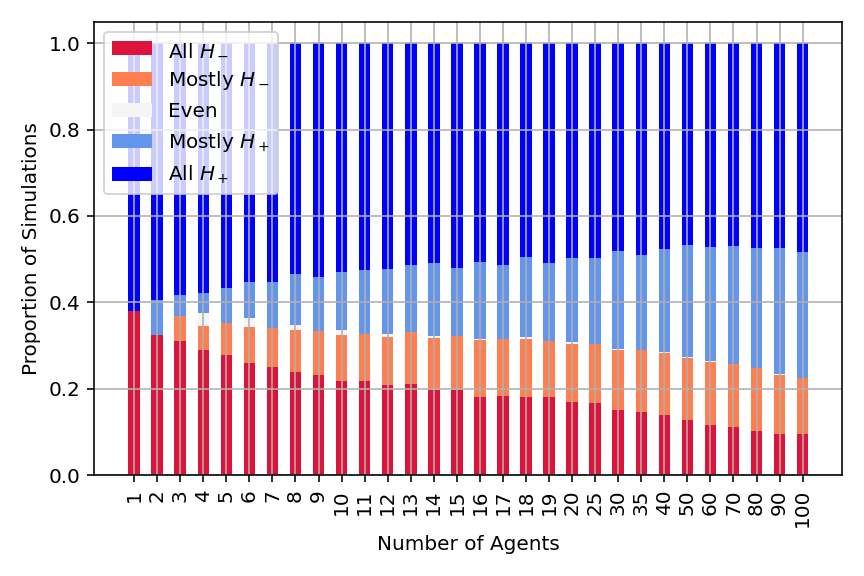}}\label{fig:AccuracyGroupAlpha}
		\end{subfigure} 
        \\
        \begin{subfigure}[][]{\includegraphics[width=0.45\textwidth]{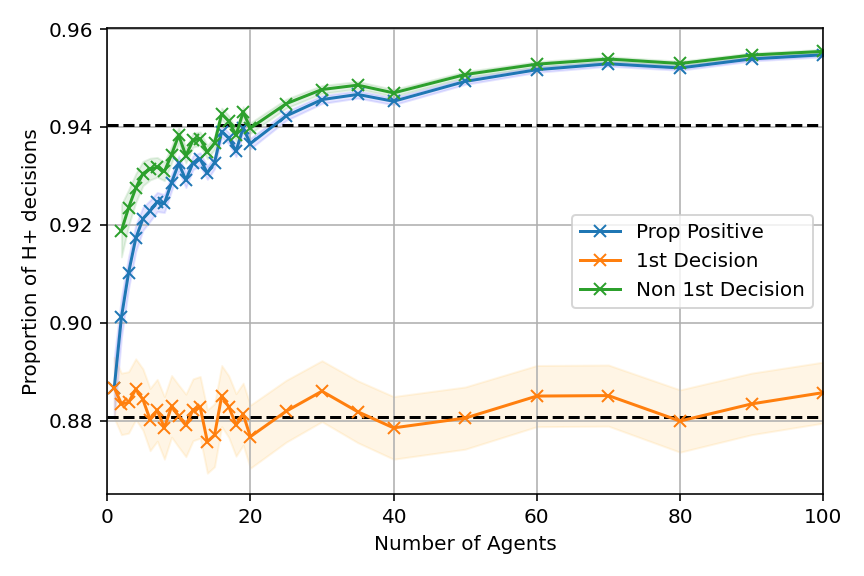}}\label{fig:AccuracyD}
		\end{subfigure} 
        \begin{subfigure}[][]{\includegraphics[width=0.45\textwidth]{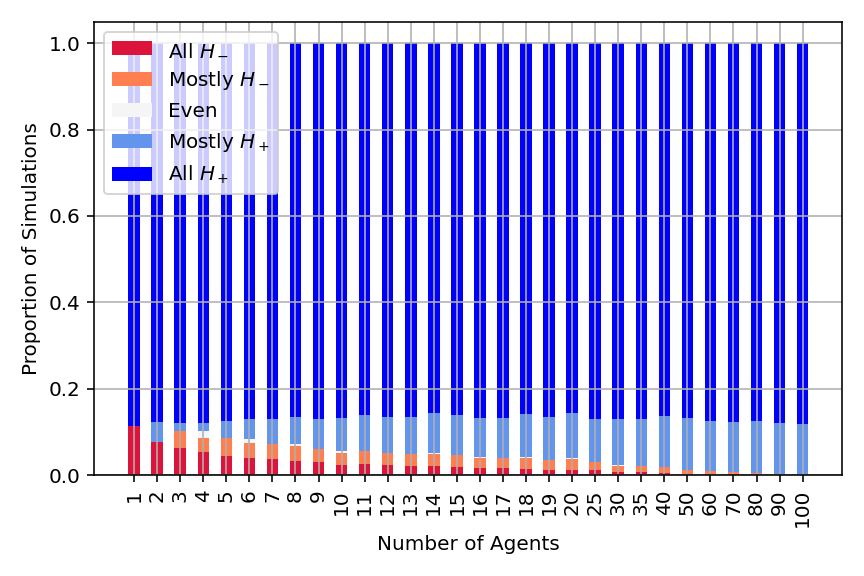}}\label{fig:AccuracyGroupD}
		\end{subfigure} 
    \caption{(a),(c)\&(e) Accuracy (proportion of $H_+$ decisions) by agents, whereas (b),(d)\&(f) are the accuracy of the group as a whole. For each value of $N$, the results of 10000 simulations are used. Dashed lines in (a),(c)\&(e) represent analytic results for single decisions (lower) and the benchmark for infinite groups (upper). In (b),(d)\&(f) ``mostly" means one or more agents disagree with the majority of the group after everyone comes to a decision, even if several resets are required. In (a)\&(b) $\theta=0.2$, (c)\&(d) $\alpha=0.5$ and (e)\&(f) $D=0.5$. Equivalent to Figures~4(a) and 4(b) in the main text.}
    \label{fig:AccuracyOther}
\end{figure}

\begin{figure}
    \centering        
         \begin{subfigure}[][]{\includegraphics[width=0.45\textwidth]{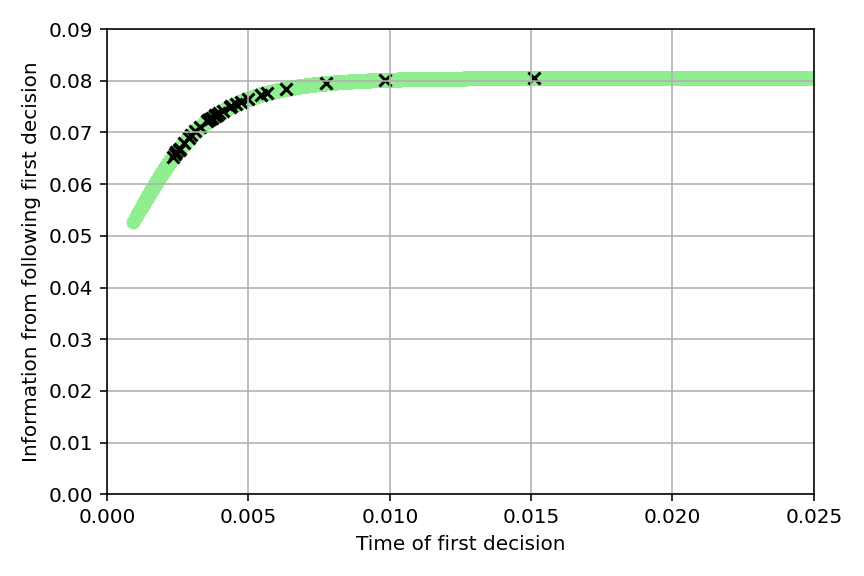}\label{fig:ScatterTheta}}
		\end{subfigure}
        \begin{subfigure}[][]{\includegraphics[width=0.45\textwidth]{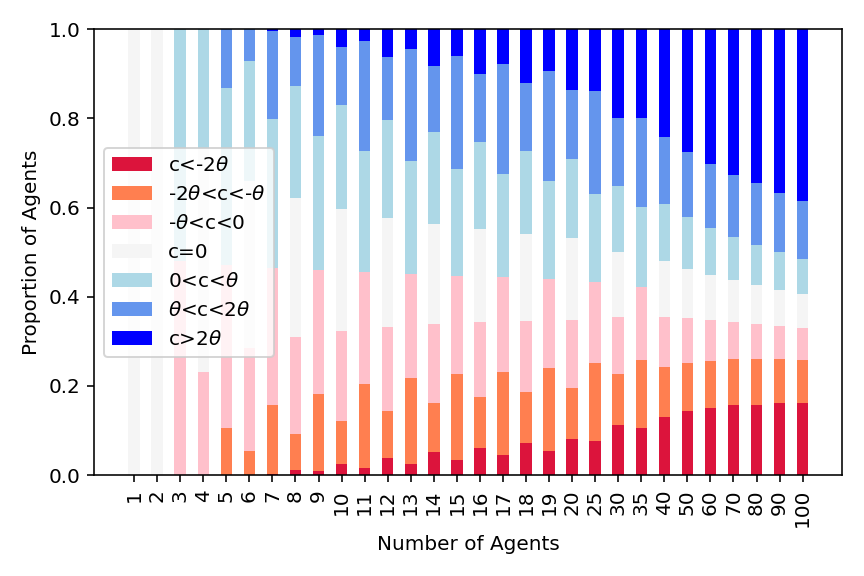}\label{fig:Wave1SocInfoCalcTheta0p2}}
		\end{subfigure}
        \\        
         \begin{subfigure}[][]{\includegraphics[width=0.45\textwidth]{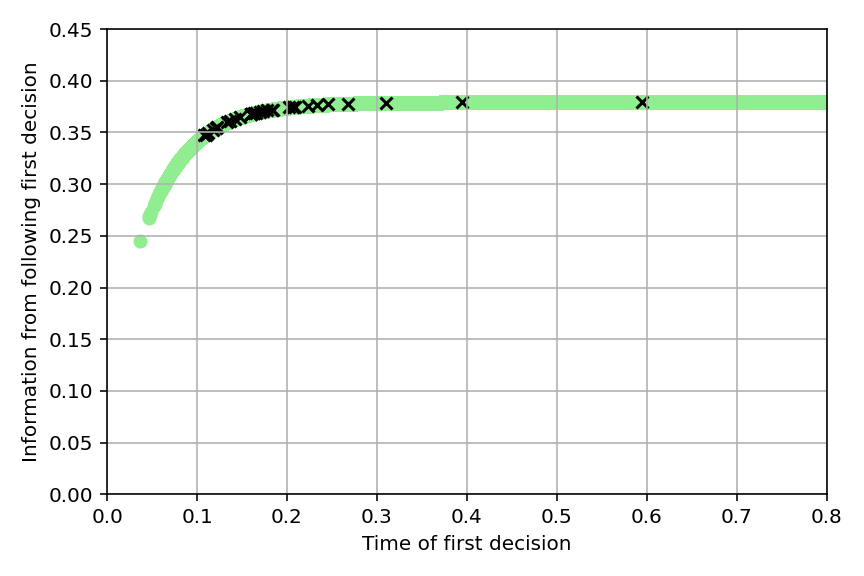}\label{fig:ScatterAlpha}}
		\end{subfigure}
        \begin{subfigure}[][]{\includegraphics[width=0.45\textwidth]{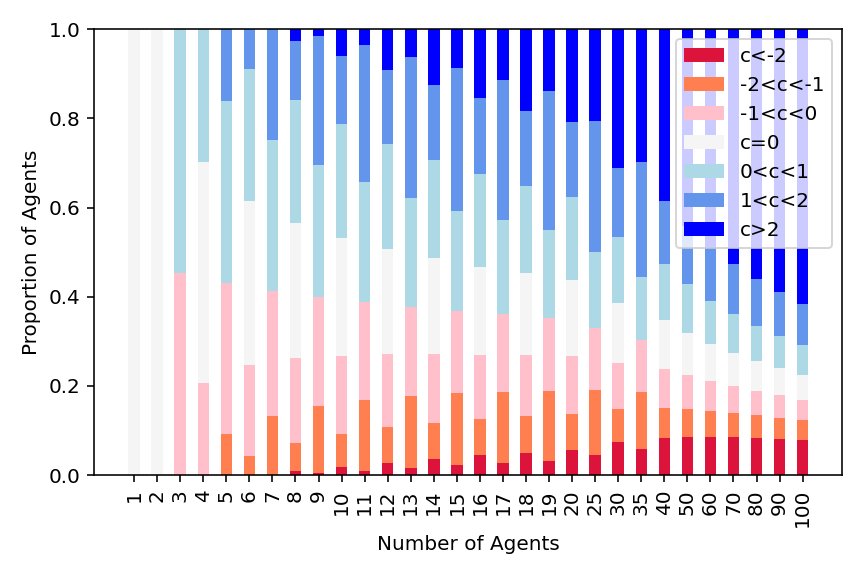}\label{fig:Wave1SocInfoCalcAlpha0p5}}
		\end{subfigure}\\
        \begin{subfigure}[][]{\includegraphics[width=0.45\textwidth]{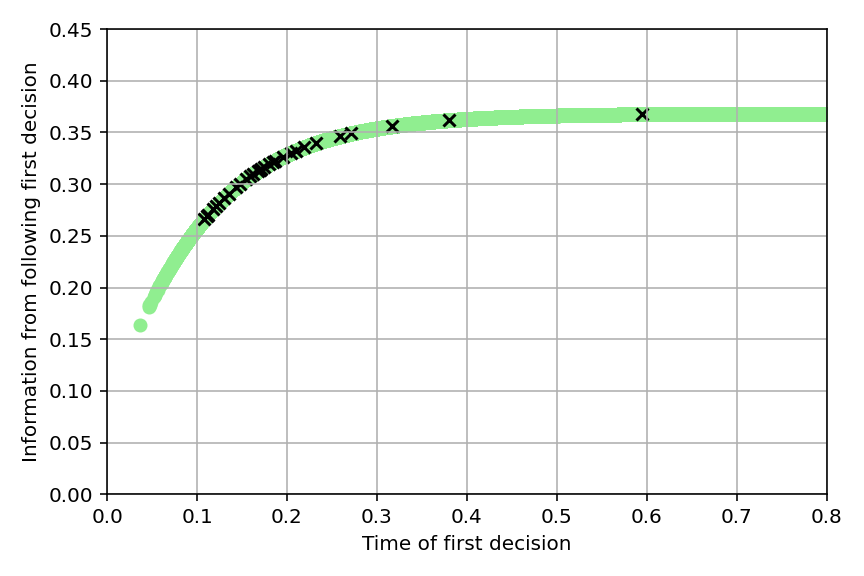}\label{fig:ScatterD}}
		\end{subfigure}   
        \begin{subfigure}[][]{\includegraphics[width=0.45\textwidth]{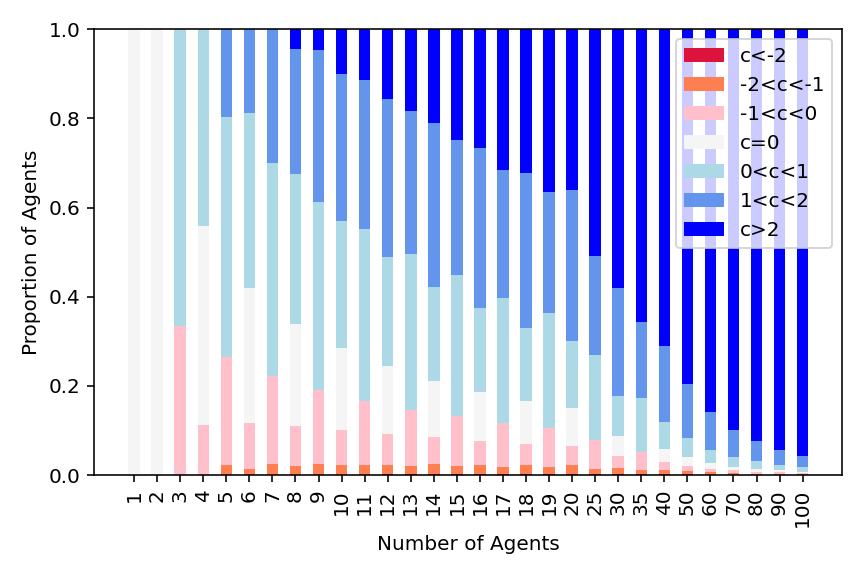}\label{fig:Wave1SocInfoCalcD0p5}}
		\end{subfigure}
    \caption{(a),(c)\&(e) Social Information gathered from a single (in)decision by time of decision as a function of time, whereas (b),(d)\&(f) is proportions of simulations with total social information from (in)decisions from wave 1 in set intervals (see Table~II in main paper). Generated by 100000 single agent simulations for time of decision, with 1000 samples of $N$ agents then taken to gather with their median time for agents (black crosses, going from top right ($N=1$) to bottom left ($N=100$)). In (a)\&(b) $\theta=0.2$, (c)\&(d) $\alpha=0.5$ and (e)\&(f) $D=0.5$. Equivalent to Figures~5(a) and 5(b) in the main text.}
    \label{fig:ScatterOther}
\end{figure}

\begin{figure}
    \centering
  
        \begin{subfigure}[][]{\includegraphics[width=0.45\textwidth]{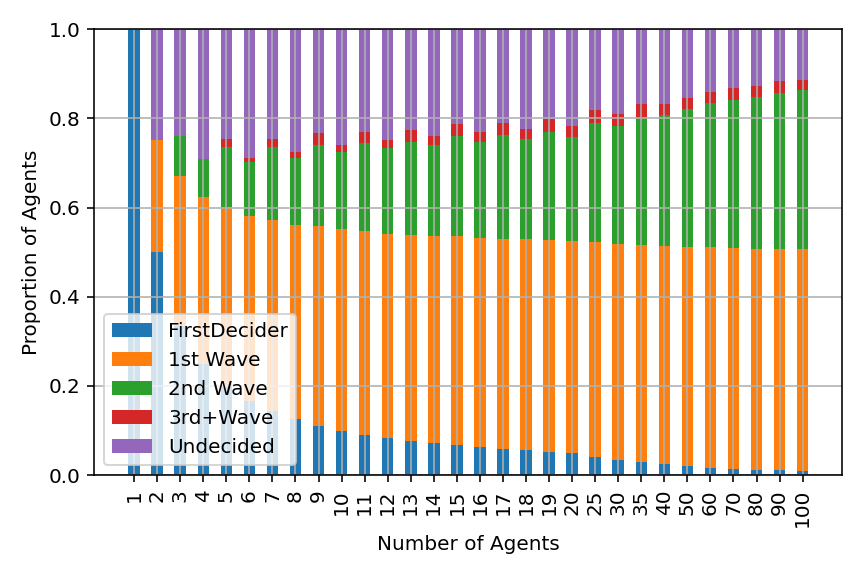}\label{fig:BarchartWaveAgentTheta0p2}}
		\end{subfigure}        
        \begin{subfigure}[][]{\includegraphics[width=0.45\textwidth]{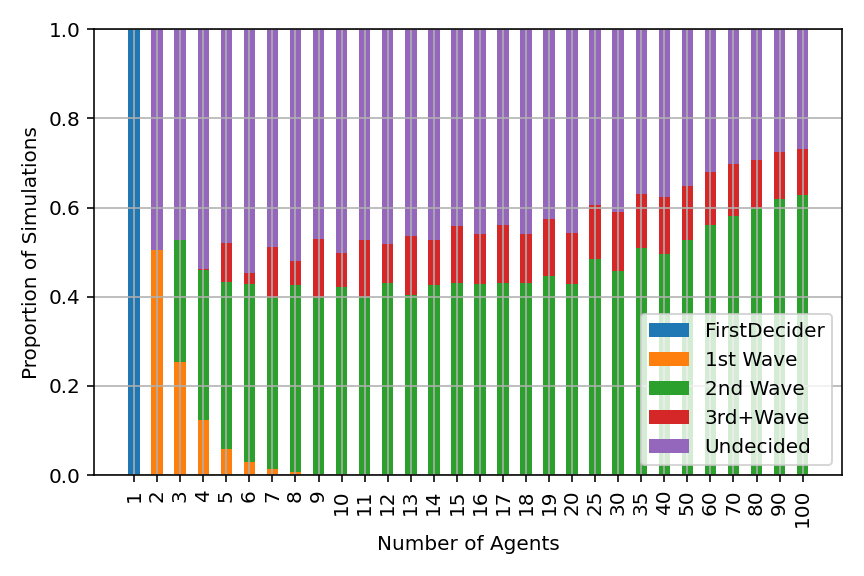}\label{fig:BarchartWaveSeedTheta0p2}}
		\end{subfigure} 
        \\
        \begin{subfigure}[][]{\includegraphics[width=0.45\textwidth]{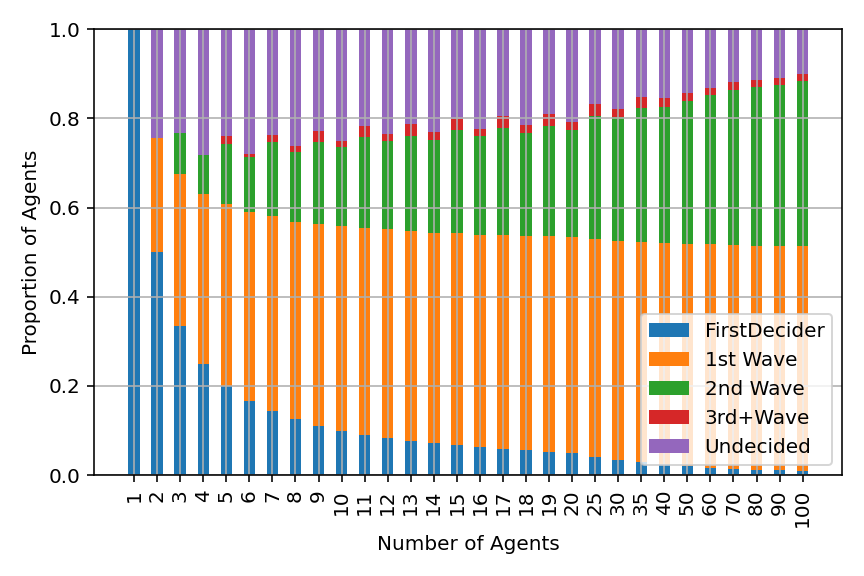}\label{fig:BarchartWaveAgentAlpha0p5}}
		\end{subfigure}        
        \begin{subfigure}[][]{\includegraphics[width=0.45\textwidth]{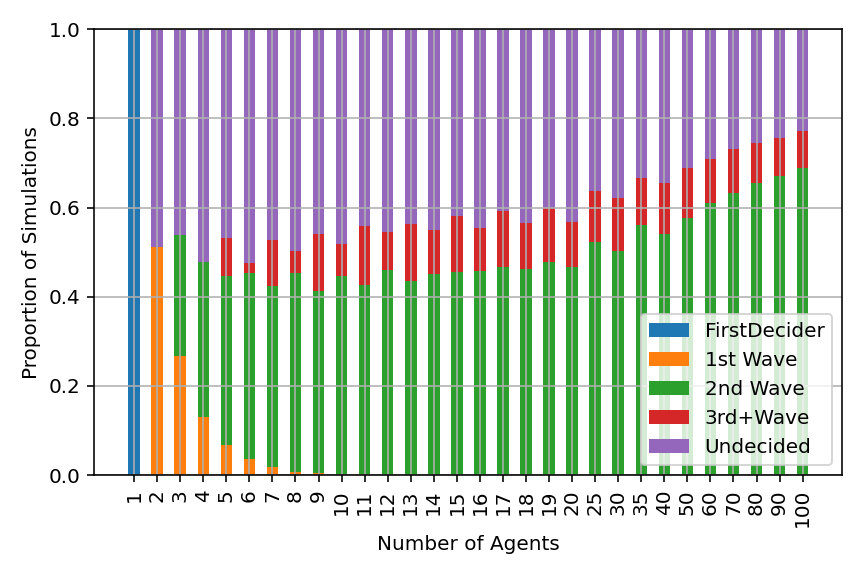}\label{fig:BarchartWaveSeedAlpha0p5}}
		\end{subfigure} 
        \\
        \begin{subfigure}[][]{\includegraphics[width=0.45\textwidth]{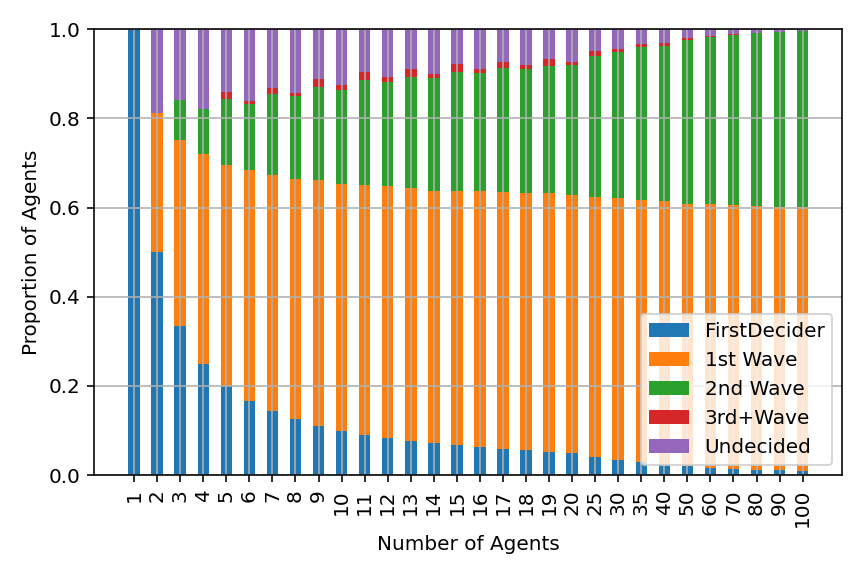}\label{fig:BarchartWaveAgentD0p5}}
		\end{subfigure}        
        \begin{subfigure}[][]{\includegraphics[width=0.45\textwidth]{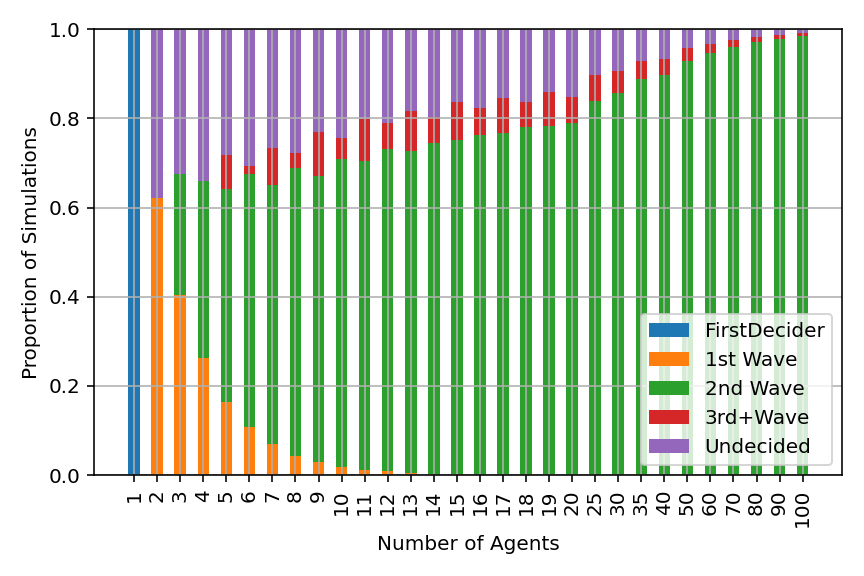}\label{fig:BarchartWaveSeedD0p5}}
		\end{subfigure} 
        \\
    \caption{Proportion of (a),(c)\&(e) agents by when they come to a decision, and (b),(d)\&(f) simulations by when the last agent comes to a decision. Similar figures are in the Supplementary Material for cases where the first decision is $H_+$ (Figure~\ref{fig:BarchartWavePlus}) or $H_-$ (Figure~\ref{fig:BarchartWaveMinus}).In (a)\&(b) $\theta=0.2$, (c)\&(d) $\alpha=0.5$ and (e)\&(f) $D=0.5$. Equivalent to Figures~6(a) and 6(d) in the main text.}
    \label{fig:BarchartWaveOther}
\end{figure}

\begin{figure}
    \centering
            \begin{subfigure}[][]{\includegraphics[width=0.45\textwidth]{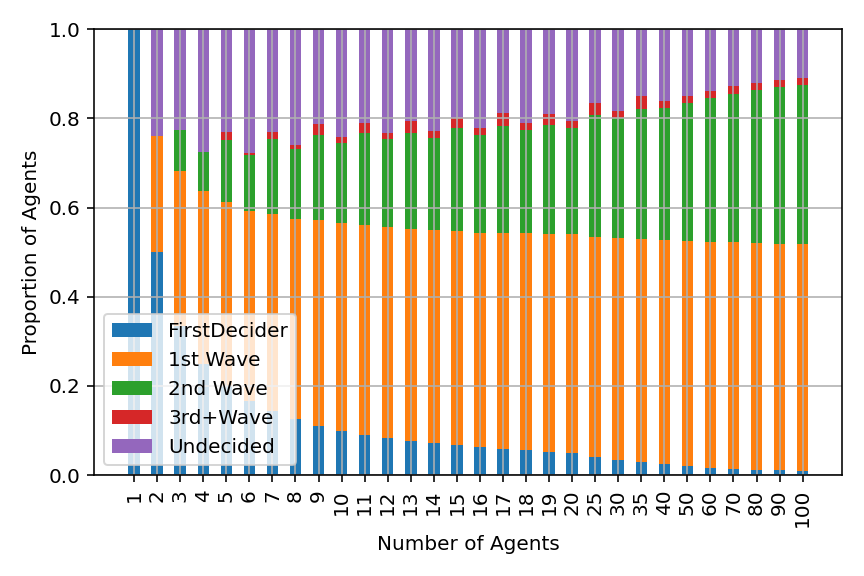}\label{fig:BarchartWaveAgentHplusTheta0p2}}
		\end{subfigure}       
        \begin{subfigure}[][]{\includegraphics[width=0.45\textwidth]{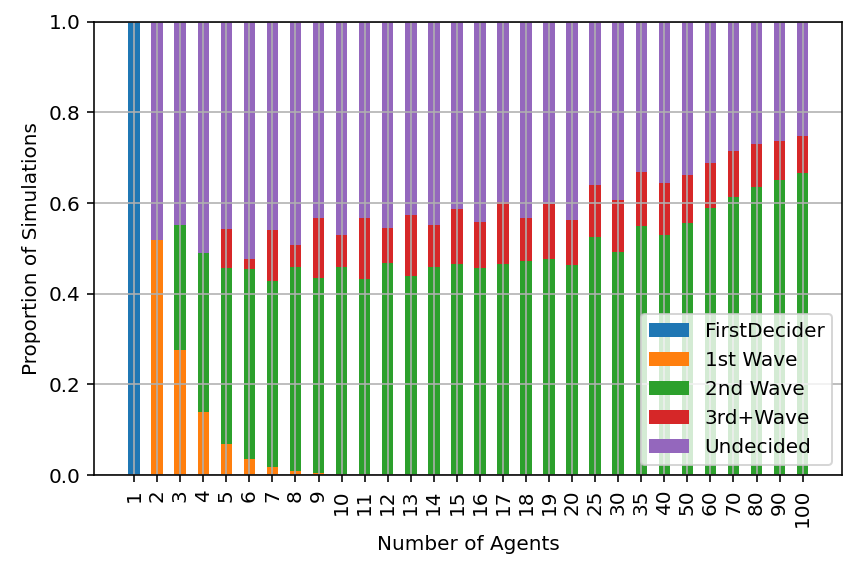}\label{fig:BarchartWaveSeedHplusTheta0p2}}
		\end{subfigure} 
\\
  \begin{subfigure}[][]{\includegraphics[width=0.45\textwidth]{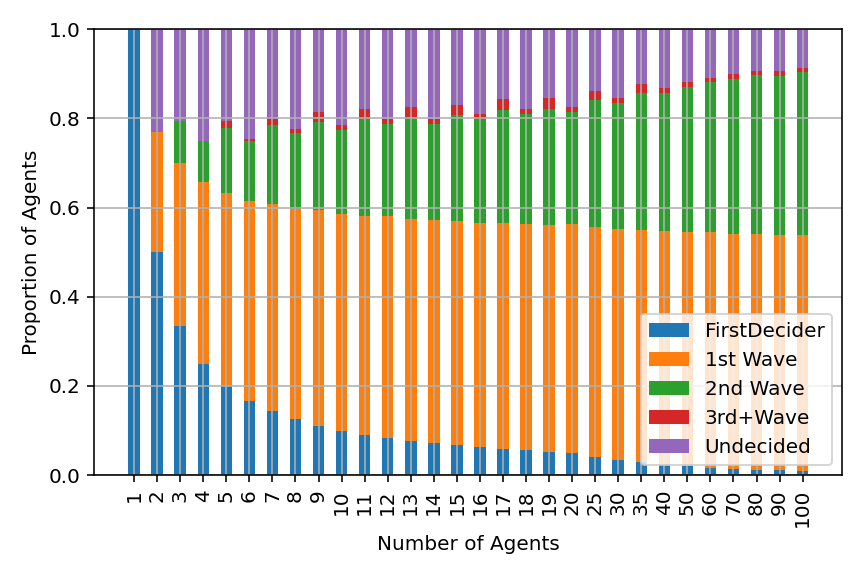}\label{fig:BarchartWaveAgentHplusalpha0p5}}
		\end{subfigure}
        \begin{subfigure}[][]{\includegraphics[width=0.45\textwidth]{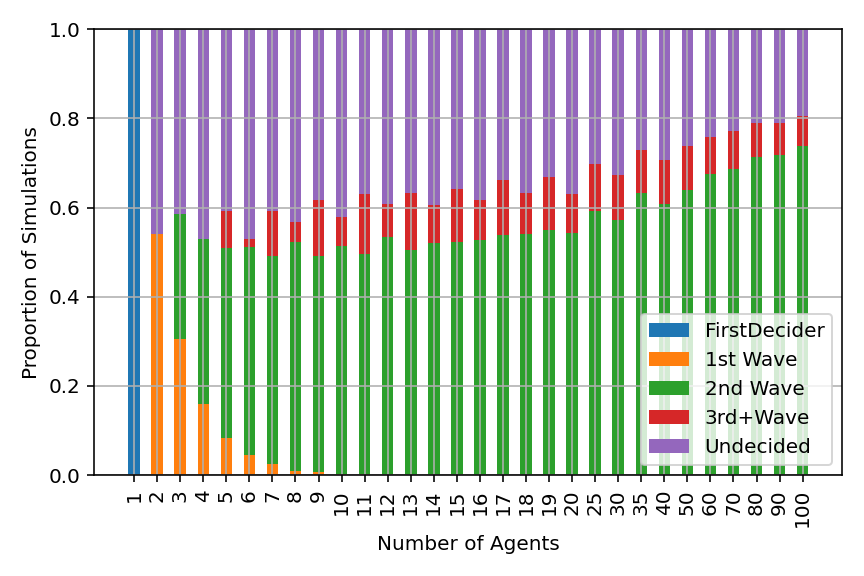}\label{fig:BarchartWaveSeedHplusalpha0p5}}
		\end{subfigure}\\
  \begin{subfigure}[][]{\includegraphics[width=0.45\textwidth]{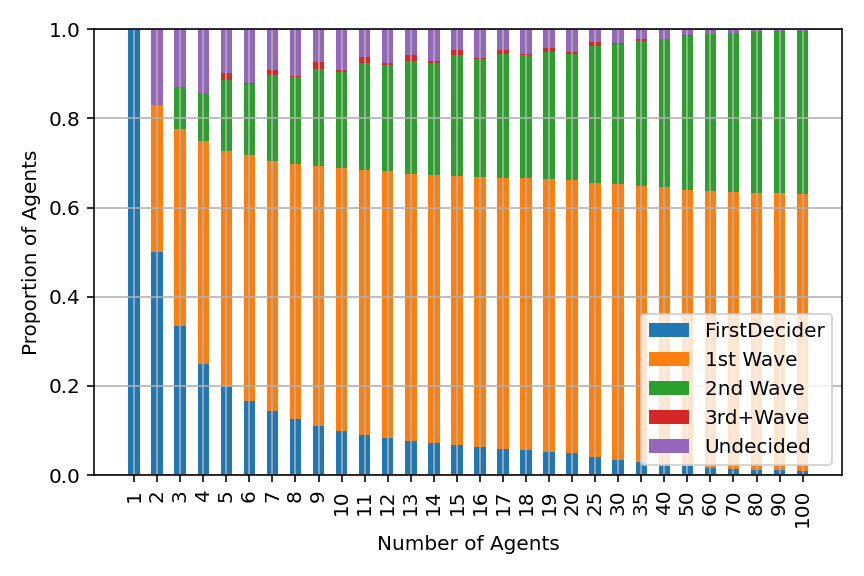}\label{fig:BarchartWaveAgentHplusDiff0p5}}
		\end{subfigure}       
        \begin{subfigure}[][]{\includegraphics[width=0.45\textwidth]{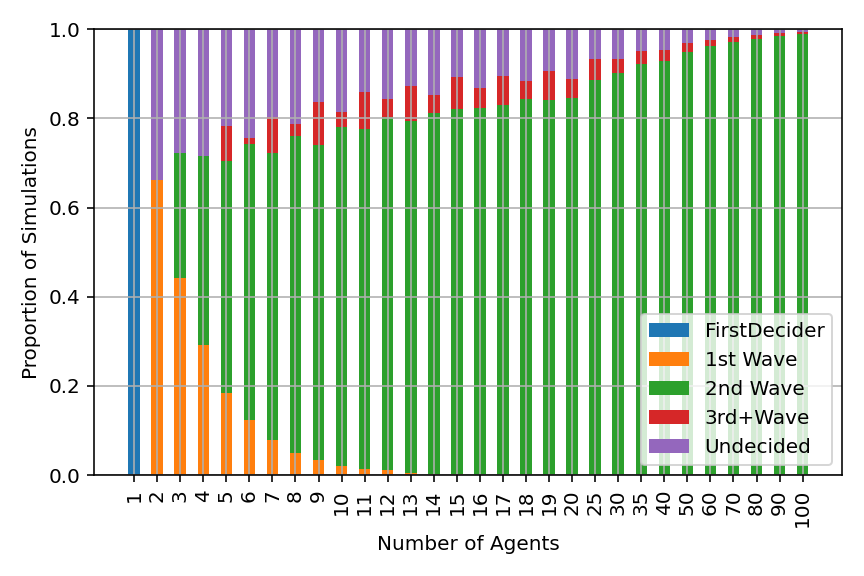}\label{fig:BarchartWaveSeedHplusDiff0p5}}
		\end{subfigure}\\
    \caption{Proportion of (a),(c)\&(e) agents by when they come to a decision given the first decision is $H_+$, and (b),(d)\&(f) simulations by when the last agent comes to a decision given the first decision is $H_+$. In (a)\&(b) $\theta=0.2$, (c)\&(d) $\alpha=0.5$ and (e)\&(f) $D=0.5$. Equivalent to Figures~6(b) and 6(e) in the main text.}
    \label{fig:BarchartWavePlus}
\end{figure}

\begin{figure}
    \centering    
            \begin{subfigure}[][]{\includegraphics[width=0.45\textwidth]{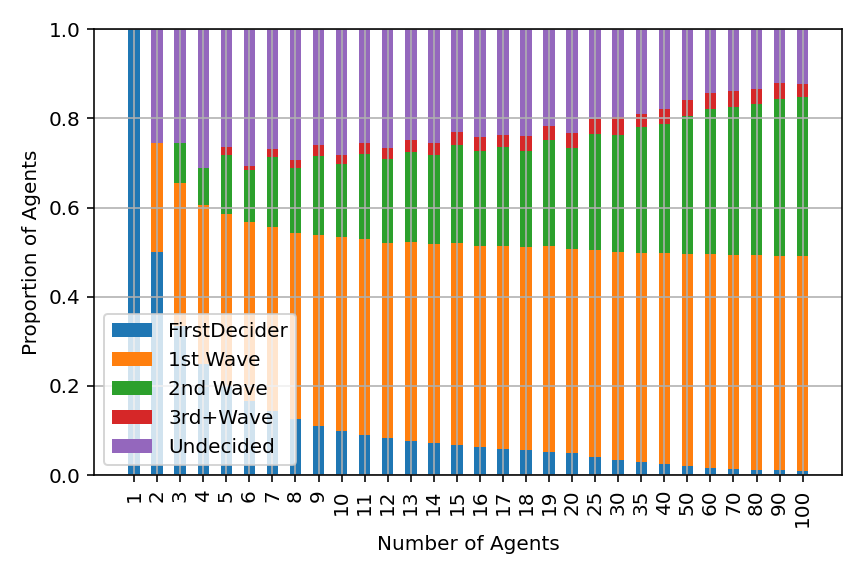}\label{fig:BarchartWaveAgentHminusTheta0p2}}
		\end{subfigure}
        \begin{subfigure}[][]{\includegraphics[width=0.45\textwidth]{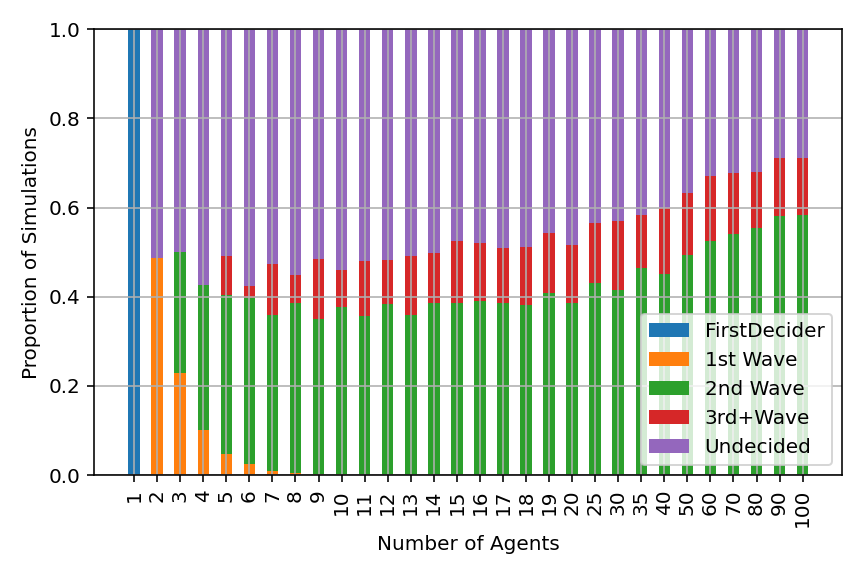}\label{fig:BarchartWaveSeedHminusTheta0p2}}
		\end{subfigure} 
        \\
        \begin{subfigure}[][]{\includegraphics[width=0.45\textwidth]{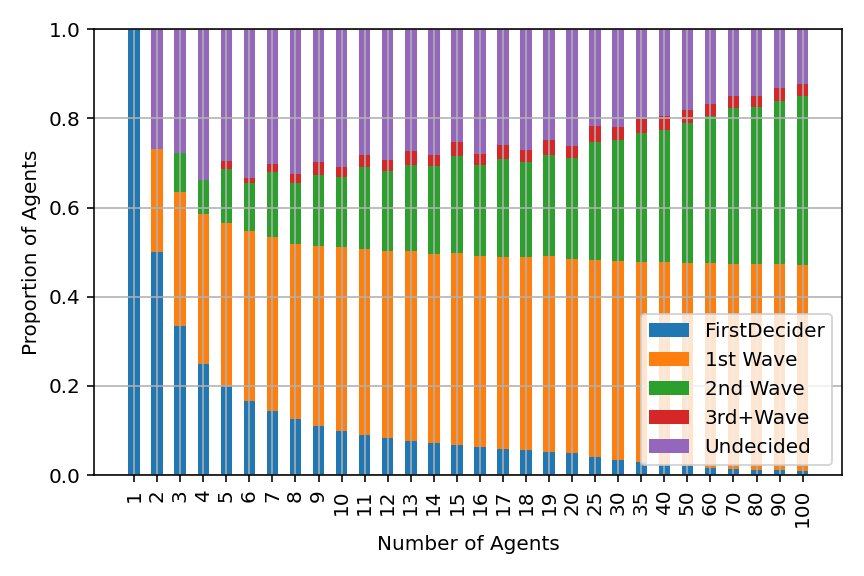}\label{fig:BarchartWaveAgentHminusAlpha0p5}}
		\end{subfigure}
        \begin{subfigure}[][]{\includegraphics[width=0.45\textwidth]{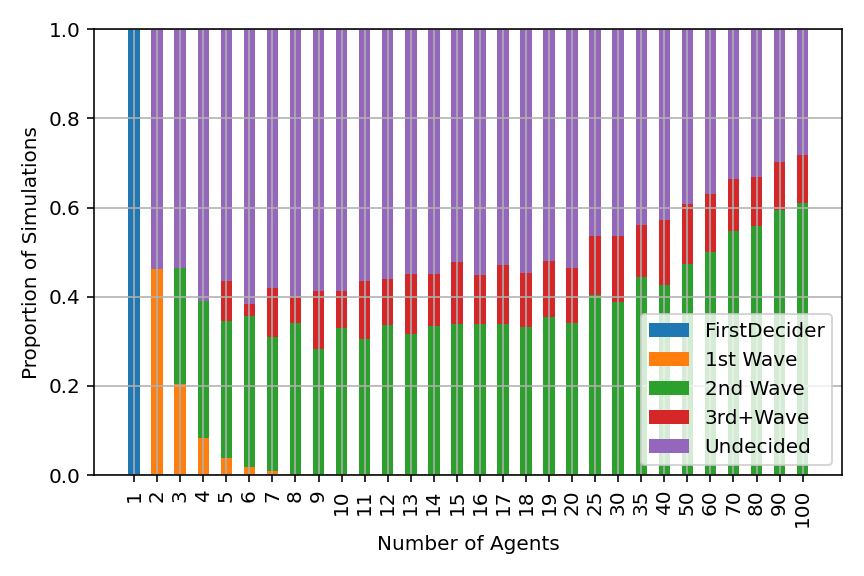}\label{fig:BarchartWaveSeedHminusAlpha0p5}}
		\end{subfigure}
        \\
            \begin{subfigure}[][]{\includegraphics[width=0.45\textwidth]{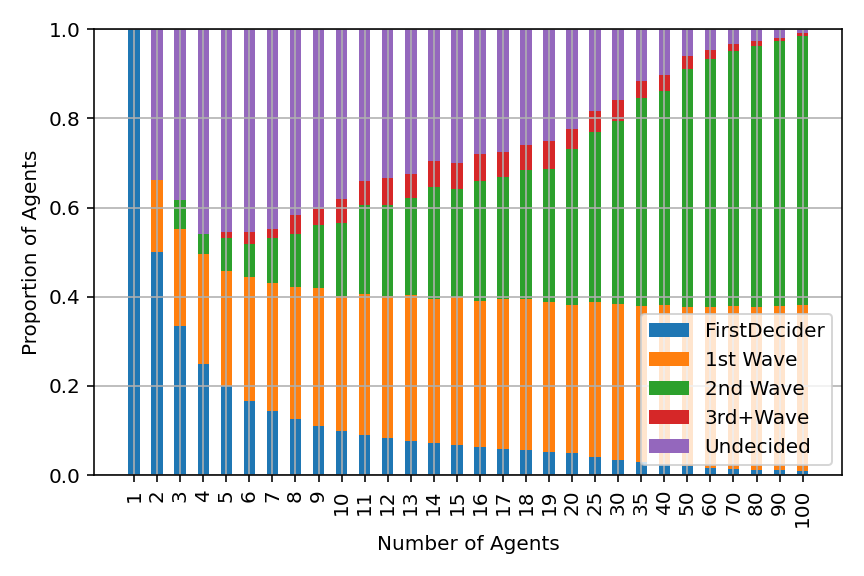}\label{fig:BarchartWaveAgentHminusDiff0p5}}
		\end{subfigure}
        \begin{subfigure}[][]{\includegraphics[width=0.45\textwidth]{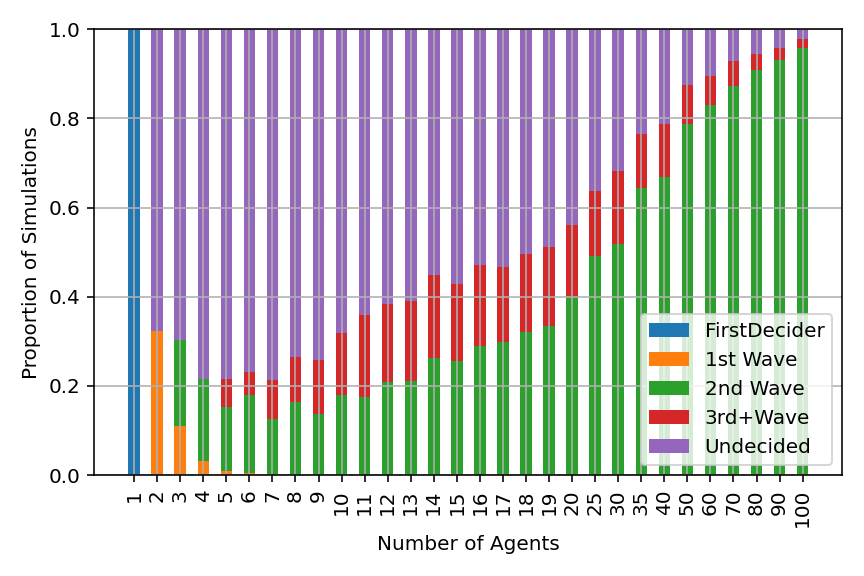}\label{fig:BarchartWaveSeedHminusDiff0p5}}
		\end{subfigure} 
    \caption{Proportion of (a),(c)\&(e) agents by when they come to a decision given the first decision is $H_-$, and (c),(d)\&(f) simulations by when the last agent comes to a decision given the first decision is $H_-$. In (a)\&(b) $\theta=0.2$, (c)\&(d) $\alpha=0.5$ and (e)\&(f) $D=0.5$. Equivalent to Figures~6(c) and 6(f) in the main text.}
    \label{fig:BarchartWaveMinus}
\end{figure}

\begin{figure}
    \centering

        \begin{subfigure}[][]{\includegraphics[width=0.45\textwidth]{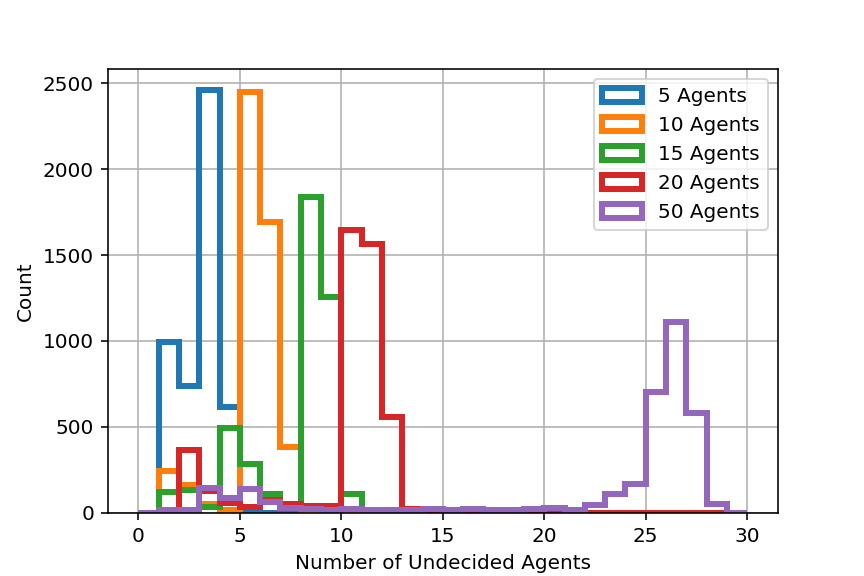}\label{fig:HistUndecidedTheta0p2}}
		\end{subfigure} 
        \begin{subfigure}[][]{\includegraphics[width=0.45\textwidth]{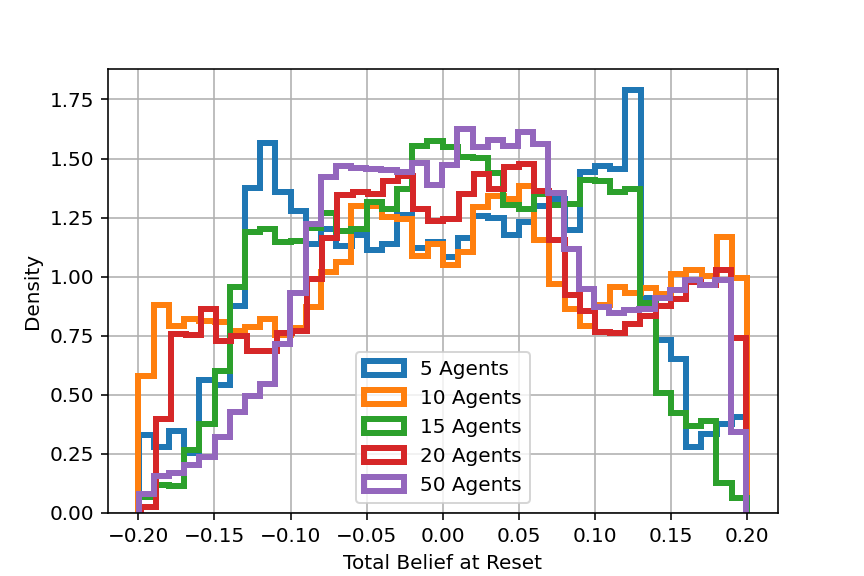}\label{fig:HistUndecidedTotalBeliefWeightedVertTheta0p2}}
		\end{subfigure}
        \\
         \begin{subfigure}[][]{\includegraphics[width=0.45\textwidth]{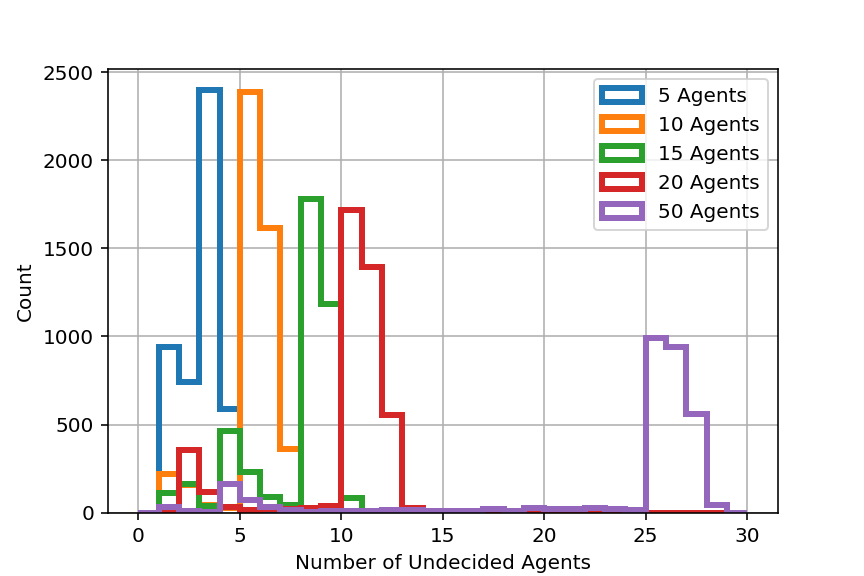}\label{fig:HistUndecidedAlpha0p5}}
		\end{subfigure} 
        \begin{subfigure}[][]{\includegraphics[width=0.45\textwidth]{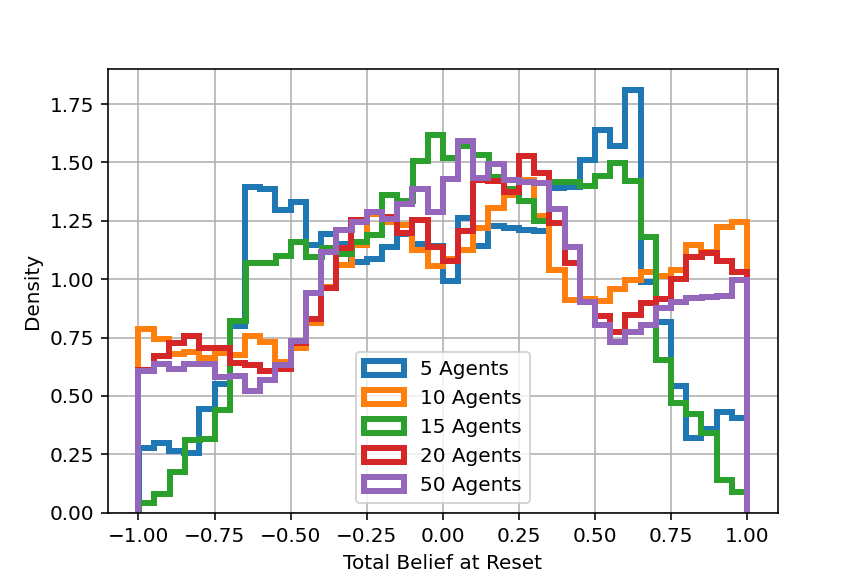}\label{fig:HistUndecidedTotalBeliefWeightedVertAlpha0p5}}
		\end{subfigure} 
        \\
         \begin{subfigure}[][]{\includegraphics[width=0.45\textwidth]{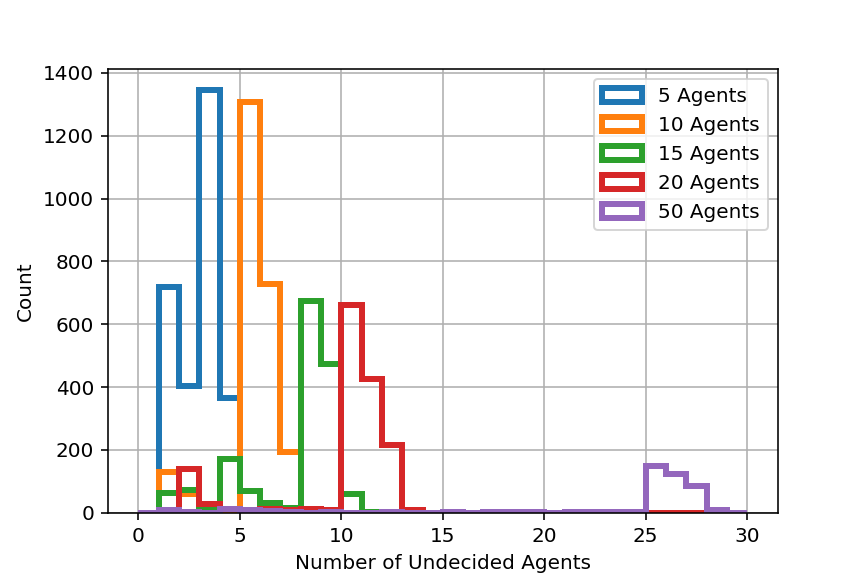}\label{fig:HistUndecidedD0p5}}
		\end{subfigure} 
        \begin{subfigure}[][]{\includegraphics[width=0.45\textwidth]{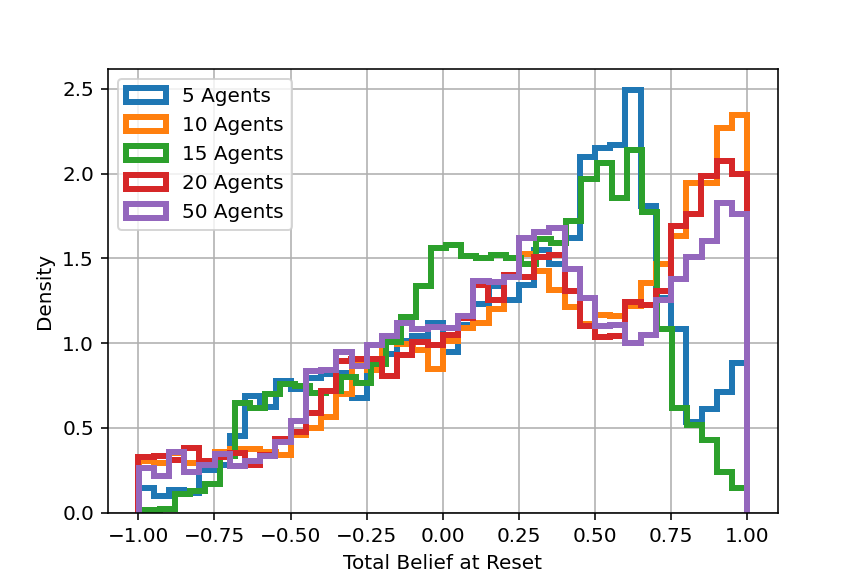}\label{fig:HistUndecidedTotalBeliefWeightedVertD0p5}}
		\end{subfigure} 
        
    \caption{Histograms on undecided agents across selected group sizes over 10000 simulations. (a),(c)\&(e) gives the non-zero number of undecided agents in the group at reset, whereas (b),(d)\&(f) gives the total belief (private plus social) at reset for undecided agents across the 10000 simulations. In (a)\&(b) $\theta=0.2$, (c)\&(d) $\alpha=0.5$ and (e)\&(f) $D=0.5$.   Equivalent to Figures~7(a) and 7(b) in the main text. See Figure~\ref{fig:HeatmapUndecidedOther} and Figure~\ref{fig:HeatmapUndecidedTotalBelief}) for 2D-Histograms covering all group sizes for size of undecided group and their total belief at reset, respectively.}
    \label{fig:HistUndecidedOther}
\end{figure}

\begin{figure}
    \centering 
            \begin{subfigure}[][]{\includegraphics[width=0.45\textwidth]{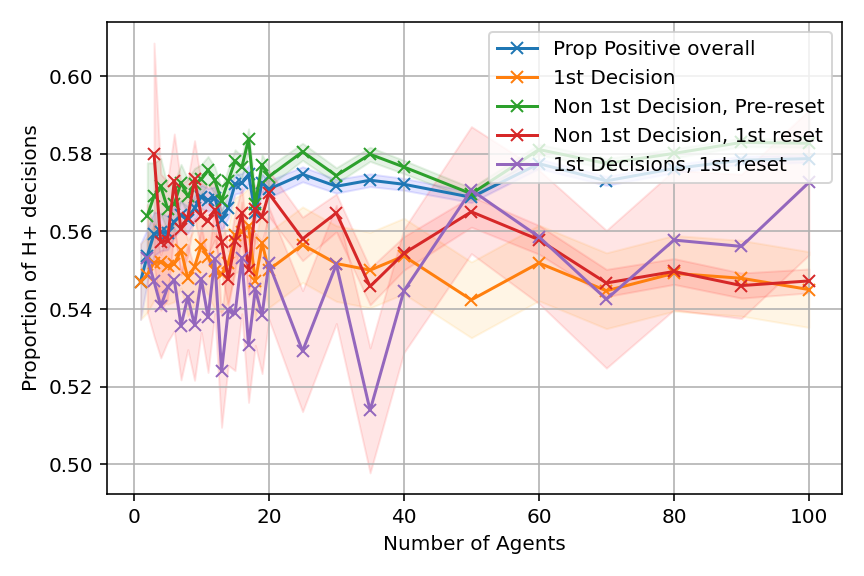}\label{fig:AccuracyPostReset1stResetTheta}}
		\end{subfigure}  
        \begin{subfigure}[][]{\includegraphics[width=0.45\textwidth]{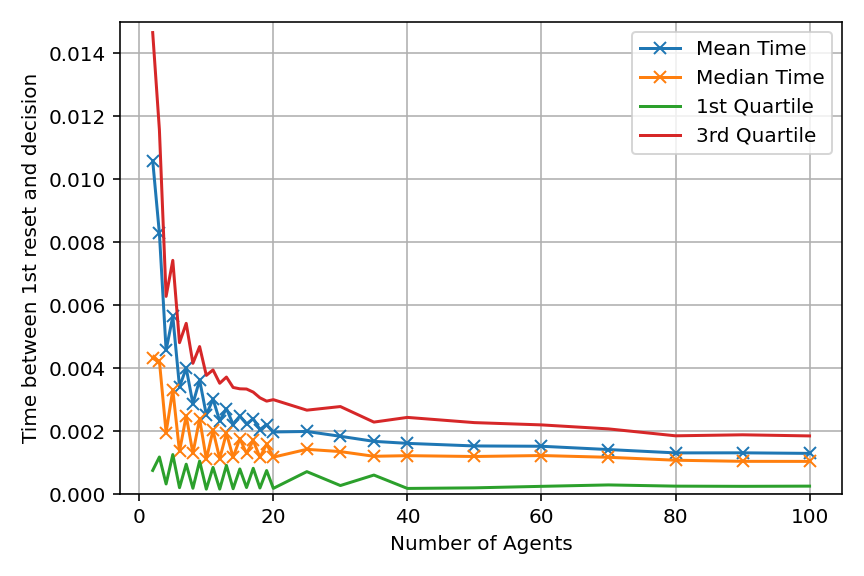}\label{fig:TimeBetweenResetsTheta}}
		\end{subfigure} 
\\
     \begin{subfigure}[][]{\includegraphics[width=0.45\textwidth]{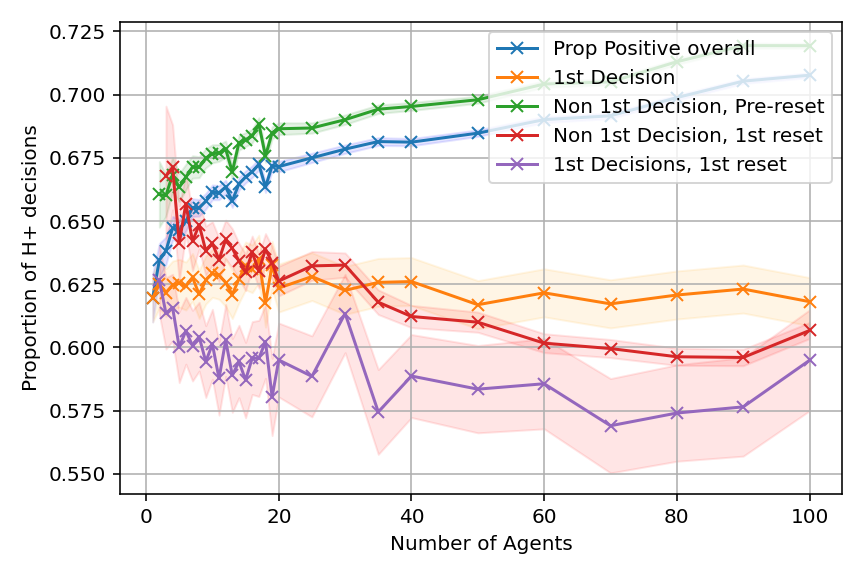}\label{fig:AccuracyPostReset1stResetAlpha}}
		\end{subfigure}  
        \begin{subfigure}[][]{\includegraphics[width=0.45\textwidth]{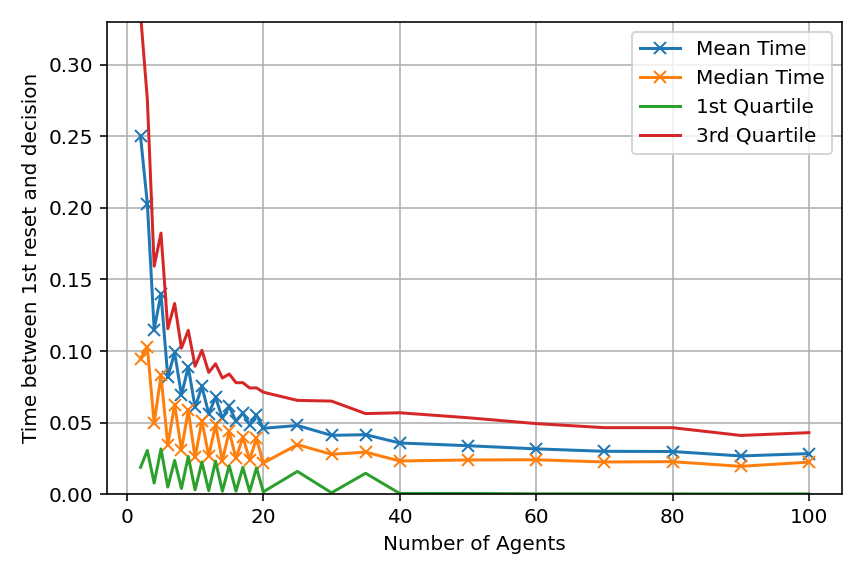}\label{fig:TimeBetweenResetsAlpha}}
		\end{subfigure} 
        \\
             \begin{subfigure}[][]{\includegraphics[width=0.45\textwidth]{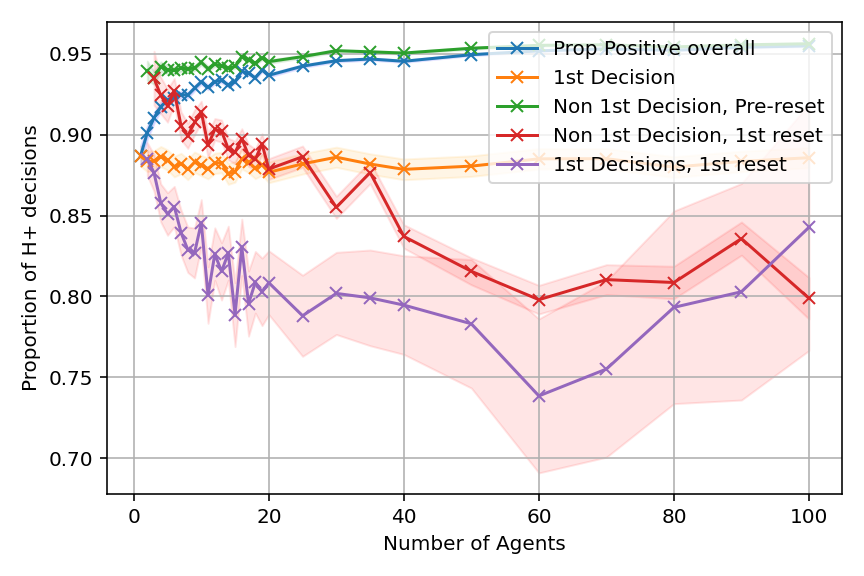}\label{fig:AccuracyPostReset1stResetD}}
		\end{subfigure}  
        \begin{subfigure}[][]{\includegraphics[width=0.45\textwidth]{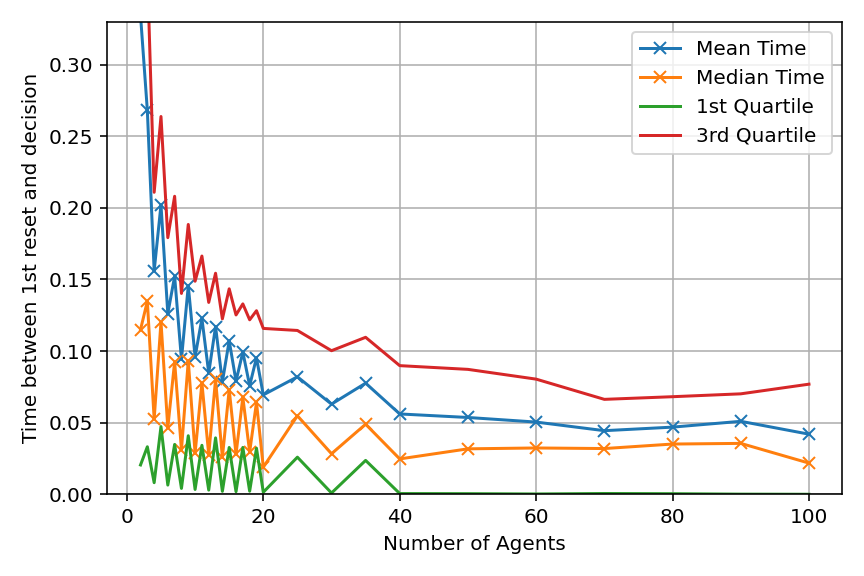}\label{fig:TimeBetweenResetsD}}
		\end{subfigure} 
    \caption{First decision after reset: (a),(c)\&(e) The accuracy of this first decision of reset and all other undecideds. (b),(d)\&(e) Time between the reset and this first decision after reset. In (b), the 1st quartile (green line) is very small but non-zero for larger, even $N$ (between $t=10^{-3}$ and $t=10^{-4}$. In (a)\&(b) $\theta=0.2$, (c)\&(d) $\alpha=0.5$ and (e)\&(f) $D=0.5$.).
}
    \label{fig:PostReset}
\end{figure}

\newpage
\section{Further Figures}

This document contains useful figures that support the main document, including results for other parameter values.

\begin{figure}[h]
    \centering    
        \begin{subfigure}[][]{\includegraphics[width=0.45\textwidth]{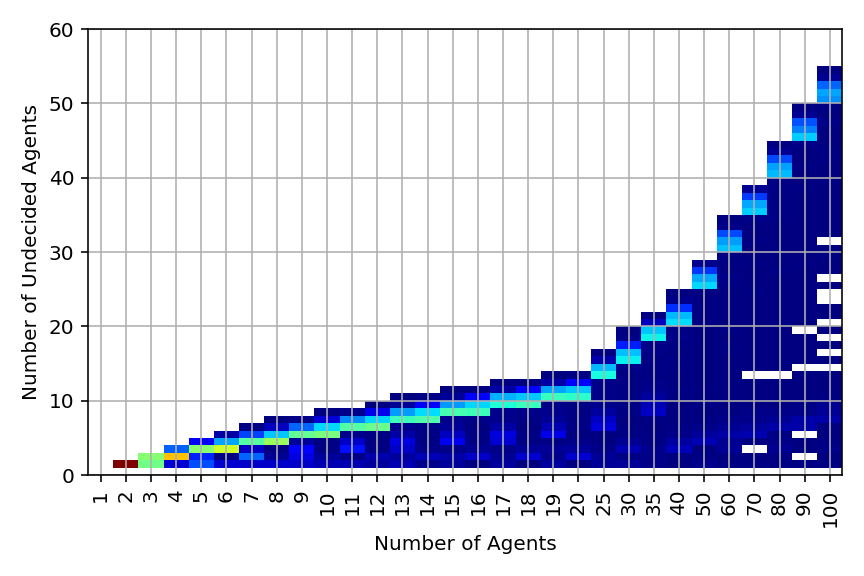}\label{fig:HeatmapUndecidedTheta1}}
		\end{subfigure}  
        \begin{subfigure}[][]{\includegraphics[width=0.45\textwidth]{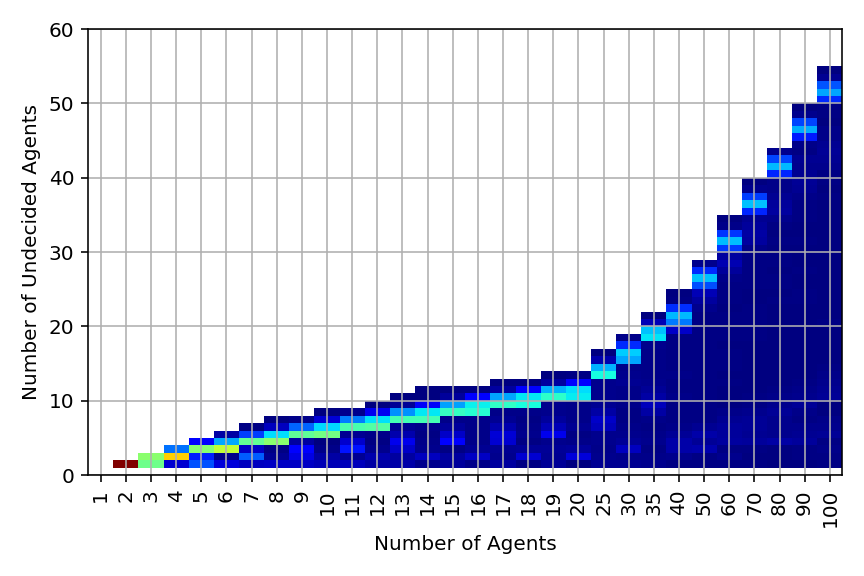}\label{fig:HeatmapUndecidedTheta0p2}}
		\end{subfigure} 
        \\
        \begin{subfigure}[][]{\includegraphics[width=0.45\textwidth]{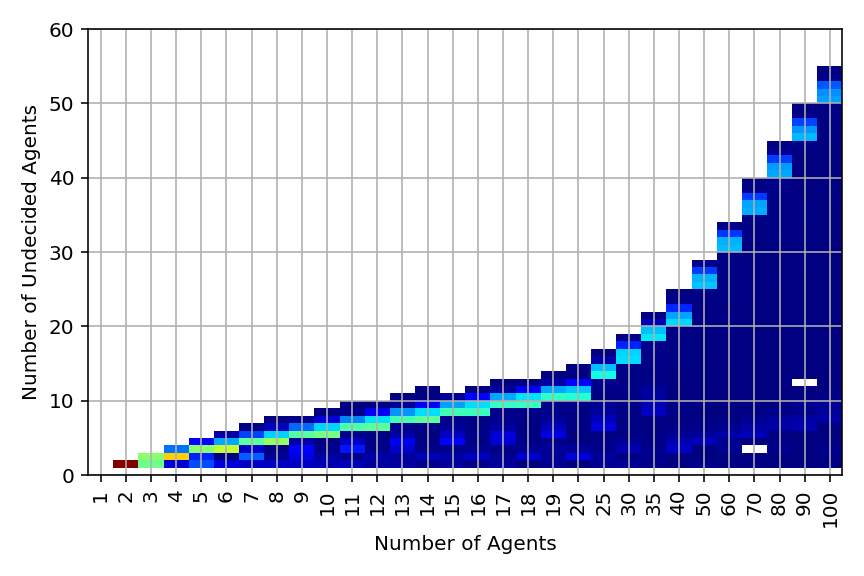}\label{fig:HeatmapUndecidedAlpha0p51}}
		\end{subfigure}  
        \begin{subfigure}[][]{\includegraphics[width=0.45\textwidth]{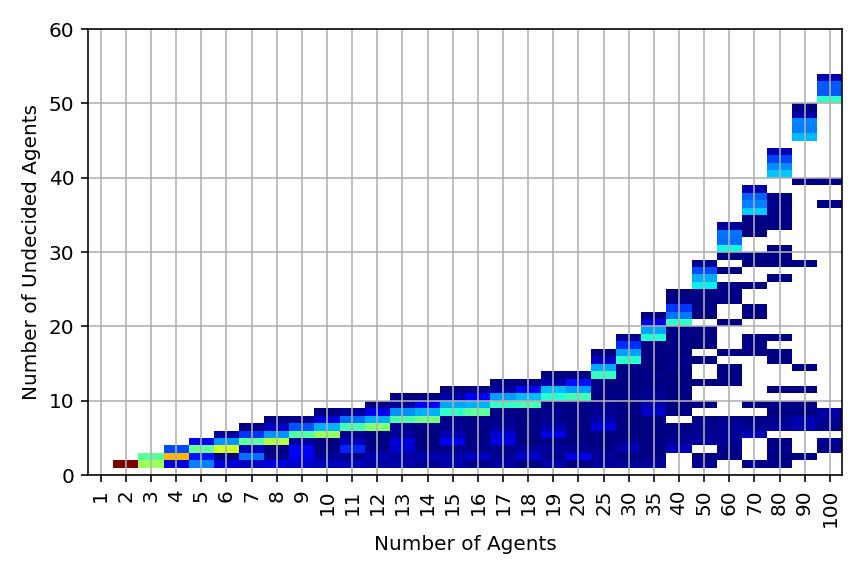}\label{fig:HeatmapUndecidedDiff0p5}}
		\end{subfigure}      
    \caption{Heatmaps correspond to the number of undecided agents on undecided agents across selected group sizes over 10000 simulations,weighted relative to count by Number of Agents (so vertically slices have the same total). In (a) $\theta=1$, (b) $\theta=0.2$, (c) $\alpha=0.5$ and (d) $D=0.5$. Histograms governing select group sizes ($N=5,10,15,20,50$, all unweighted) available in Figure~7(a) in the main text (for $\theta=1$) and Figure~\ref{fig:HistUndecidedOther}(a),(c)\&(e) within this Supplementary Material.  
    }
    \label{fig:HeatmapUndecidedOther}
\end{figure}

\begin{figure}
    \centering
         \begin{subfigure}[][]{\includegraphics[width=0.45\textwidth]{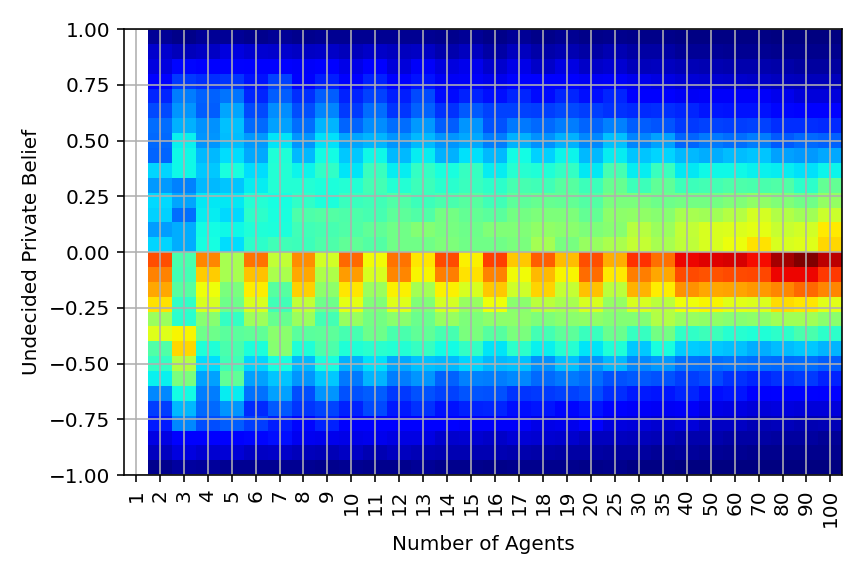}\label{fig:HeatmapUndecidedPrivateBeliefWeightedVertTheta1}}
		\end{subfigure}  \begin{subfigure}[][]{\includegraphics[width=0.45\textwidth]{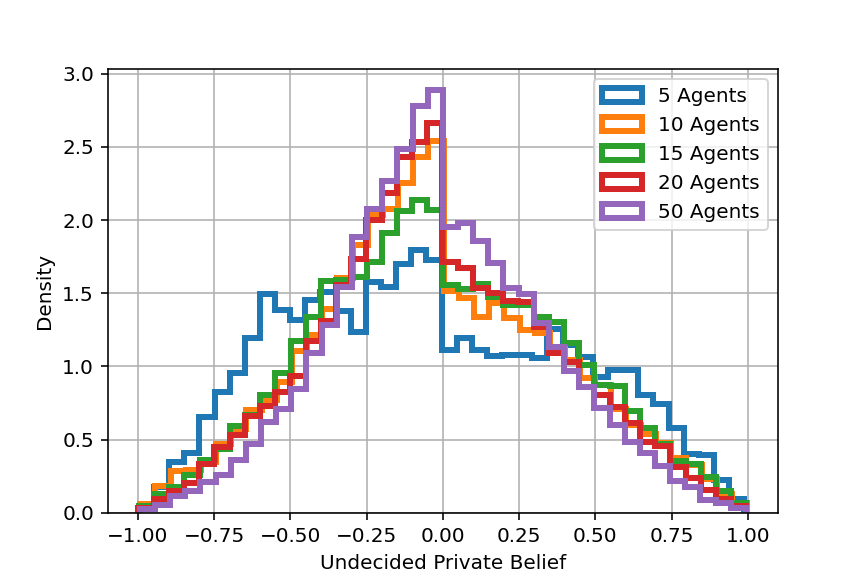}\label{fig:HistUndecidedPrivateBeliefWeightedVertTheta1}}
		\end{subfigure} 
         \\
         \begin{subfigure}[][]{\includegraphics[width=0.45\textwidth]{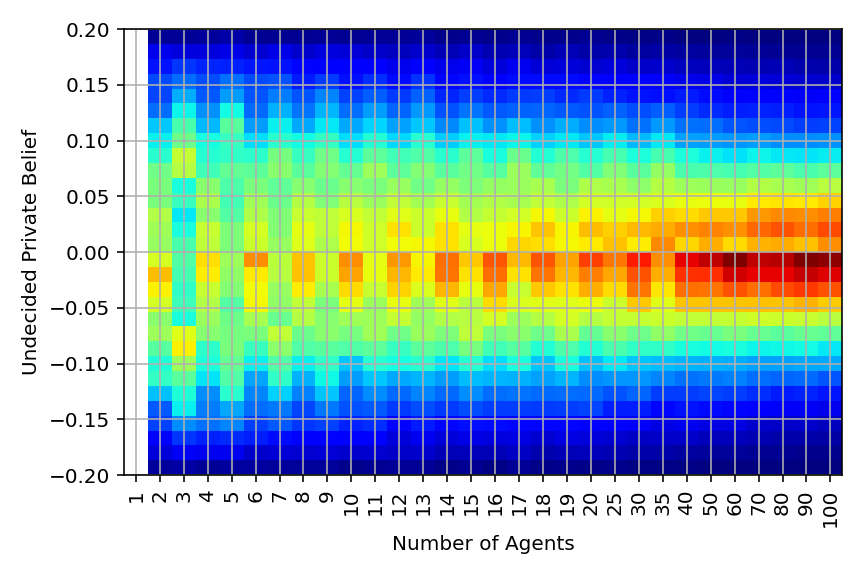}\label{fig:HeatmapUndecidedPrivateBeliefWeightedVertTheta0p2}}
		\end{subfigure}
        \begin{subfigure}[][]{\includegraphics[width=0.45\textwidth]{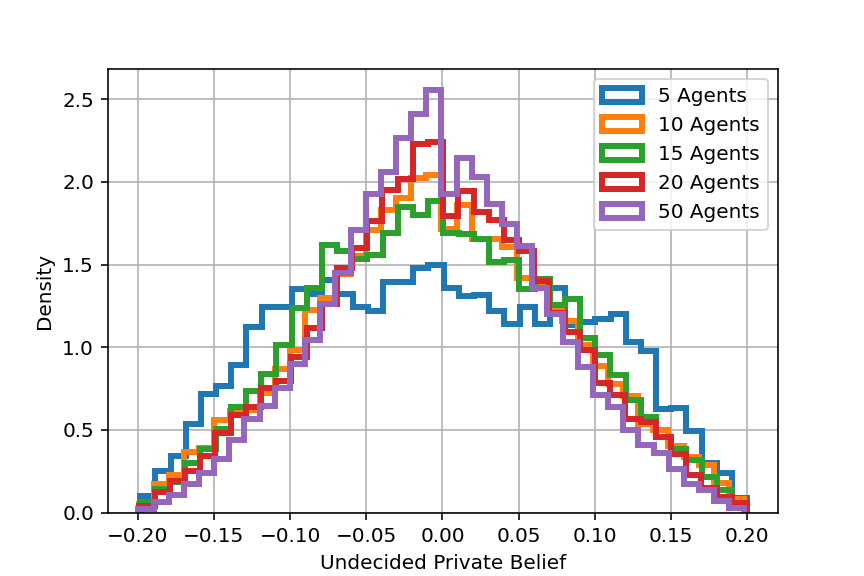}\label{fig:HistUndecidedPrivateBeliefWeightedVertTheta0p2}}
		\end{subfigure}
        \\
        \begin{subfigure}[][]{\includegraphics[width=0.45\textwidth]{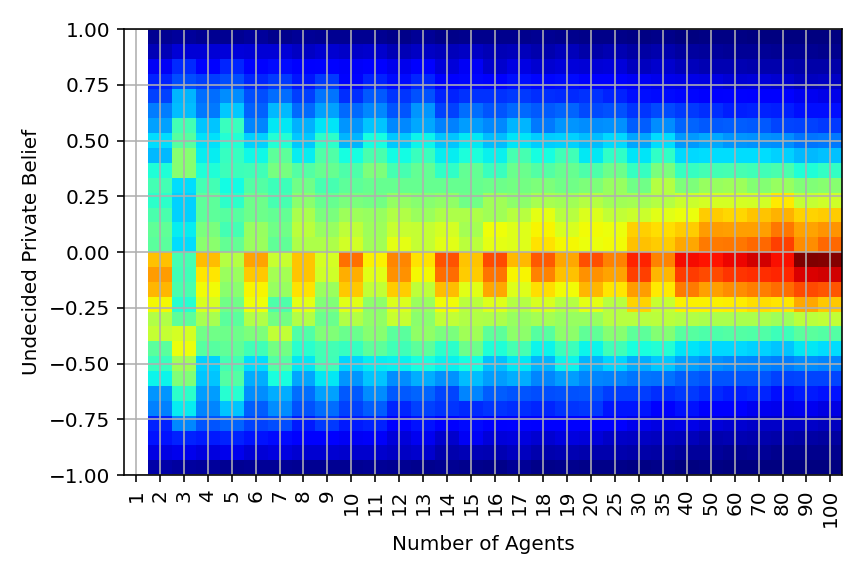}\label{fig:HeatmapUndecidedPrivateBeliefWeightedVertalpha0p5}}
		\end{subfigure}
        \begin{subfigure}[][]{\includegraphics[width=0.45\textwidth]{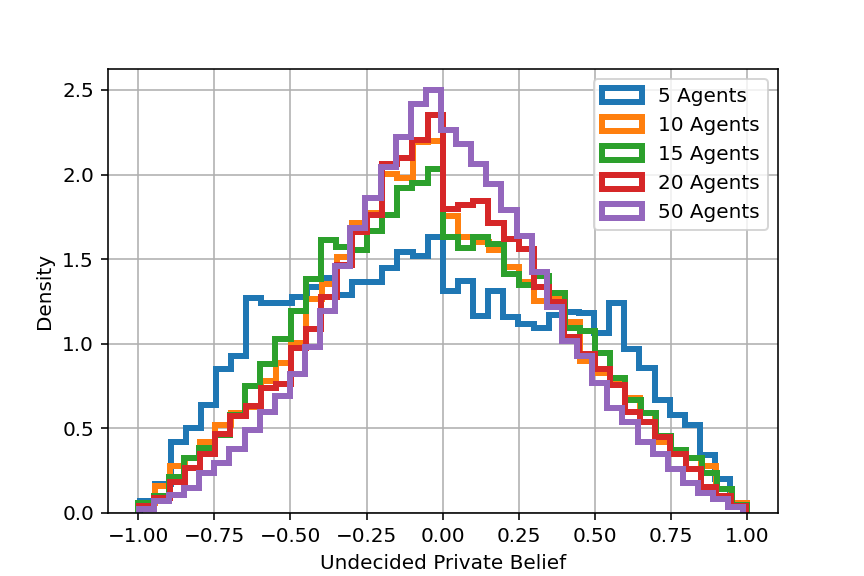}\label{fig:HistUndecidedPrivateBeliefWeightedVertalpha0p5}}
		\end{subfigure}
        \\
        \begin{subfigure}[][]{\includegraphics[width=0.45\textwidth]{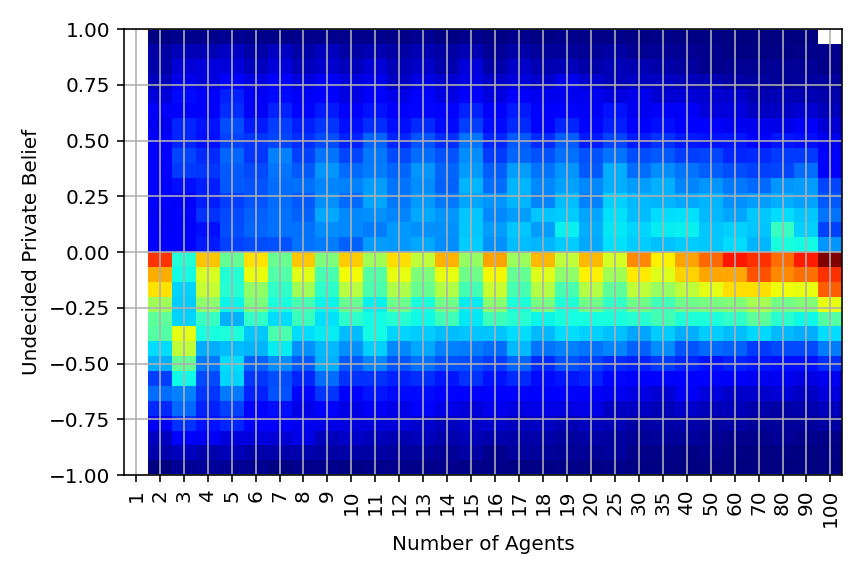}\label{fig:HeatmapUndecidedPrivateBeliefWeightedVertDiff0p5}}
		\end{subfigure}
        \begin{subfigure}[][]{\includegraphics[width=0.45\textwidth]{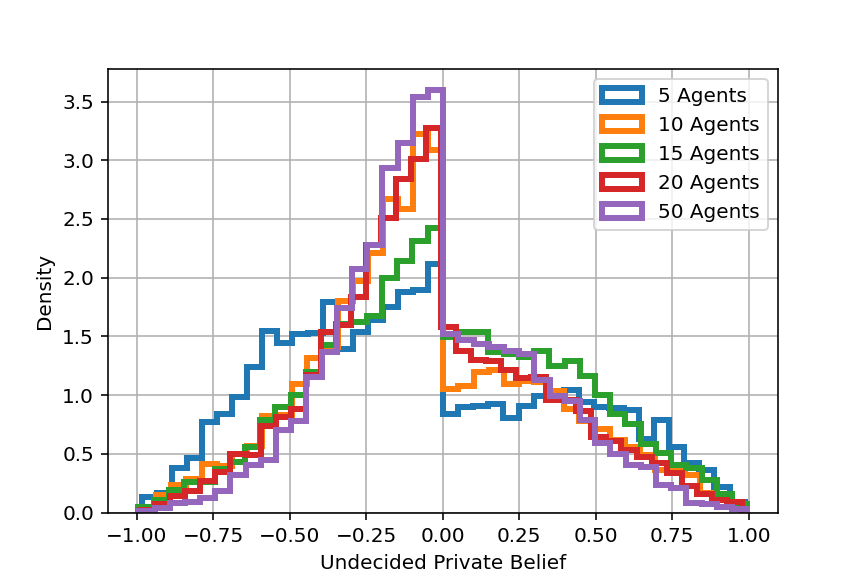}\label{fig:HistUndecidedPrivateBeliefWeightedVertDiff0p5}}
		\end{subfigure}

    \caption{(a),(c),(e)\&(g) Heatmaps of private belief
    weighted relative to count by Number of Agents. (b),(d),(f)\&(h) Corresponding histograms for selected group sizes, making it clear what the vertical slices in in the heatmaps are. (a)\&(b) $\theta=1$, (b)\&(d) $\theta=0.2$, (e)\&(f)$\alpha=0.5$ and (g)\&(h) $D=0.5$. Note, all agents at the first decision should be distributed as $p_+(t)$, like Figure~2(a) in the main paper.  $p_+(t)$ has a peak somewhere above zero, whereas these histograms have peaks just below zero and have large drops around zero; this drop is less clear for odd $N$ and for $\theta=0.2$. Distributions are peaker/narrower for larger $N$ since decision times are faster. 
    }
    \label{fig:HeatmapUndecidedPrivateBelief}
\end{figure}

\begin{figure}
    \centering
       \begin{subfigure}[][]{\includegraphics[width=0.45\textwidth]{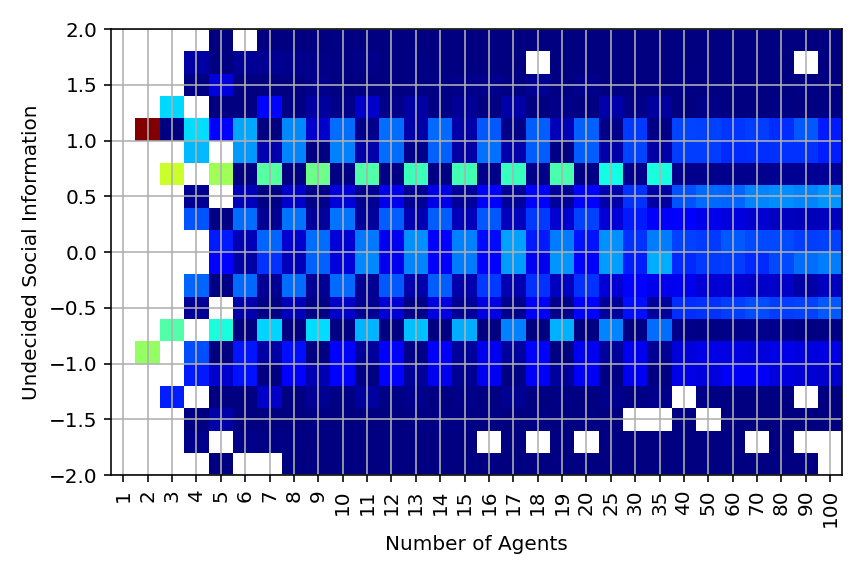}\label{fig:HeatmapUndecidedSocialBeliefWeightedVertTheta1}}
		\end{subfigure}  
              \begin{subfigure}[][]{\includegraphics[width=0.45\textwidth]{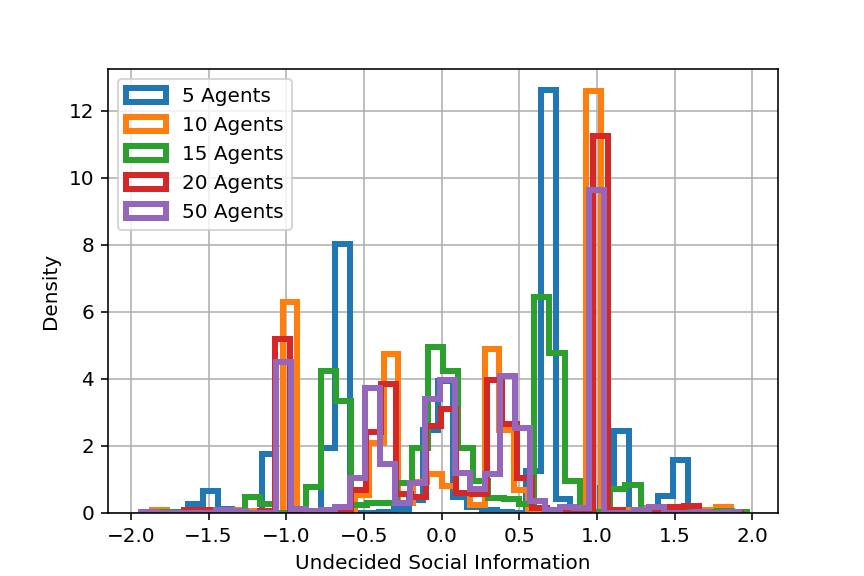}\label{fig:HistUndecidedSocialBeliefWeightedVertTheta1}}
		\end{subfigure}  \\
        \begin{subfigure}[][]{\includegraphics[width=0.45\textwidth]{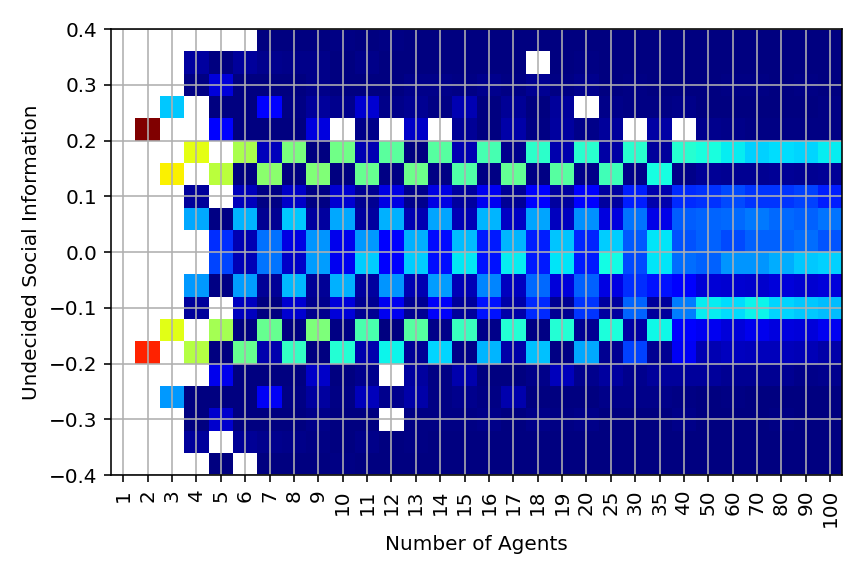}\label{fig:HeatmapUndecidedSocialBeliefWeightedVertTheta0p2}}
		\end{subfigure}      
        \begin{subfigure}[][]{\includegraphics[width=0.45\textwidth]{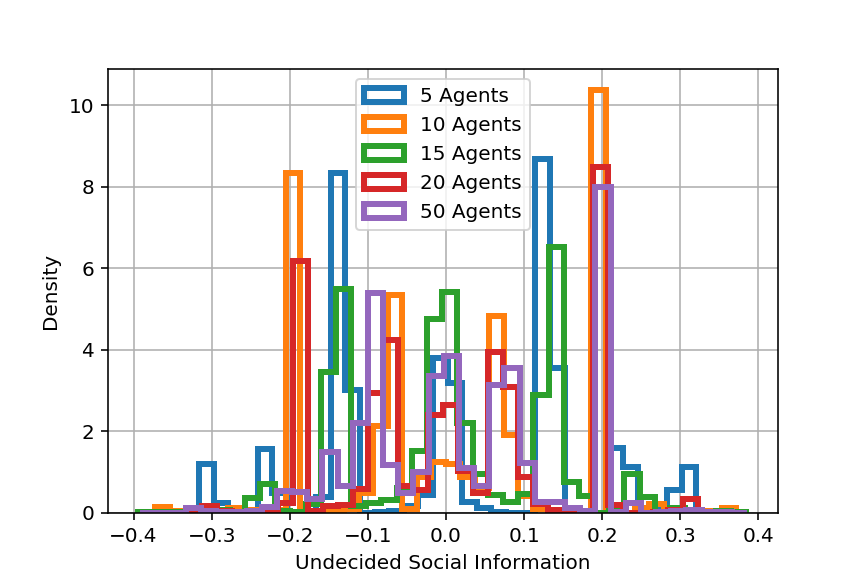}\label{fig:HistUndecidedSocialBeliefWeightedVertTheta0p2}}
		\end{subfigure} 
        \\     
        \begin{subfigure}[][]{\includegraphics[width=0.45\textwidth]{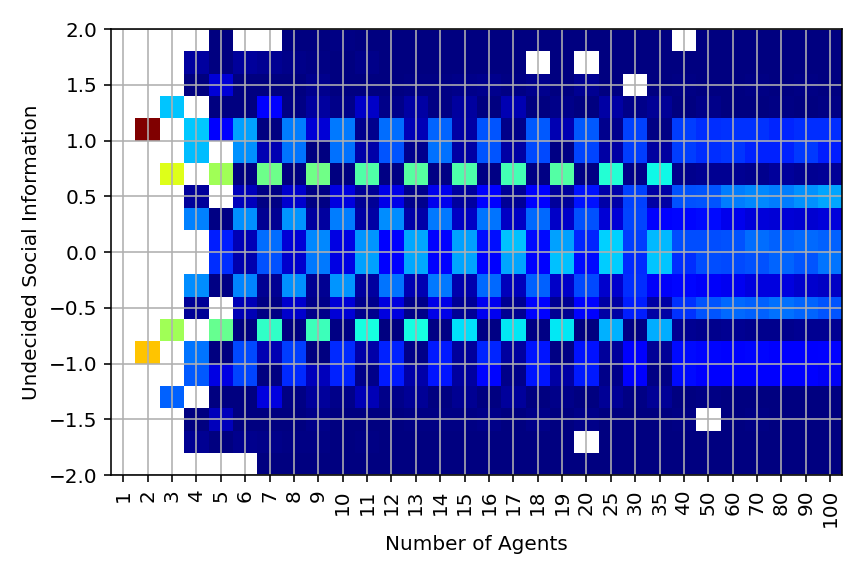}\label{fig:HeatmapUndecidedSocialBeliefWeightedVertAlpha0p5}}
		\end{subfigure}      
        \begin{subfigure}[][]{\includegraphics[width=0.45\textwidth]{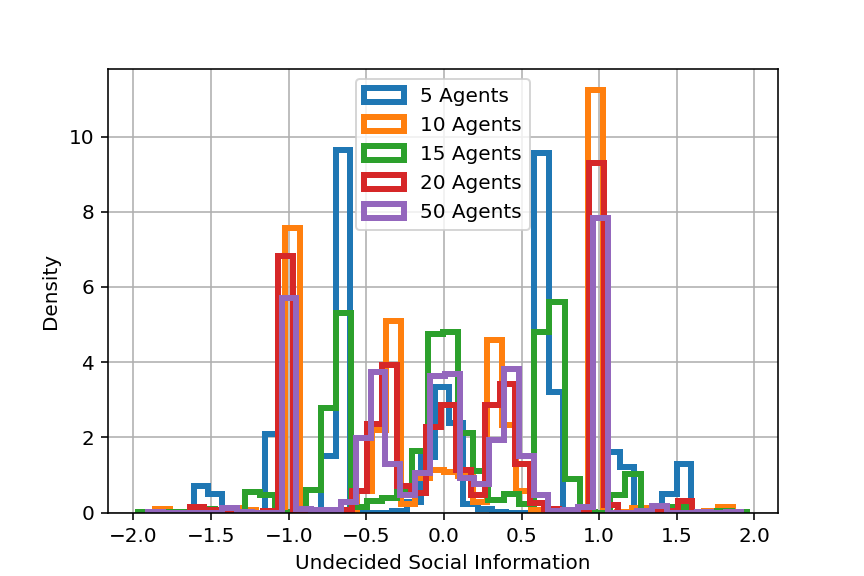}\label{fig:HistUndecidedSocialBeliefWeightedVertAlpha0p5}}
		\end{subfigure} 
        \\
        \begin{subfigure}[][]{\includegraphics[width=0.45\textwidth]{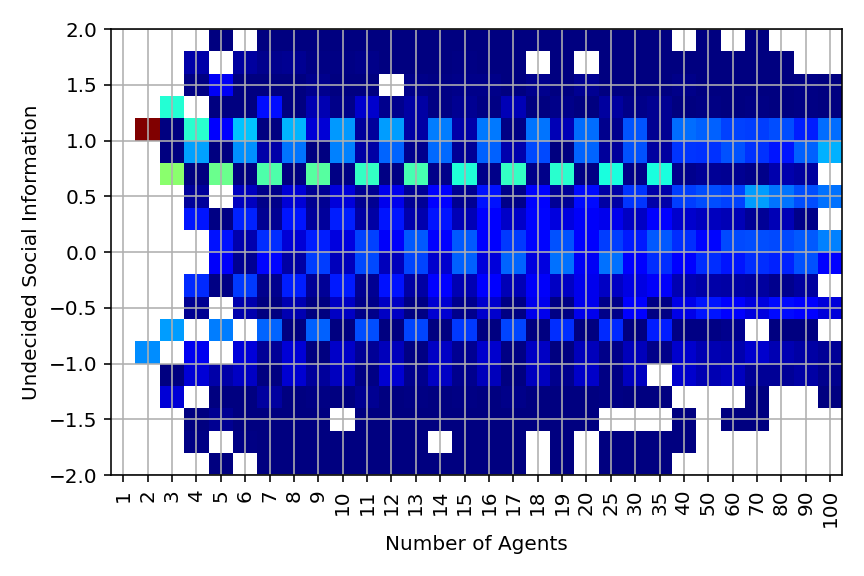}\label{fig:HeatmapUndecidedSocialBeliefWeightedVertDiff0p5}}
		\end{subfigure}      
        \begin{subfigure}[][]{\includegraphics[width=0.45\textwidth]{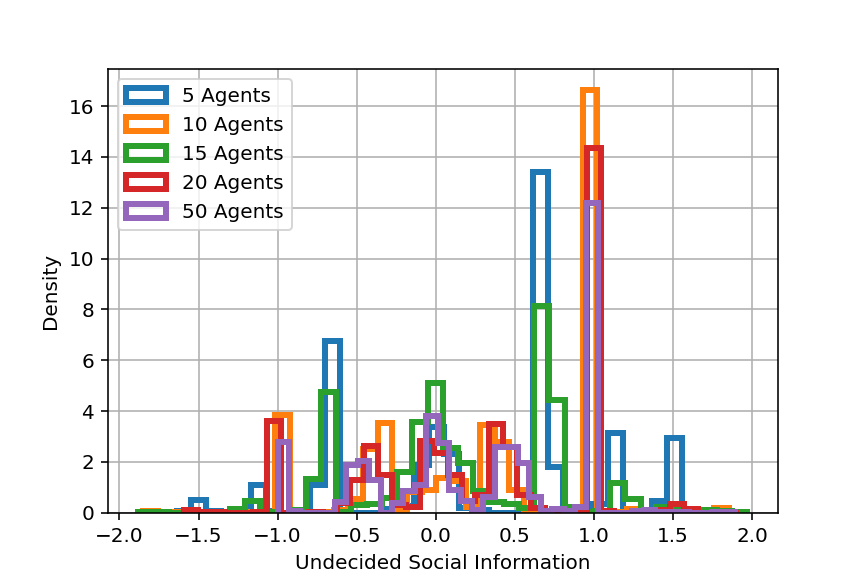}\label{fig:HistUndecidedSocialBeliefWeightedVertDiff0p5}}
		\end{subfigure}         
    \caption{(a),(c),(e)\&(g) Heatmaps of received Social Information of Undecided agents weighted relative to count by Number of Agents (so vertically slices have the same total). (b),(d),(f)\&(h) Corresponding histograms for selected group sizes. (a)\&(b) $\theta=1$, (b)\&(d) $\theta=0.2$, (e)\&(f)$\alpha=0.5$ and (g)\&(h) $D=0.5$. An odd/even $N$ pattern exists: even $N$ having similar peaks around $\pm\theta$ shifted by (in)decision majorities from wave 1 of $0,\pm2,\pm4...$); odd $N$ have peaks corresponding to $\pm\theta$ then majorities from wave 1 of $\pm 1,\pm3,\pm5,$... In particular, for even $N$, the peaks around $\pm1$ correspond to $H_\pm$ decision followed by a tied wave 1 (with small amounts of $H_\mp$ followed by a $\pm 6$ or $\pm8$ majority). White areas have zero instances within the 10000 simulations. }   \label{fig:HeatmapUndecidedSocialBelief}
\end{figure}

\begin{figure}
    \centering
        \begin{subfigure}[][]{\includegraphics[width=0.45\textwidth]{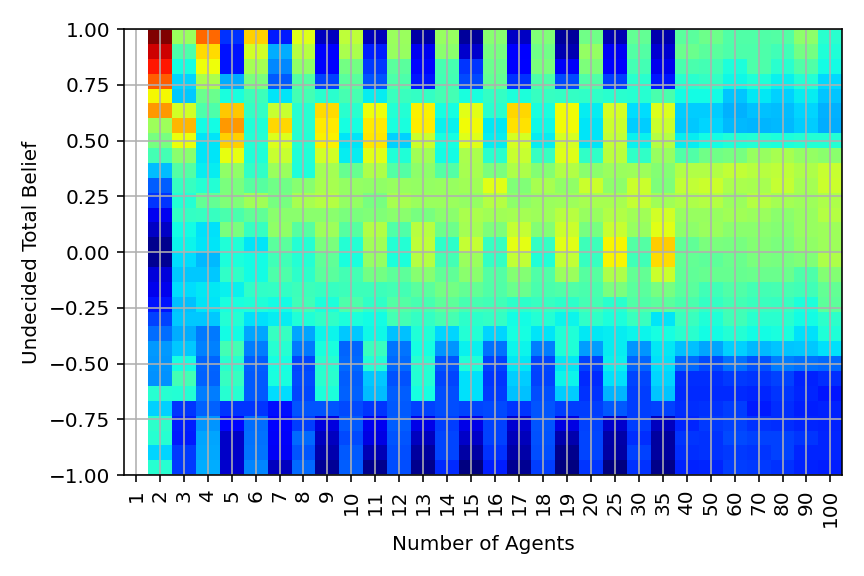}\label{fig:HeatmapUndecidedTotalBeliefWeightedVertTheta1}}
		\end{subfigure}  
        \begin{subfigure}[][]{\includegraphics[width=0.45\textwidth]{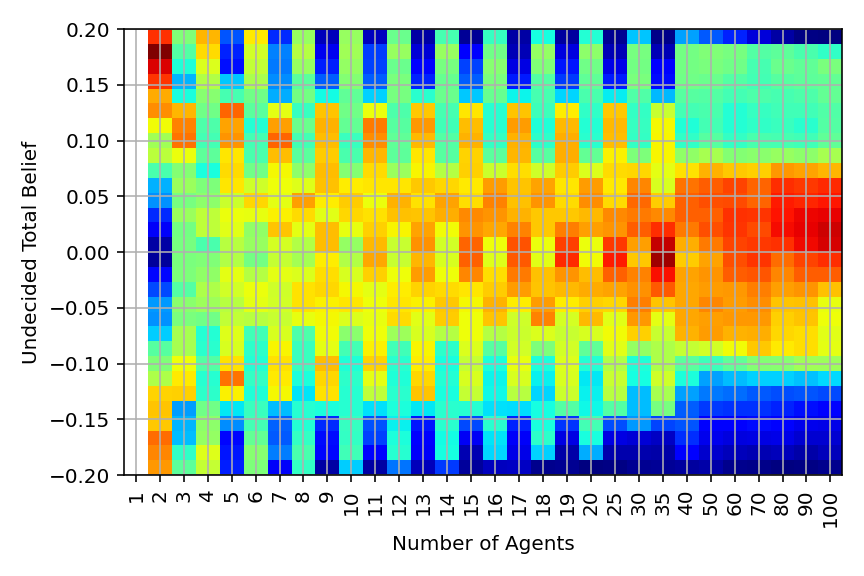}\label{fig:HeatmapUndecidedTotalBeliefWeightedVertTheta0p2}}
		\end{subfigure} \\
     \begin{subfigure}[][]{\includegraphics[width=0.45\textwidth]{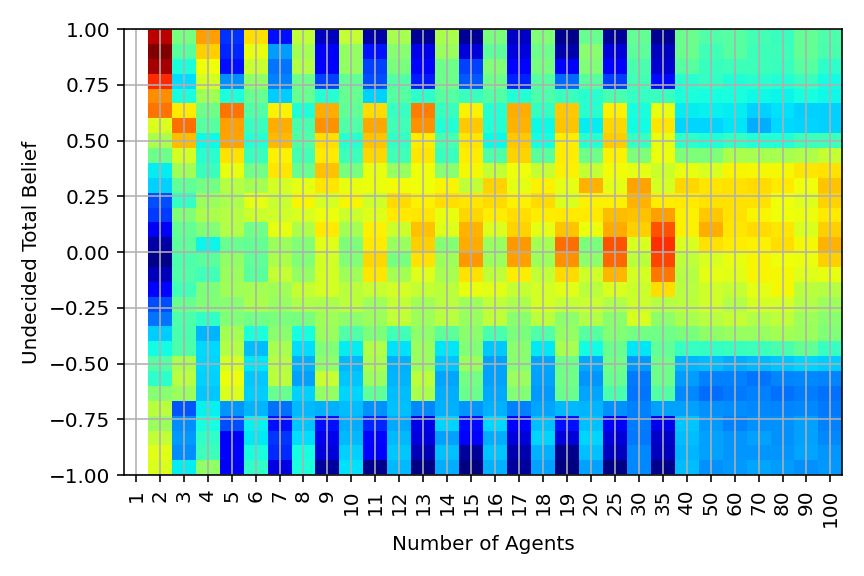}\label{fig:HeatmapUndecidedTotalBeliefWeightedVertalpha0p5}}
		\end{subfigure} 
         \begin{subfigure}[][]{\includegraphics[width=0.45\textwidth]{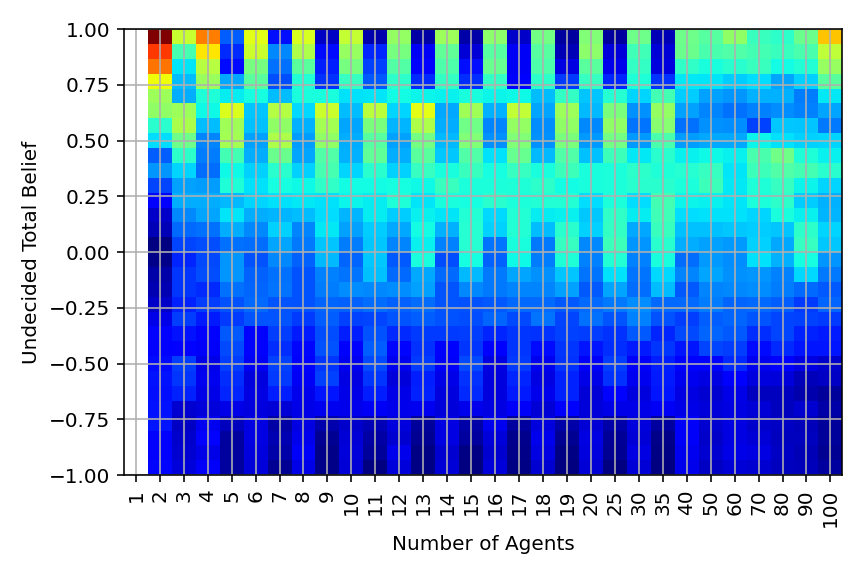}\label{fig:HeatmapUndecidedTotalBeliefWeightedVertDiff0p5}}
		\end{subfigure} \\    
    \caption{Heatmaps of Total belief of Undecided agents for (a) $\theta=1$, (b) $\theta=0.2$, (c) $\alpha=0.5$ and (d) $D=0.5$. 
    The heatmap is weighted relative to count by Number of Agents (so vertically slices have the same total). There are corresponding histograms for select group sizes/vertical slices ($N=5,10,15,20,50$) in Figure~7(b) in main paper (for (a)) and Figure~\ref{fig:HistUndecidedOther} within this Supplementary Material (for (b)-(d)).}
    \label{fig:HeatmapUndecidedTotalBelief}
\end{figure}